\documentclass[10pt]{article}
\usepackage{geometry}
\usepackage{graphicx}
\usepackage{amssymb}
\usepackage{epstopdf}
\usepackage[section]{placeins}
\usepackage[square,authoryear]{natbib}
\title{East-west faults due to planetary contraction}

\author{Mikael Beuthe\\
\it Royal Observatory of Belgium,\\
\it Avenue Circulaire 3, 1180 Brussels, Belgium\\
\it E-mail: mbeuthe@oma.be}
\date{June 12, 2010} 

\begin{document}

\maketitle

\begin{abstract}
Contraction, expansion and despinning have been common in the past evolution of Solar System bodies.
These processes deform the lithosphere until it breaks along faults.
Their characteristic tectonic patterns have thus been sought for on all planets and large satellites with an ancient surface.
While the search for despinning tectonics has not been conclusive, there is good observational evidence on several bodies for the global faulting pattern associated with contraction or expansion, though the pattern is seldom isotropic as predicted.
The cause of the non-random orientation of the faults has been attributed either to regional stresses or to the combined action of contraction/expansion with another deformation (despinning, tidal deformation, reorientation).
Another cause of the mismatch may be the neglect of the lithospheric thinning at the equator or at the poles due either to latitudinal variation in solar insolation or to localized tidal dissipation.
Using thin elastic shells with variable thickness, I show that the equatorial thinning of the lithosphere transforms the homogeneous and isotropic fault pattern caused by contraction/expansion into a pattern of faults striking east-west, preferably formed in the equatorial region.
By contrast, lithospheric thickness variations only weakly affect the despinning faulting pattern consisting of equatorial strike-slip faults and polar normal faults.
If contraction is added to despinning, the despinning pattern first shifts to thrust faults striking north-south and then to thrust faults striking east-west.
If the lithosphere is thinner at the poles, the tectonic pattern caused by contraction/expansion consists of faults striking north/south.
I start by predicting the main characteristics of the stress pattern with symmetry arguments.
I further prove that the solutions for contraction and despinning are dual if the inverse elastic thickness is limited to harmonic degree two, making it easy to determine fault orientation for combined contraction and despinning.
I give two methods for solving the equations of elasticity, one numerical and the other semi-analytical.
The latter method yields explicit formulas for stresses as expansions in Legendre polynomials about the solution for constant shell thickness.
Though I only discuss the cases of a lithosphere thinner at the equator or at the poles, the method is applicable for any latitudinal variation of the lithospheric thickness.
On Iapetus, contraction or expansion on a lithosphere thinner at the equator explains the location and orientation of the equatorial ridge.
On Mercury, the combination of contraction and despinning makes possible the existence of zonal provinces of thrust faults differing in orientation (north-south or east-west), which may be relevant to the orientation of lobate scarps.
\end{abstract}


{\it
\noindent
The final version of this preprint is published in Icarus (www.elsevier.com/locate/icarus)\\
doi:10.1016/j.icarus.2010.04.019}

\section{Introduction}

A great variety of tectonic features is found on nearly all solid planets and large satellites of the Solar System: ridges and scarps, rifts and grabens, furrows and grooves etc.
Their origin on Earth mainly lies in the movement of nearly rigid plates but other mechanisms must be found elsewhere since plate tectonics is unique to Earth.
Though tectonic features can often be explained by a regional effect, such as an impact or the emplacement of a volcanic load, a subclass of them sometimes form a global pattern on the surface.
In such a case, their cause must be the global deformation of the whole lithosphere which generates a global stress distribution resulting in a characteristic faulting pattern at the surface.

Several processes affect the planetary figure.
Though a change in the mean planetary radius (contraction or expansion) is the simplest conceivable deformation,
its underlying causes can be complex since they depend on the complicated physics of the interior of a self-gravitating differentiated body (e.g. \citet{andrewshanna2008} for Mars, \citet{squyres1986} and \citet{collins2009} for icy satellites).
On a planet with a lithosphere of constant thickness, contraction and expansion bring about a homogeneous distribution of randomly oriented compressional and extensional faults, respectively \citep{melosh1977}.
The surfaces of Mars \citep{anderson2001,anderson2008,knapmeyer2006} and Mercury \citep{watters2009} show widespread compressional tectonic features, termed wrinkle ridges and lobate scarps.
Their distribution is however far from uniform and their orientation is not random.
On Mars, regional effects such as the Tharsis rise have strongly influenced the global pattern of wrinkle ridges and the possible role of a global contraction event remains under discussion \citep{mangold2000,nahm2010}.
On Mercury, the lobate scarps that have been identified in Mariner~10 images have a greater cumulative length in the southern latitudes, generally trend within $50^\circ$ of the north-south direction and preferably dip northward below $50^\circ\rm S$ \citep{watters2004,wattersnimmo2009}.
The anisotropy of lobate scarps is often attributed to the additional contribution of despinning (see below), but the regional effect of the Caloris basin has also been invoked \citep{thomas1988b}.
A more complete and less illumination-biased fault catalog should result from the analysis of Messenger images and altimetry data; results from the first flyby are promising \citep{solomon2008,watters2009}.
Among large icy satellites, Ganymede is a showcase for extensional tectonics \citep{pappalardo2004}.
Global expansion is also thought to have played a role in the formation of the global tectonic grids on Rhea \citep{consolmagno1985,moore1985,thomas1988a}, Dione \citep{moore1984,consolmagno1985} and Ariel \citep{plescia1987,nyffenegger1997}, though the orientation of faults has been shown to be non-random in each case.
Global contraction may also have occurred on Rhea and Dione.
New tectonic analyses of Rhea \citep{moore2007,wagner2007} and Dione \citep{goffpochat2009,wagner2009,stephan2010} are underway with Cassini data.
While partial imaging makes it difficult to identify a global tectonic grid on Iapetus \citep{singer2008}, its huge equatorial ridge must be related to a global deformation.
Contraction \citep{castillo2007} has been cited as a possible culprit, but the corresponding faulting pattern neither predicts the equatorial location nor the east-west orientation of the ridge.

The next simplest deformation is a change in the planetary flattening due to a decrease in the rotation rate, or despinning, that results from tidal effects leading to a resonant or synchronous rotation \citep{murraydermott}.
Despinning tectonics on a planet with a thin lithosphere of constant thickness consist of an equatorial zone of strike-slip faults (striking at about 60 degrees from the north) and of polar zones of extensional faults or joints striking east-west \citep{burns1976,melosh1977}.
Despun bodies are common in the Solar System: Mercury is in resonant rotation with the Sun \citep{goldreich1968} and nearly all large satellites in the Solar System are in synchronous rotation with their parent body \citep{peale1999}.
Simultaneous despinning and contraction have been used to justify the predominant north-south orientation of lobate scarps on Mercury \citep{melosh1978,pechmann1979,dombardhauck2008}.
Beside the pattern of young lobate scarps, Mercury exhibits a global grid of more ancient lineaments which \citet{melosh1978} and \citet{pechmann1979} interpreted as evidence of despinning (plus contraction), though the case is not closed  \citep{melosh1988,thomas1988b}.
While despinning is a generic phenomenon for large icy satellites, none exhibits unambiguous evidence of the corresponding global tectonic pattern, either because of later resurfacing or because despinning occurred before the surface was fully formed \citep{squyres1986}.
The location of Iapetus' ridge suggests a relation with despinning \citep{porco2005,castillo2007} but it was immediately noted that despinning tectonics cannot produce east-west features at the equator.
Global tectonic patterns can also be produced by other types of deformations not examined in this paper: reorientation relative to the spin axis \citep{melosh1980b,leith1996,matsuyama2008}, recession \citep{melosh1980a,helfenstein1983}, diurnal tides and non-synchronous rotation \citep{helfenstein1985, greenberg1998,wahr2009}.
These deformations can be seen as a superposition of biaxial deformations and thus share the basics of the despinning model.

Up to now the distribution and orientation of tectonic features have always been computed with the assumption of a lithosphere of constant thickness, one reason being the lack of data about the lithospheric thickness and its variations.
The choice of a specific variation of lithospheric thickness may thus seem ad hoc if its cause is a variation in the internal heating flux.
Two external phenomena provide a better motivation for a model of lithospheric thickness variation.
On the one hand, the latitudinal variation in the solar insolation elevates the lithospheric isotherm at the equator of a planet without atmosphere.
For Mercury, the lithosphere could be thinner by a factor of two at the equator in comparison with the poles \citep{mckinnon1981,melosh1988};
a longitude variation is also expected because of the difference in average temperature between the so-called equatorial hot and warm poles.
On the other hand, tidal heating acting on the satellites with an eccentric orbit creates additional lateral variations in the lithospheric thickness.
\citet{nimmo2007} however found no evidence in the global shape of Europa for the thickness variations predicted by \citet{ojakangas1989} (see also \citet{tobie2003}).
Since large satellites are at present in a synchronous state of rotation, tidal heating is localized in longitude.
Nevertheless tidal heating is evenly distributed in longitude before tidal locking is achieved, so that the lithospheric thickness only depends on latitude during the despinning phase.

The goal of this paper is to predict global tectonic patterns due to contraction and despinning on a lithosphere with latitudinal thickness variation.
Though I will assume in the numerical examples that the thickness variation is symmetrical about the equatorial plane, the method is also valid for non-symmetrical variations.
The prediction of tectonic patterns follows the following procedure.
First the lithosphere is modeled as a thin elastic spherical shell \citep{turcotte1981,beuthe2008}.
Second the contraction or despinning is represented as a static deformation of the shell given by the difference between the initial and final figure of the planet.
The stresses caused by the deformation of the thin elastic spherical shell are computed with the equations of equilibrium and elasticity.
Third the style and orientation of faults is predicted from the stresses with Anderson's theory of faulting \citep{jaeger}.

The most significant result of this paper concerns the contraction of a planet with a lithosphere thinner at the equator (predictions for expansion follow by changing the sign of the stresses).
If the contracting lithosphere is of constant thickness, stresses are homogeneous and isotropic, which means that their orientation and magnitude are independent of position and that the principal tangential stresses are equal.
The thinning of the lithosphere at the equator has two effects: (1) the stress becomes most compressive at the equator and (2) the meridional component of the stress becomes more compressive than the azimuthal component.
This situation, as well as the corresponding expansion case, favors the development of faults striking east-west and starting preferably at the equator.
The contraction of a shell thinner at the equator is thus complementary to despinning, for which the azimuthal stress is always more compressive than the meridional stress.
If the inverse thickness is at most of harmonic degree two I show that this complementarity is described by an exact relation between a contracting planet and a planet that is spinning up.
As for despinning, its tectonic pattern is not much affected by the variation of the lithospheric thickness: the main effect is the reduction in size of the polar provinces of tensional stress as the lithosphere thickens at the poles.
The combination of contraction and despinning renders possible the existence of provinces of thrust faults differing in their orientation (north-south or east-west), but this only occurs for a specific ratio between contraction and despinning.
If the lithosphere is thinner at the poles, global contraction (resp. expansion) leads to thrust (resp. normal) faults striking north-south and preferably formed in the polar regions, whereas the despinning pattern is weakly affected.

As for applications, I show that contraction or expansion combined with lithospheric thinning at the equator provides an explanation for the location and orientation of Iapetus' equatorial ridge.
I also attempt to account for the orientation of lobate scarps on Mercury.
Tectonic patterns resulting from simultaneous contraction and despinning on a lithosphere thinner at the equator are consistent with the latitudinal dependence of lobate scarps.
The amount of required contraction is however much larger than what has been estimated from observations, making it necessary to resort to more contrived scenarios involving fault reactivation.

The paper can be read at two levels: I present general results and geophysical interpretations in the main text, while the semi-analytical method of solution is described in detail in the appendices, which also include explicit solutions as well as a discussion of the relationship between thin shells, thick shells and Love numbers.
Section \ref{section2} states the membrane equations that must be solved in order to compute the stresses generated by the deformation of a thin elastic shell with variable thickness.
Remarkably, assumptions of axial and equatorial symmetry (presented in Section \ref{section3}) are sufficient for the prediction of the basic characteristics of stress and strain.
These symmetries lead to dualities between the contraction and despinning solutions when the inverse thickness is at most of degree two, reducing by half the computational load and simplifying the discussion of combined contraction and despinning.
I also compute a first order approximation of the contraction stresses from duality arguments.
In Section~\ref{section4}, I show how to solve the membrane equation.
After discussing the parameterization of the thickness variation, I frame the problem for a numerical solution with Mathematica, before explaining a semi-analytical method in which truncated Legendre expansions serve to rewrite the membrane equation as a matrix equation.
In Section~\ref{section5}, I use the contraction and despinning solutions of the membrane equation in order to predict the extent of tectonic provinces, i.e. areas with faults of given style and orientation.
The cases of expansion and spinning-up are related to contraction and despinning by a sign change.
In Section \ref{section6}, I discuss the application of the predicted tectonic patterns to Iapetus' ridge and Mercury's lobate scarps.

\section{Membrane stress in a deformed spherical shell}
\label{section2}

The style and orientation of faults due to contraction and despinning can be predicted from knowledge of the stresses in the lithosphere.
The lithosphere is modeled as a thin elastic spherical shell (in this paper, lithospheric thickness and elastic thickness are synonymous).
The important assumptions that the shell is thin and that it is elastic are analyzed by \citet{melosh1977} for a shell of constant thickness.

First, Melosh shows that models with a thin or a thick shell predict very similar despinning faulting patterns, except the possible presence of an equatorial thrust fault province when the shell is thick (the contraction tectonic pattern is not affected by the thickness of the shell).
Despinning stresses at the surface of a thick shell can easily be computed if the secular tidal Love numbers $h_2^T$ and $l_2^T$ are known (see Appendix \ref{Lovenumbers}).
At a given latitude, the stress magnitude mainly depends on $h_2^T$ whereas the relative size of the principal stresses depends on the ratio $l_2^T/h_2^T$.
When the shell thickness increases, $h_2^T$ decreases so that despinning stresses decrease in magnitude.
While Melosh computes thin shell stresses under the assumption of hydrostatic flattening, \citet{matsuyama2008} extend the domain of application of thin shell formulas by using a non-hydrostatic flattening parameterized by $h_2^T$.
This latter procedure is equivalent to keeping the overall $h_2^T$-dependence in thick shell formulas while setting the ratio $l_2^T/h_2^T$ to its membrane value $(1+\nu)/(5+\nu)$, where $\nu$ is Poisson's ratio (see Appendix \ref{Lovenumbers}).

Second, Melosh argues that the faulting pattern predicted by the elastic model is not substantially modified when the lithosphere becomes plastic, though the numerical values of the stresses are bounded by the yield stress.
The elastic model is thus limited to the prediction of the faulting style (including its orientation) but does not say whether faulting occurs.

Stresses in a shell satisfy equilibrium equations which relate them to the load deforming the shell.
However the load causing the contraction or the flattening change during despinning is a priori unknown.
Thus the full formulation of elastic equations involving stress-strain and strain-displacement relations becomes necessary.
For an elastic shell of constant thickness, contraction stresses were obtained by Lam\'e \citep[see][p.~142]{love} whereas despinning stresses were computed by \citet{vening1947}.
There are no analytic formulas giving equivalent results for a thin elastic shell with variable thickness but \citet{beuthe2008} recently derived the scalar equations governing the deformation of such a shell under an arbitrary load.

Since displacements of the shell surface due to contraction and despinning have a large wavelength, bending moments in the shell are negligible: the shell mainly responds by stretching and is said to be in the membrane regime \citep{kraus,turcotte1981}.
In the membrane limit, bending moments are set to zero: deformation equations take a simplified form which can be obtained from the full deformation equations by setting to zero the bending rigidity $D$.
In that case, the membrane equations for a spherical shell of radius $R$ and variable thickness $h$ are
\begin{eqnarray}
\Delta' F &=& - R \, q \, ,
\label{flexA}
\\
  {\cal C} \left( \alpha \,; F \right) - (1+\nu) \, {\cal A} \left( \alpha \,; F \right) &=& \frac{1}{R} \, \Delta' w \, .
 \label{flexB}
\end{eqnarray}
where $q$ is the transverse load (positive inward), $w$ is the transverse displacement (positive outward) and $F$ is an auxiliary function called the {\it stress function}.
Tensile stress is positive.
The tangential load has been set to zero.
The differential operators $\Delta'$, ${\cal C}$ and ${\cal A}$ are defined in Appendix \ref{DifferentialOperators} by Eqs.~(\ref{defD1})-(\ref{defA1}).
The thickness $h$ is included in the elastic parameter $\alpha$:
\begin{equation}
\alpha = \frac{1}{Eh} \, .
\end{equation}
$E$ is Young's modulus (which is allowed to be variable), $\nu$ is Poisson's ratio (which must be constant).

In the membrane limit, the stress integrated over the thickness of the shell, or {\it stress resultant}, is given by
\begin{equation}
\left( N_{\theta\theta} , N_{\varphi\varphi}, N_{\theta\varphi} \right) = \left( {\cal O}_2  , {\cal O}_1  , - {\cal O}_3   \right) F \, .
\label{stressInt}
\end{equation}
with the operators ${\cal O}_i$ being defined by Eqs.~(\ref{opO1})-(\ref{opO3}).
The stress as expressed in Eq.~(\ref{stressInt}) automatically satisfies the two tangential equilibrium equations while the radial equilibrium equation is encoded in Eq.~(\ref{flexA}).
In the membrane limit, there is thus only one equation where elastic parameters appear, Eq.~(\ref{flexB}), which results from an equation of compatibility between the strains.

In the membrane limit, the stress is constant through the shell thickness. It is thus related to the stress resultant by
\begin{equation}
\sigma_{ij} = \frac{1}{h} N_{ij} \, .
\label{stressSurf}
\end{equation}

The strain is given by
\begin{eqnarray}
\epsilon_{\theta\theta} &=& \frac{1}{E}\left( \sigma_{\theta\theta} -\nu \sigma_{\varphi\varphi} \right) \, ,
\label{strainSurf1}\\
\epsilon_{\varphi\varphi} &=& \frac{1}{E}\left( \sigma_{\varphi\varphi} -\nu \sigma_{\theta\theta} \right) \, ,
\label{strainSurf2}\\
\epsilon_{\theta\varphi} &=& \frac{1+\nu}{E} \,  \sigma_{\theta\varphi} \, .
\label{strainSurf3}
\end{eqnarray}

If the load is known, the stress function $F$ can be computed from the first membrane equation (\ref{flexA}) and the stress from Eqs.~(\ref{stressInt})-(\ref{stressSurf}).
In that case, the elastic thickness only appears in Eq.~(\ref{stressSurf}), which means that the solution for variable thickness is directly obtained from the solution of constant thickness by the scaling $h_0/h$ ($h_0$ is the mean elastic thickness).
This procedure implies that the load is the same whether the elastic thickness is constant or not.
However the load is a priori unknown when modeling contracting and despinning events; instead, the transverse displacement $w$ is the known input.
The stress function $F$ must thus be computed from the second membrane equation, Eq.~(\ref{flexB}), which is linear in $F$ and $w$ with variable coefficients depending on $1/Eh$.
As a corollary, the load causing contraction or despinning is not the same in the cases of constant and variable elastic thickness.

It should be noted that the variation of the lithospheric thickness has an effect on the deformation.
Contraction is not a problem if it is defined as a uniform change in radius, but it is another matter if the definition involves physical processes such as an internal thermal contraction or an expansion due to water-ice transition.
Despinning causes deformations of harmonic degrees zero and two (the former is very small and usually neglected) if the lithospheric thickness is constant but other harmonic degrees appear if it is variable.
In principle the effect of internal contraction and despinning on the shape of a planet with a variable lithospheric thickness should be computed with a three-dimensional model of the interior.
Love numbers quantify the response of a planet with a spherically symmetric internal structure (see Appendix \ref{Lovenumbers}), but there is no standard technique dealing with the non-spherical case.
Extreme cases of a very thin or a very thick lithosphere are not a problem.
On the one hand, the lithosphere does not affect the shape of the body in the limit of vanishing thickness.
On the other hand, thickness variations of a thick lithosphere are expected to be relatively small with respect to the average thickness:
latitudinal variations in solar insolation, for example, induce smaller thickness variations (in percentage) in a thick lithosphere than in a thin one.
In both cases, the response of the planet is thus adequately described by using a model with a spherically symmetric internal structure.
Between these extremes, the size of the deformations with harmonic degrees not equal to zero (for contraction) or two (for despinning) should be investigated, but this question is beyond the scope of this paper.
I thus choose to define contraction and despinning as phenomena associated with deformations of harmonic degrees zero and two, respectively.

How good an approximation is the membrane limit?
If bending terms are not neglected, Eq.~(\ref{flexB}) remains the same but new terms appear in Eq.~(\ref{stressInt}) for the stress resultants.
These new terms have the form $(D/R^3)\,{{\cal O}_i}w$ (with $D=Eh^3/12(1-\nu^2)$) so that they are smaller than the dominant term ${\cal O}_i\,F$ by a factor $\xi=12R^2/h^2$.
Neglected bending terms are thus of the same order of magnitude as other small terms present in the equations before the `thin shell' limit of large $\xi$ is taken \citep{beuthe2008}.
It is possible to solve the complete equations (with bending terms and no large $\xi$ limit) with the same method as proposed in this paper, but the gain in precision would only be of order $1/\xi$.
Even so the theory remains a thin shell theory in the sense that the radial stress is neglected.

The membrane and stress equations (\ref{flexB})-(\ref{stressSurf}) become nondimensional with the following definitions:
\begin{eqnarray}
\bar \alpha &=& \frac{\alpha}{\alpha_0} \, ,
\label{adim1} \\
\bar F &=& \alpha_0 \, F \, ,
\label{adim2} \\
\bar w &=& \frac{w}{R} \, ,
 \label{adim3} \\
\bar N_{ij} &=& \alpha_0 \, N_{ij} \, ,
 \label{adim4} \\
\bar \sigma_{ij} &=& \frac{1}{E} \, \sigma_{ij} \, ,
\label{adim5}
\end{eqnarray}
where $\alpha_0$ is the average of $\alpha$ on the sphere, i.e. the coefficient of degree zero in the expansion of $\alpha$ in Legendre polynomials ($\alpha_0\neq\frac{1}{Eh_0}$).
The membrane equation (\ref{flexB}) then becomes
\begin{equation}
 {\cal C} \left( \bar \alpha \,; \bar F \right) - (1+\nu) \, {\cal A} \left( \bar \alpha \,; \bar F \right) =  \Delta' \bar w \, .
 \label{flexBadim}
\end{equation}
The nondimensional stress is computed with the nondimensional version of Eqs.~(\ref{stressInt})-(\ref{stressSurf}):
\begin{eqnarray}
\left( \bar N_{\theta\theta} , \bar N_{\varphi\varphi}, \bar N_{\theta\varphi} \right) &=& \left( {\cal O}_2  , {\cal O}_1  , - {\cal O}_3   \right) \bar F \, ,
\label{stressIntAdim} \\
\bar\sigma_{ij} &=& \bar \alpha \, \bar N_{ij} \, .
\label{stressSurfAdim}
\end{eqnarray}
The strain is already nondimensional:
\begin{eqnarray}
\epsilon_{\theta\theta} &=& \bar\alpha \left( {\cal O}_2 - \nu {\cal O}_1 \right) \bar F \, ,
\label{strainSurfAdim1} \\
\epsilon_{\varphi\varphi} &=& \bar\alpha \left( {\cal O}_1 - \nu {\cal O}_2 \right) \bar F \, ,
\label{strainSurfAdim2} \\
\epsilon_{\theta\varphi} &=& - \bar\alpha \left(1+\nu\right) {\cal O}_3 \, \bar F \, .
\label{strainSurfAdim3}
\end{eqnarray}
The solution of the nondimensional membrane equation (\ref{flexBadim}) when $\bar w$ is only of degree zero ($\bar w=\bar w_0P_0$) will be called the {\it contraction solution} $\bar F^C$ (whatever the sign of $\bar w_0$), whereas the solution when $\bar w$ is only of degree two ($\bar w=\bar w_2P_2$) will be called the {\it despinning solution} $\bar F^D$ (whatever the sign of $\bar w_2$).
Negative (resp. positive) values of $\bar w_0$ correspond to contracting (resp. expanding) planets, whereas positive (resp. negative) values of $\bar w_2$ correspond to despinning (resp. spinning-up) planets.
Solutions for a constant elastic thickness are computed in Appendix \ref{ConstantElasticThickness}.

\section{Symmetry and duality}
\label{section3}

As discussed in the introduction, I make the simplifying assumption that the value of the elastic thickness only depends on the latitude.
The contraction and despinning problems are then axially symmetric.
Though the numerical method developed in Section \ref{section4} does not require it, I also assume equatorial symmetry, i.e. mirror symmetry about the equatorial plane.
These two assumptions approximately hold for the physical causes of the thickness variation considered here: variation in solar insolation and tidal heating during despinning, provided that the obliquity of the body with respect to the Sun (in the former case) or to the tidal source (in the latter case) is small.

Using symmetry and duality arguments, I describe in this section the characteristics of stress and strain on contracting or despinning shells thinner (or thicker) at the equator.
The duality relation will also serve to obtain a numerical approximation of the contraction solution.

\subsection{Axial and equatorial symmetry}
\label{symmetry}

\begin{figure}
   \centering
   \includegraphics[width=6.9cm]{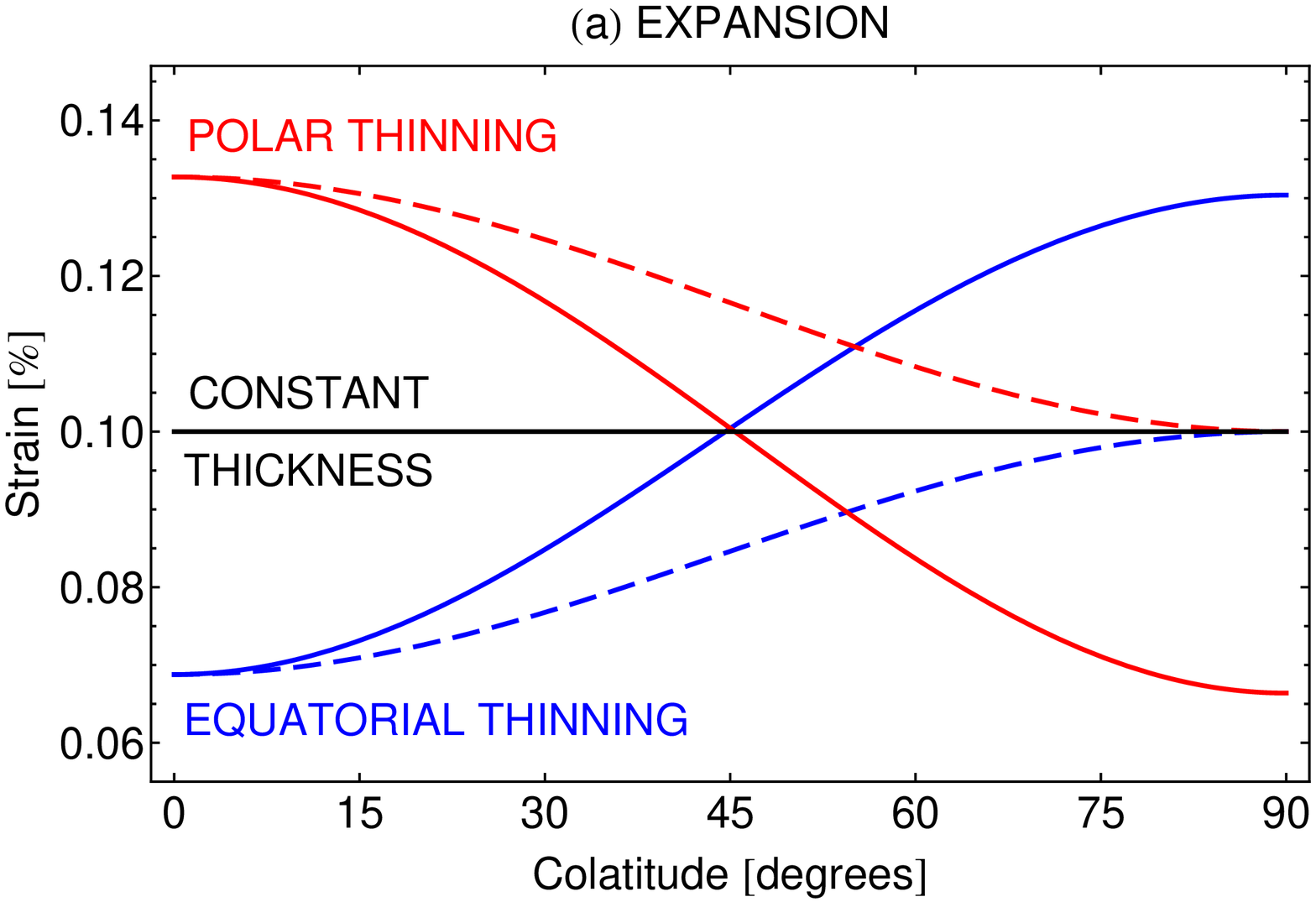}
   \hspace{0.5cm}
     \includegraphics[width=6.9cm]{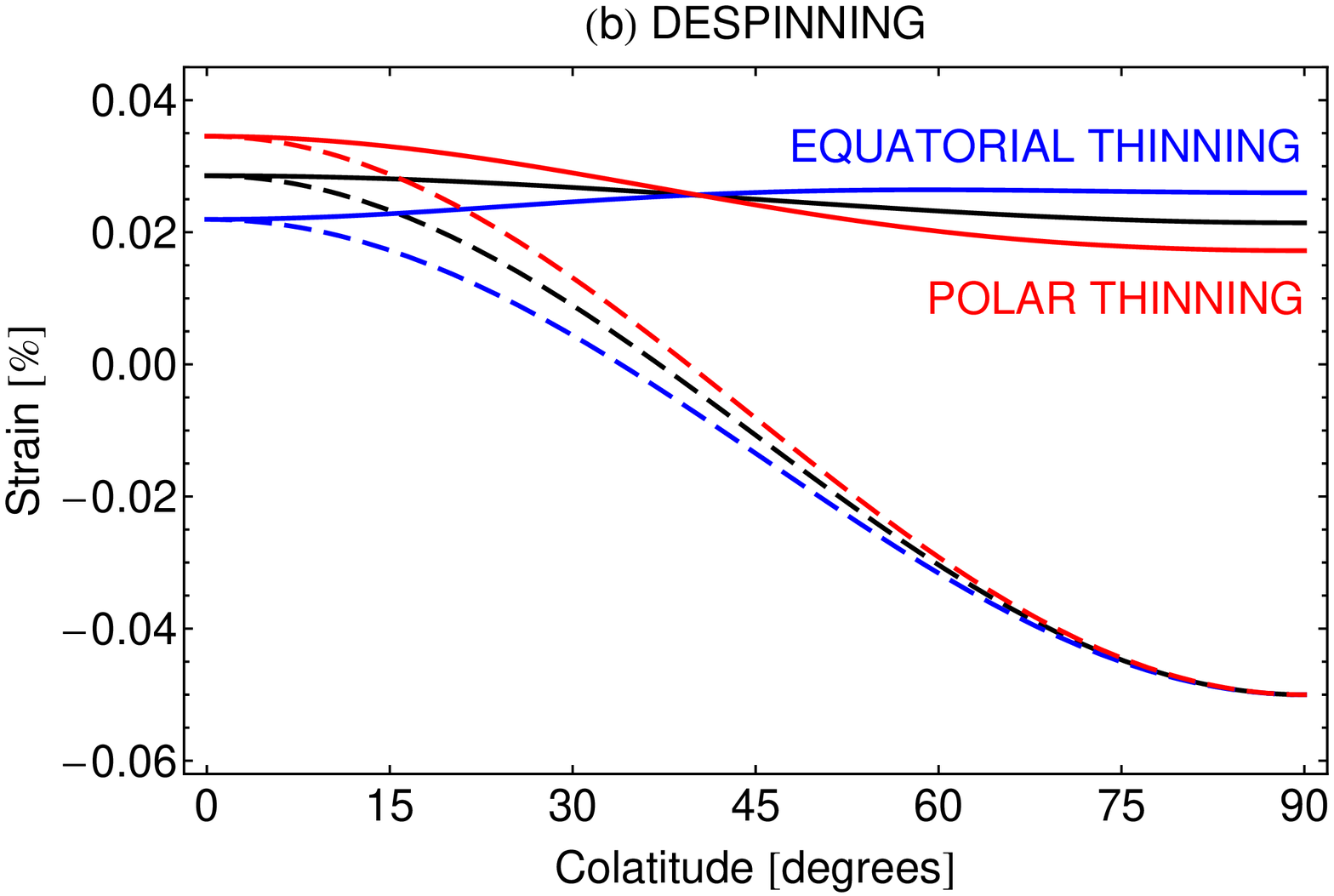}
   \caption{\footnotesize
   Strain as predicted from symmetry arguments: (a) expansion, (b) despinning.
   Continuous curves refer to $\epsilon_{\theta\theta}$ while dashed curves refer to $\epsilon_{\varphi\varphi}$.
   Three cases are shown: constant thickness, equatorial thinning, polar thinning.
   The unlabeled middle curves in the despinning case are for constant elastic thickness.
   The numerical values used to draw the curves are:  $\bar w_0\!=\!0.001$ for expansion, $\bar w_2\!=\!0.001$ for despinning, $r=1/2$ for equatorial thinning, $r=2$ for polar thinning
   ($r$ is defined by Eq.~(\ref{etptr})), Poisson's ratio equal to $0.25$.
   Extensional strain is positive. }
   \label{StrainSym}
\end{figure}

The main characteristics of the contraction solution can be determined from symmetry arguments, without solving the membrane equation.
Axial symmetry gives the following constraints on stress and strain (see also Fig.~\ref{StrainSym}):
\begin{enumerate}
\item
the meridional and azimuthal stresses (resp. strains) are principal stresses (resp. strains), i.e. $\sigma_{\theta\varphi}=0$ (resp. $\epsilon_{\theta\varphi}=0$).
\item
the meridional and azimuthal stresses are equal at the poles;
this is also true of the strains.
\item
the slopes of the stress and strain components vanish at the poles (it is a consequence of the previous point combined with the equations of equilibrium);
they also vanish at the equator if there is equatorial symmetry.
\item
the average of the meridional strain is independent of the elastic thickness (see Eq.~(\ref{average})): 
$<\epsilon_{\theta\theta}>=\bar w_0$ for contraction, $<\epsilon_{\theta\theta}>=\bar w_2/4$ for despinning.
\item
the azimuthal strain at the equator is independent of the elastic thickness (see Eq.~(\ref{straineq})):
$\epsilon_{\varphi\varphi}|_{\theta=\frac{\pi}{2}}=\bar w_0$ for contraction, $\epsilon_{\varphi\varphi}|_{\theta=\frac{\pi}{2}}=-\bar w_2/2$ for despinning.
\end{enumerate}

These constraints allow me to predict the response of the shell to a radius variation.
The stresses for a constant elastic thickness are given in Appendix \ref{ConstantElasticThickness}.
Fig.~\ref{StrainSym} illustrates the following discussion.
I suppose that the shell is expanding ($\bar w_0>0$) so that stresses and strains are positive (the solution for a contracting shell is obtained by changing the sign of stresses and strains).
I also assume equatorial symmetry with a monotonous variation of the thickness from the pole to the equator, so that stress and strain components can be expected to vary monotonously.
Stress and strain are concentrated where the shell is weaker.
In other words, the membrane stretching is maximum where the shell is thinnest.
If the shell is thinner at the equator, the azimuthal strain is thus maximum at the equator, where it takes the value $\bar w_0$ imposed by the equatorial length variation, and minimum at the poles, where it is equal to the meridional strain.
The meridional strain is also maximum at the equator and minimum at the poles, taking somewhere in between its average value $\bar w_0$.
Therefore, the meridional strain always exceeds the azimuthal strain when the shell is thinner at the equator.
This ordering is conserved for the stresses: the meridional stress always exceeds the azimuthal stress when the shell is thinner at the equator, as shown by the inversion of Eqs.~(\ref{strainSurf1})-(\ref{strainSurf3}) with $\nu<1$.
If the shell is thinner at the poles, a similar analysis leads to the conclusion that the strain (resp. stress) is maximum at the poles and that the azimuthal strain (resp. stress) always exceeds the meridional strain (resp. stress).

The despinning shell case ($\bar w_2>0$) is not as simple since the combined effects of the non-spherical deformation and thickness variation can lead to local extrema of the stress and strain components between the poles and the equator.
However the ordering of the strains ($\epsilon_{\varphi\varphi}$ more compressive than $\epsilon_{\theta\theta}$) is expected to be the same as in case of constant elastic thickness, since the meridional strain has an average positive value ($\bar w_2/4$) while the azimuthal strain varies between the value of the meridional strain at the poles and the negative value $-\bar w_2/2$ at the equator.
The location of the shell thinning has thus less influence on the stress/strain pattern than in the contraction case.
Because of stress/strain concentration in weaker areas, the stress and strain components are expected to come closer to (resp. farther from) zero at the poles if the shell becomes thinner at the equator (resp. at the poles), as shown in Fig.~\ref{StrainSym}b.
The meridional strain curve will accordingly move so as to keep its average constant.
The principle of stress/strain concentration does not apply to the azimuthal strain at the equator since its value is fixed by the geometry.
As above, the ordering is conserved when going from strains to stresses: $\sigma_{\varphi\varphi}$ is more compressive than $\sigma_{\theta\theta}$.
Another argument for ordering will be obtained from the contraction-despinning duality in Section \ref{duality}.

\subsection{Contraction-despinning duality}
\label{duality}

Still assuming axial and equatorial symmetry, I now show that the contraction and despinning solutions are not independent if the harmonic expansion of the inverse thickness is limited to degree two.
Even if this assumption is not strictly valid, the harmonic of degree two will dominate if the thickness variation is monotonous between the poles and the equator.
Under these assumptions, $\bar\alpha$ can be written in terms of Legendre polynomials of degrees zero and two:
\begin{equation}
\bar\alpha = 1 + \bar\alpha_2 \, P_2 \, ,
\label{expansionalpha}
\end{equation}
where $P_2$ is the Legendre polynomial of degree two defined by Eq.~(\ref{defP2}).
The parameter $\bar\alpha_2$ is in the range $[-1,2]$ and is negative (resp. positive) for shells thinner at the equator (resp. poles).
The parameterization of the thickness variation is discussed in more detail in Section \ref{ThicknessVariation}.

The stress functions for contraction and despinning then satisfy the following duality:
\begin{equation}
\frac{\bar F^C}{\bar w_0}  = \frac{1}{1-\nu} -\bar\alpha_2 \, \frac{\bar F^D}{\bar w_2} \, .
\label{relationCD1}
\end{equation}
The first term in the right-hand side is the contraction solution when the elastic thickness is constant.
The duality~(\ref{relationCD1}) can be checked by substitution in Eq.~(\ref{flexBadim}) and using the identity (\ref{flexureidentity2}).
Since spherical harmonics of degree one belong to the kernel of $\Delta'$ (see Appendix \ref{DifferentialOperators}), the duality remains valid if $\bar\alpha$ includes such harmonics.
If $\bar\alpha$ contains harmonics up to degree $k>2$, a similar relation exists (see Eq.~(\ref{GeneralizedDuality})) between the solutions for right-hand sides of higher degree ($\bar w=\bar w_3\,P_3$, ... , $\bar w=\bar w_k\,P_k$);
I use it in Appendix \ref{ContractionFirstOrder} for the computation of the contraction solution at first order.
I also give below a relation of fourth degree between the stresses in order to discuss the combination of contraction and despinning.

Eq.~(\ref{relationCD1}) leads to a duality between the stress resultants for the contraction and despinning solutions:
\begin{equation}
\frac{ \bar N_{ij}^C }{\bar w_0} = \frac{\delta_{ij}}{1-\nu} - \bar\alpha_2 \, \frac{ \bar N_{ij}^D }{\bar w_2} \, .
\label{relationCD2}
\end{equation}
The stress resultants for a contracting shell are thus the same as those for a despinning shell with $\bar w_2=-\bar\alpha_2\bar w_0$, plus the stress resultants for a contracting shell of constant thickness.
If the contracting shell ($\bar w_0<0$) is thinner at the equator ($\bar\alpha_2<0$), the equivalent `despinning' shell is actually spinning faster since $\bar w_2<0$.
This duality is clearly visible by comparing Figs.~\ref{StressIntContraction}a and \ref{StressIntContraction}b.
A similar duality holds for the stresses, but the first term in the right-hand side now depends on latitude:
\begin{equation}
\frac{ \bar \sigma_{ij}^C }{\bar w_0} = \bar\alpha  \, \frac{\delta_{ij}}{1-\nu} - \bar\alpha_2 \, \frac{ \bar \sigma_{ij}^D }{\bar w_2} \, .
\label{relationCD3}
\end{equation}
Finally the duality relation for the strains reads
\begin{equation}
\frac{\epsilon_{ij}^C}{\bar w_0} = \bar\alpha \, \delta_{ij} - \bar\alpha_2 \, \frac{\epsilon_{ij}^D}{\bar w_2} \, .
\label{relationCD4}
\end{equation}

These dualities are extremely useful.
Their most obvious use is to generate both contraction and  despinning solutions by computing only one of them, but they also have other advantages.
First, Eq.~(\ref{relationCD3}) shows that the ordering in size of the components of the stress is either the same (if $\bar\alpha_2>0$) or reversed (if $\bar\alpha_2<0$) when going from the contraction to the despinning solution.
Thus the azimuthal stress is more compressive than the meridional stress for a despinning shell whether the shell is thinner at the equator or at the poles.
While the isotropic term $\bar\alpha\delta_{ij}/(1-\nu)$ in Eq.~(\ref{relationCD3}) does not affect the ordering of the stress components, its spatial dependence influences the positions of the maxima which must thus be computed from the numerical solution.

Second, the dualities serve to compute a first approximation of the contraction solution on a shell of variable thickness, knowing the despinning solution on a shell of constant thickness.
The trick comes from the fact that the term linear in $\bar\alpha_2$ in the contraction solution is related by duality to the term independent of $\bar\alpha_2$ in the despinning solution.
The contraction solution is computed in such a way in Appendix \ref{ContractionFirstOrder} (see Eqs.~(\ref{stressC1})-(\ref{strainC2})).
In dimensional notation, the contraction stresses at first order in $\bar\alpha_2$ read
\begin{eqnarray}
\sigma^C_{\theta\theta} &\cong&   \sigma_0 \left( 1 +  3 \, \bar\alpha_2 \, \frac{\nu+(2+\nu)\cos2\theta}{2(5+\nu)} \right) \, ,
\label{stressC1dim}
\\
\sigma^C_{\varphi\varphi} &\cong&  \sigma_0 \left( 1 + 3 \, \bar\alpha_2 \, \frac{1+(1+2\nu)\cos2\theta}{2(5+\nu)} \right) \, ,
\label{stressC2dim}
\end{eqnarray}
where $\sigma_0$ is the stress for a radial contraction of $\delta R$ ($\delta R>0$) when the thickness is constant:
\begin{equation}
\sigma_0 = - \frac{E}{1-\nu} \frac{\delta R}{R} \, .
\end{equation}
Fig.~\ref{StressFirstOrder}a shows that the first order approximation is good if $\bar\alpha_2=-2/5$ and rather bad if $\bar\alpha_2$ is close to its lower bound $-1$.
In Appendix \ref{ContractionFirstOrder}, the method is generalized to an $\bar\alpha$ having an arbitrary Legendre expansion (see Eqs.~(\ref{stressC1bis})-(\ref{stressC2bis})).
Though the basic characteristics of stress and strain for a despinning shell could be predicted by the symmetry arguments of Section \ref{symmetry}, duality cannot serve for the computation of their numerical values.
Their formulas at first order in $\bar\alpha_2$ are given in Appendix \ref{DespinningFirstOrder}.

A third interesting consequence of the dualities concerns the analysis of combined contraction and despinning.
If the amounts of contraction and despinning are related by
\begin{equation}
\bar w_2 = \bar\alpha_2 \, \bar w_0 \, ,
\label{threshold}
\end{equation}
the total stress computed with the duality (\ref{relationCD3}) is isotropic (though not homogeneous):
\begin{equation}
\bar\sigma^C_{ij} + \bar\sigma^D_{ij} = \bar\alpha \, \bar w_0 \, \frac{ \delta_{ij}}{1-\nu} \, .
\label{isotropicstress}
\end{equation}
Consider a contracting ($\bar w_0<0$) and despinning ($\bar w_2>0$) body with a lithosphere thinner at the equator ($\bar\alpha_2<0$).
If there is less (resp. more) despinning than the threshold (\ref{threshold}), the meridional stress is more (resp. less) compressive than the azimuthal stress.
This fact is useful when determining the orientation of faults for contraction plus despinning.
The threshold is reached for a contraction $\delta R$ ($\delta R>0$) of
\begin{equation}
\delta R = \frac{h_2^T}{\bar\alpha_2} \, \frac{(\Omega_f^2-\Omega_i^2) R^2}{3g} \, ,
\label{physicalthreshold}
\end{equation}
where
$\Omega_i$ (resp. $\Omega_f$) is the initial (resp. final) angular rate, $g$ is the surface gravity and $h_2^T$ is the degree-two displacement Love number.
The above equation results from the substitution of Eqs.~(\ref{w2flat}) and (\ref{flat}) into Eq.~(\ref{threshold}).

The threshold (\ref{threshold}) becomes dependent on latitude when the inverse thickness expansion includes higher harmonics.
If $\bar\alpha=1+\bar\alpha_2\,P_2+\bar\alpha_4\,P_4$, the generalization of the stress duality reads
\begin{equation}
\frac{ \bar \sigma_{ij}^C }{\bar w_0} + \bar\alpha_2 \, \frac{ \bar \sigma_{ij}^D }{\bar w_2} + \bar\alpha_4 \, \frac{ \bar \sigma_{ij}^E}{\bar w_4} = \bar\alpha  \, \frac{\delta_{ij}}{1-\nu}  \, ,
\label{relationCD3bis}
\end{equation}
where the index $E$ denotes the solution of the membrane equation (\ref{flexBadim}) for $\bar w=\bar w_4P_4$, which is given at zeroth order in ($\bar\alpha_2$,$\bar\alpha_4$) by Eqs.~(\ref{degree4Const1})-(\ref{degree4Const2}).
Since there is actually no deformation of degree four, the third term in the left-hand side of Eq.~(\ref{relationCD3bis}) should not be interpreted as a physical stress but rather as a deviation from the duality (\ref{relationCD3}).
At the threshold (\ref{threshold}), the total stress $\bar\sigma^T_{ij}=\bar\sigma^C_{ij} + \bar\sigma^D_{ij}$ is thus not isotropic anymore:
\begin{equation}
\bar\sigma^T_{\theta\theta} - \bar\sigma^T_{\varphi\varphi} \cong  \frac{\bar\alpha_4 \, \bar w_0}{19+\nu} \, \left( 12 P_4 -10 P_2 - 2 \right) \,  .
\label{stressdiff}
\end{equation}
If $\bar\alpha_4\bar w_0>0$, the meridional stress is more (resp. less) compressive than the azimuthal stress for high (resp. low) latitudes, with the turning point being given by a latitude of $\arcsin(1/\sqrt{7})\cong22.2^\circ$ (this value is only valid if $\bar\alpha_2,\bar\alpha_4\ll1$).
If $\bar\alpha_4\bar w_0<0$, the high and low latitudes zones are exchanged.
In Section \ref{section4}, I propose a parameterization of the inverse thickness which has an unlimited expansion in Legendre polynomials (see Eq.~(\ref{expansionalphaTer})).
Fig.~\ref{ThrustFaultOrientation} then shows the latitudinal dependence of the threshold between north-south and east-west thrust faults for different types of thickness variations.
The case described by Eq.~(\ref{stressdiff}) corresponds to a value of the parameter $k$ close to zero.

\section{Solving the membrane equation}
\label{section4}

Symmetry and duality arguments have served us well for the determination of the main characteristics of the stress and strain curves, as well as for a first numerical approximation of the contraction solution.
It remains however necessary to solve the full membrane equation in order to compute both the contraction and despinning solutions at an arbitrary degree of precision.
I will present two methods which are valid for an arbitrary deformation of the surface.
The first one is numerical and has the advantage of minimizing the amount of programming.
The second one is semi-analytical, in the sense that the solution can be expressed as a perturbation expansion in Legendre polynomials about the solution for constant shell thickness.
It has the advantage of producing explicit formulas for the stresses in which the influence of parameters describing thickness variations clearly appears.
Before tackling the methods of resolution, I briefly discuss the parameterization of the thickness variation.

\subsection{Thickness variation}
\label{ThicknessVariation}

\begin{figure}
   \centering
   \includegraphics[width=6.9cm]{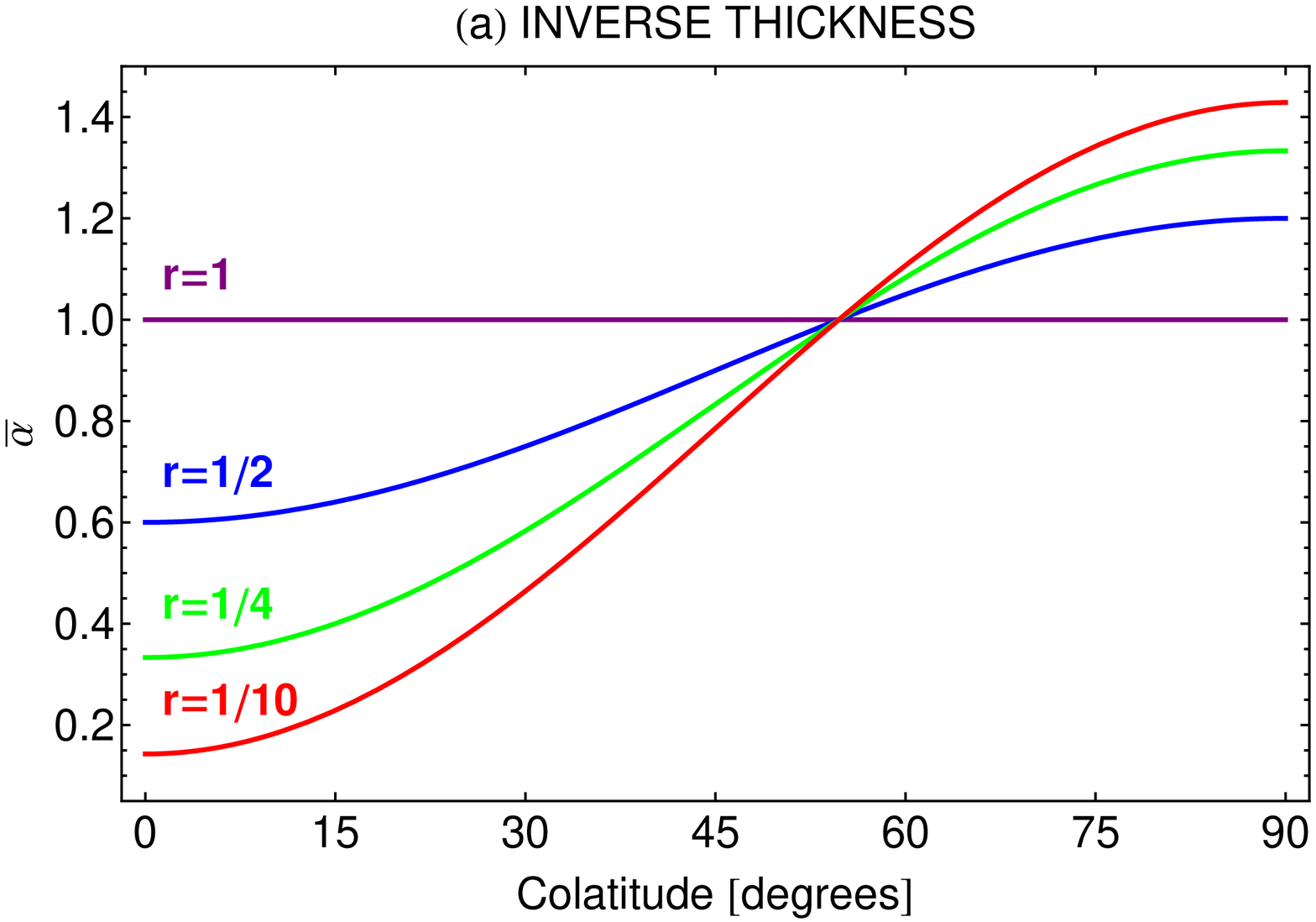}
   \hspace{0.5cm}
    \includegraphics[width=6.9cm]{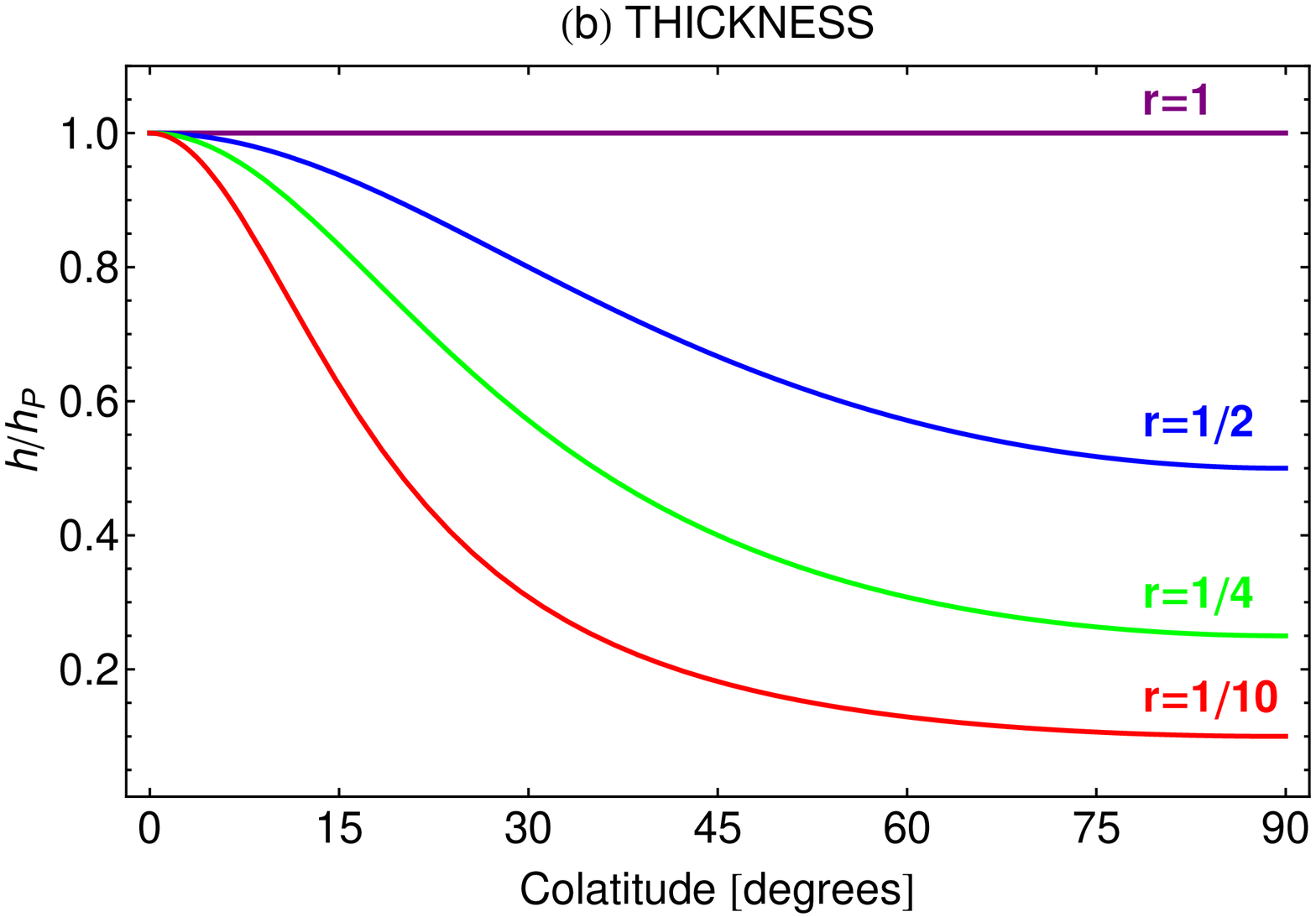} 
   \caption{\footnotesize
   Equatorial thinning of the lithosphere as parameterized by Eqs.~(\ref{expansionalphaBis})-(\ref{alpha2}): (a) profiles of $\bar\alpha$ (proportional to the inverse thickness), (b) thickness profiles normalized by the polar thickness.
   The four profiles correspond to equator-to-pole thickness ratios $r$ equal to $(1,1/2,1/4,1/10)$.}
   \label{AlphaCurves}
\end{figure}

Since my aim is to predict tectonic patterns without assuming a specific lithospheric structure, I will work with a simple parameterization of the variation of the shell thickness.
I choose to parameterize $\bar\alpha$ instead of the thickness $h$ because it is more convenient for the semi-analytical method.
A few assumptions constrain the possible form of the thickness variation.
Variations caused by solar insolation and tidal heating are of very large wavelength, so that only slow-varying functions of the colatitude are permissible.
Axial symmetry means that the thickness only depends on the colatitude $\theta$, but it also imposes that the derivative of the thickness vanishes at the poles.
If there is equatorial symmetry, the derivative of the thickness vanishes at the equator (this restriction can be lifted without changing the methods of resolution).
A simple parameterization of the thickness compatible with these constraints is given by Eq.~(\ref{expansionalpha}):
\begin{equation}
\bar\alpha(\theta) = 1 + \bar\alpha_2 \, P_2(\cos\theta) \, ,
\label{expansionalphaBis}
\end{equation}
where the dependence on the colatitude $\theta$ has been made explicit.
I define the equator-to-pole thickness ratio $r$ by
\begin{equation}
r = \frac{h_E}{h_P} \, ,
\label{etptr}
\end{equation}
where $h_E$ is the equatorial thickness and $h_P$ the polar thickness.
The coefficient $\bar\alpha_2$ is then given by
\begin{equation}
\bar\alpha_2 = 2 \, \frac{r-1}{r+2} \, .
\label{alpha2}
\end{equation}
The thickness is positive everywhere if $-1\leq\bar\alpha_2\leq 2$.
Negative and positive values of $\bar\alpha_2$ describe elastic shells that are thinner at the equator and at the poles, respectively.
Fig.~\ref{AlphaCurves} shows various profiles of $\bar\alpha$ and $h/h_P$ when the lithosphere is thinner at the equator: the values $r=(1,1/2,1/4,1/10)$ correspond to $\bar\alpha_2=(0,-2/5,-2/3,-6/7)$.

Although the parameterization (\ref{expansionalphaBis}) is sufficient for the determination of the most important features of tectonic patterns, it suffers from a weakness revealed by Fig.~\ref{AlphaCurves}b.
When the equator-to-pole thickness ratio decreases, the zone of equatorial thinning expands toward the pole.
In other words, the parameter $\bar\alpha_2$ not only affects the equator-to-pole thickness ratio, but it also modifies the relative extension of the thin and thick zones.
Therefore, the parameterization (\ref{expansionalphaBis}) is not appropriate for the analysis of the influence of the thin zone size.
I thus define a new parameterization in which the interval $[0,\pi/2]$ is non-linearly stretched by a function $\psi_k$.
As a result, the extension of the thin zone can be modulated without affecting the equator-to-pole thickness ratio:
\begin{equation}
\tilde\alpha(\theta) = 1 + \tilde\alpha_2 \, P_2(\cos \psi_k(\theta)) \, ,
\label{expansionalphaTer}
\end{equation}
where $-1\leq\tilde\alpha_2\leq 2$.
The `tilde' notation indicates that the normalization differs from the one used for $\bar\alpha$ because the Legendre coefficient of degree zero differs from one.
The term $P_2(\cos \psi_k(\theta))$ has generally an unlimited expansion in Legendre polynomials, including a term of degree zero.
Nonetheless the normalization of $\alpha$ does not matter for the computation of stress and strain; it only affects the stress resultants.

\begin{figure}
   \centering
   \includegraphics[width=6.9cm]{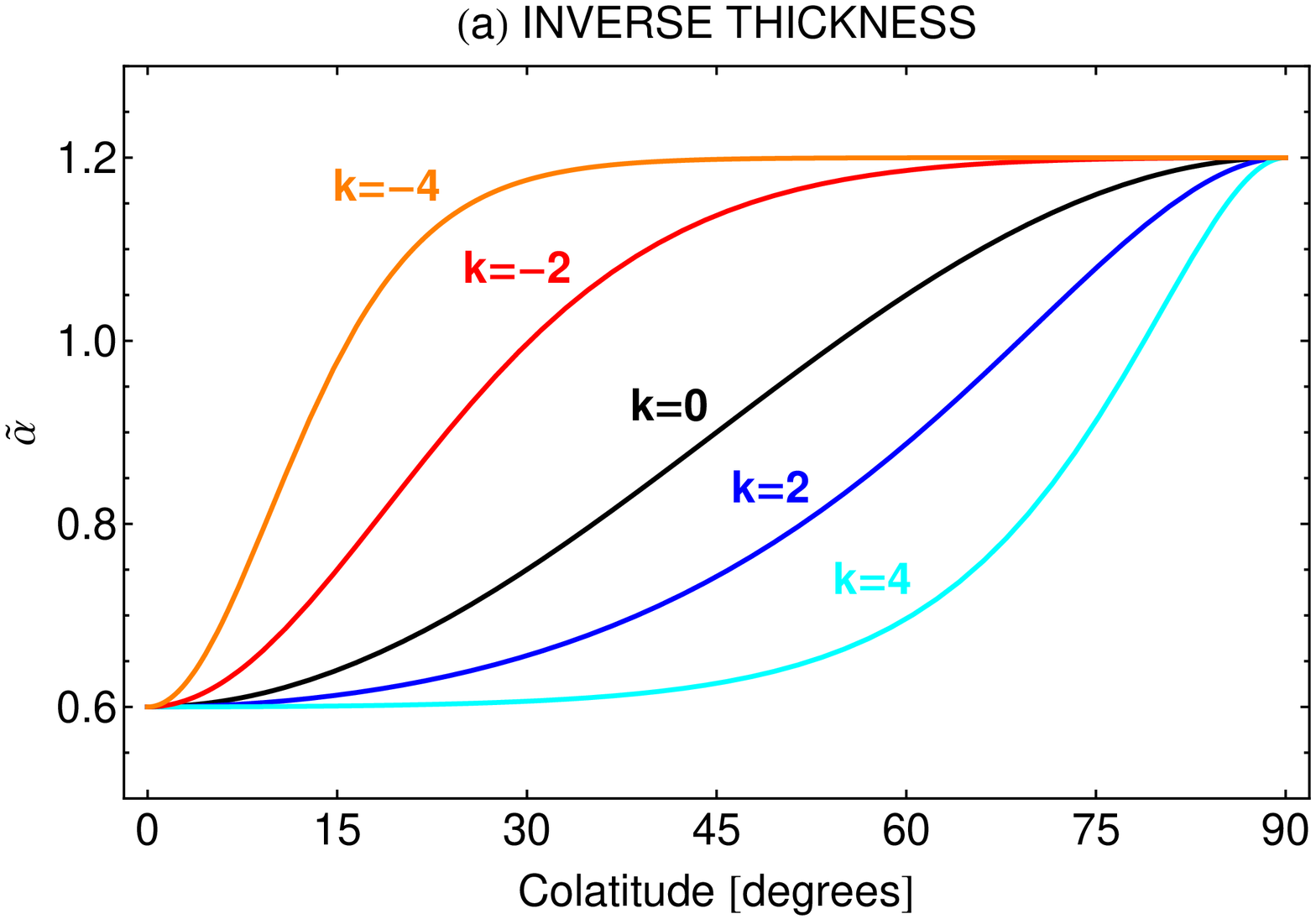}
   \hspace{0.5cm}
    \includegraphics[width=6.9cm]{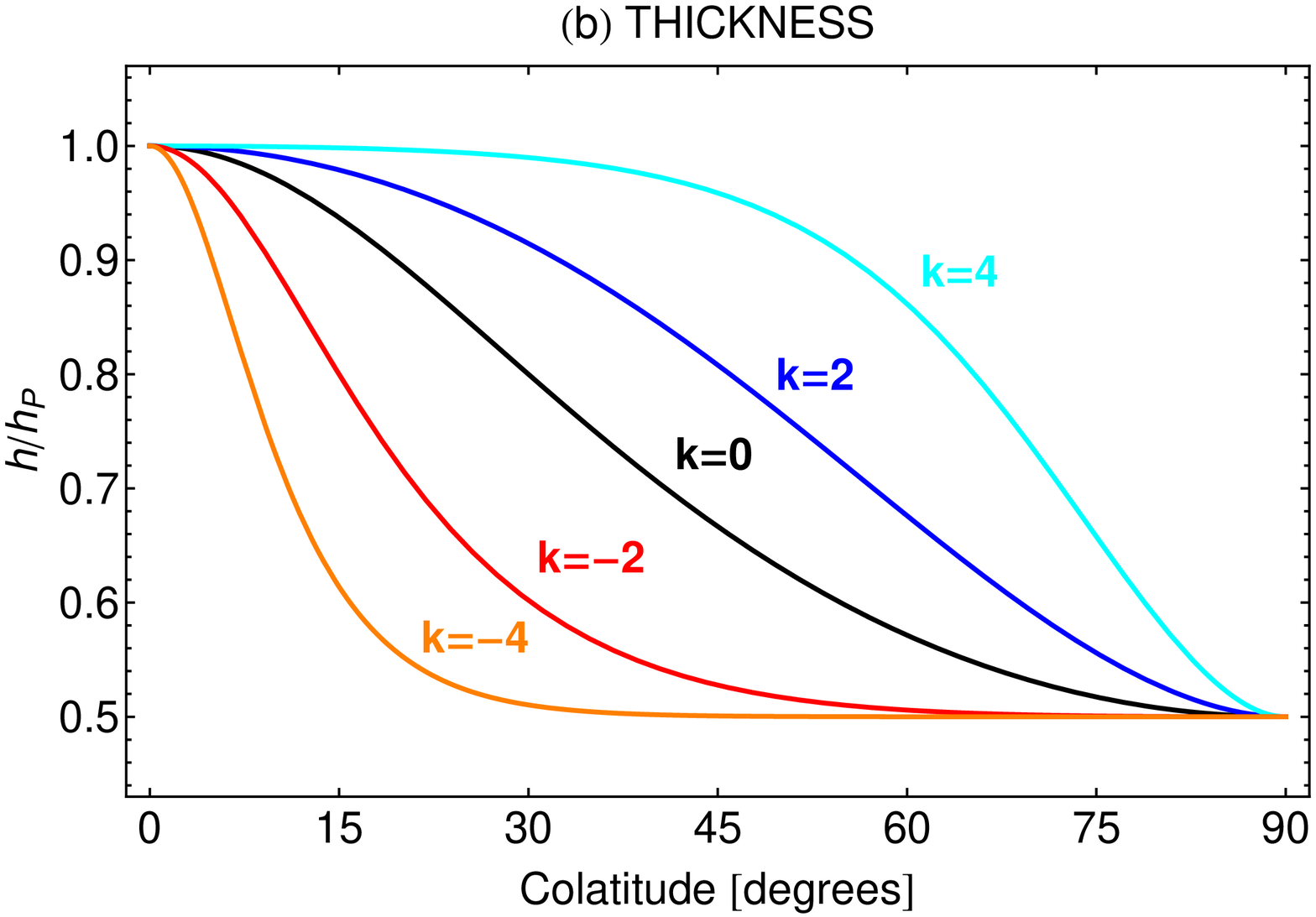} 
   \caption{\footnotesize
   Equatorial thinning of the lithosphere as parameterized by Eqs.~(\ref{expansionalphaTer})-(\ref{nlimit}):
   (a) profiles of $\tilde\alpha$ (proportional to the inverse thickness), 
   (b) thickness profiles normalized by the polar thickness. 
   The equator-to-pole thickness ratio $r$ is equal to $1/2$: the curves $k=0$ are the same as the curves $r=1/2$ in Fig.~\ref{AlphaCurves}.
   Curves $k=2,4$ correspond to a reduction of the thin zone whereas curves $k=-2,-4$ correspond to an extension of the thin zone.}
   \label{DeformedCurves}
\end{figure}

The function $\psi_k(\theta)$ is defined on the interval $[0,\pi/2]$ by
\begin{equation}
 \psi_k(\theta) = \left\{
 \begin{tabular}{ll}
$\frac{\pi}{2} \, \frac{\sinh(2k\theta/\pi)}{\sinh(k)}$ & if  $k>0$ \, , \\
$\frac{\pi}{2} \, \frac{\tanh(2k\theta/\pi)}{\tanh(k)}$ & if  $k<0$ \, .
 \end{tabular}
 \right.
\end{equation}
It can extended to the interval $[0,\pi]$ by mirror symmetry about the equatorial plane.
The limit $k\rightarrow0$ yields the parameterization (\ref{expansionalphaBis}):
\begin{equation}
\lim_{k\rightarrow0} \psi_k(\theta) = \theta \, .
\label{nlimit}
\end{equation}
Fig.~\ref{DeformedCurves} shows that positive values of $k$ correspond to a reduction in the size of the thin zone.
The equator-to-pole thickness ratio is left unchanged since it is controlled by $\tilde\alpha_2$.
Conversely, negative values of $k$ correspond to an extension of the thin zone.

\subsection{Numerical method}

The software Mathematica \citep{wolfram} has a powerful command called `NDSolve' for the numerical solution of differential equations.
When there is axial symmetry, the second membrane equation (\ref{flexBadim}) is an ordinary differential equation of order four.
I found it convenient to solve it as a system of two differential equations of order two:
\begin{eqnarray}
\Delta' \bar F &=& - \bar q \, ,
\label{flexAreduced}
\\
 - \Delta' \left( \bar \alpha \, \bar q \right) - (1+\nu) \, {\cal A} \left( \bar \alpha \,; \bar F \right) &=& \Delta' \bar w \, .
 \label{flexBreduced}
\end{eqnarray}
where the nondimensional load $\bar q$ is defined by
\begin{equation}
\bar q= R \, \alpha_0 \, q \, .
\end{equation}
The various terms in the above equations can be expanded with the formulas of Appendix \ref{DifferentialOperators}:
\begin{eqnarray}
\Delta' \bar F &=& \Delta \bar F + 2 \bar F
\, , \\
\Delta' \left( \bar \alpha \, \bar q \right) &=&
\left( \Delta \bar \alpha \right) \bar q + 2 \frac{\partial \bar \alpha}{\partial \theta} \, \frac{\partial \bar q}{\partial \theta} + \bar \alpha \left( \Delta \bar q \right) + 2 \bar \alpha \, \bar q
\, , \\
{\cal A} \left( \bar \alpha \,; \bar F \right) &=& \left( \frac{\partial^2 \bar \alpha}{\partial \theta^2} + \bar \alpha \right) \left(  \cot \theta \, \frac{\partial \bar F}{\partial \theta} + \bar F \right)
+ \left(  \cot \theta \, \frac{\partial \bar \alpha}{\partial \theta} + \bar \alpha \right) \left( \frac{\partial^2 \bar F}{\partial \theta^2} + \bar F \right)
\, ,
\end{eqnarray}
with
\begin{equation}
\Delta f = \frac{\partial^2 f}{\partial \theta^2} + \cot \theta \, \frac{\partial f}{\partial \theta} \, .
\end{equation}

In principle the system (\ref{flexAreduced})-(\ref{flexBreduced}) should be solved on the segment $[0,\pi]$ but the points $0$ and $\pi$
must be excluded from the interval because of the apparent polar singularities in spherical coordinates.
I thus restrict the interval to $[a,b]$ with $a=\epsilon$ and $b=\pi-\epsilon$ ($\epsilon$ is a small positive number, for example 0.001).

Four boundary conditions are required.
Smooth solutions with axial symmetry satisfy
\begin{eqnarray}
\frac{\partial \bar F}{\partial \theta} |_{\theta=a} &=& \frac{\partial \bar q}{\partial \theta} |_{\theta=a} = 0 \, ,
\label{bc1} \\
\frac{\partial \bar F}{\partial \theta} |_{\theta=b} &=& \frac{\partial \bar q}{\partial \theta} |_{\theta=b} = 0 \, .
\label{bc2}
\end{eqnarray}
Before applying these equations as boundary conditions, I should make a caveat.
The stress function has the particularity that it can be redefined at will because of the degree-one freedom mentioned in Appendix \ref{DifferentialOperators}:
\begin{equation}
\bar F \rightarrow \bar F + \beta \cos\theta \, ,
\end{equation}
where $\beta$ is an arbitrary real number.
Since $\cos\theta$ satisfies Eqs.~(\ref{bc1})-(\ref{bc2}) in the limit $\epsilon\rightarrow0$, another condition must be specified in order to fully determine $\bar F$, for example:
\begin{equation}
\bar F|_{\theta=a} = 0 \, .
\label{bc3}
\end{equation}
This last condition being arbitrary, one should keep in mind that the same physical problem can be solved with various boundary conditions.
The resulting stress functions only differ by their degree-one content, i.e. by $\cos\theta$ multiplied by some constant.
The corresponding stresses are of course identical.

Suitable boundary conditions for the interval  $[a,b]$ consist of Eq.~(\ref{bc3}), plus three conditions  chosen among Eqs.~(\ref{bc1})-(\ref{bc2}).
I found it best to specify three conditions on $\bar F$ and one condition on $\bar q$.
In most cases NDSolve also works well with only Eqs.~(\ref{bc1})-(\ref{bc2}) as boundary conditions, meaning that its algorithm chooses one possible solution for $\bar F$ among an infinity (note that the freedom in the definition of $\bar F$ only appears in the limit $\epsilon\rightarrow0$).
Though it is not necessary, one can get rid of the $\cos\theta$ component of $\bar F$ by orthogonalization:
\begin{equation}
\bar F \rightarrow \bar F - \frac{3}{2} \int_0^\pi \,  \bar F \cos\theta \, \sin\theta \, d\theta \, .
\end{equation}

If there is equatorial symmetry, it is sufficient to consider the interval $[a,b]$ with $a=\epsilon$ and $b=\pi/2$.
In that case,  Eqs.~(\ref{bc1})-(\ref{bc2}) form a suitable set of boundary conditions since $\cos\theta$ violates Eq.~(\ref{bc2}) when $b=\pi/2$.
The $\cos\theta$ component of $\bar F$ then vanishes.

\subsection{Semi-analytical method}

If the elastic thickness is constant, the membrane equation (\ref{flexBadim}) is diagonal in the basis of spherical harmonics.
The contraction solution is then of harmonic degree zero whereas the despinning solution is of degree two (see Appendix \ref{ConstantElasticThickness}).
This straightforward method cannot be used when the elastic thickness is spatially variable because of the coupling of the different harmonic degrees and orders.
Here I show that the system of coupled differential equations remains manageable under the assumption of axial symmetry.
Since the imposed deformation and the elastic thickness do not depend on the longitude, only zonal spherical harmonics contribute.

The problem thus consists in solving the membrane equation (\ref{flexBadim}) with expansions in zonal spherical harmonics, i.e. Legendre polynomials: $\bar \alpha=\sum\bar\alpha_\ell\,P_\ell$, $\bar F=\sum\bar F_\ell\,P_\ell$ and $\bar w=\sum\bar w_\ell\,P_\ell$.
The action of the operators ${\cal C}$ and ${\cal A}$ on Legendre polynomials produces a finite sum of Legendre polynomials which is computed in Appendices~\ref{OperatorsOnHarmonics} and \ref{OperatorsOnPoly} (final results are embodied in Eqs.~(\ref{Cpoly})-(\ref{Apoly})).
If the expansions are truncated at degree $n$, the membrane equation takes a matrix form with each row corresponding to a harmonic degree on which the membrane equation is projected, except for degree one:
\begin{equation}
{\mathbf M}^{(n)} \, {\mathbf {\bar F}} = {\mathbf \Delta' \mathbf  {\bar w}}  \, ,
\label{flexBmat}
\end{equation}
with the {\it membrane matrix} ${\mathbf M}^{(n)}$ being defined by
\begin{equation}
{\mathbf M}^{(n)}={\mathbf C}^{(n)} - (1+\nu) \, {\mathbf A}^{(n)} \, .
\label{membranematrix}
\end{equation}
${\mathbf C}^{(n)}$ and ${\mathbf A}^{(n)} $ are matrices approximating at degree $n$ the operators ${\cal C}$ and ${\cal A}$, respectively.
${\mathbf {\bar F}}$ and ${\mathbf {\bar w}}$ are vectors containing the coefficients $\bar F_\ell$ and $\bar w_\ell$, respectively.
${\mathbf \Delta'}$ is the diagonal matrix with elements $\delta'_\ell=-\ell(\ell+1)+2$.
The absence of a row for degree one is due to the fact that spherical surface harmonics of degree one do not belong to the images of the operators ${\cal C}$ and ${\cal A}$ (see Appendix \ref{OperatorsOnPoly}).
Correspondingly, there is no coefficient of degree one in the vectors ${\mathbf {\bar F}}$ and ${\mathbf {\bar w}}$ since the membrane equation does not constrain them (see Appendix \ref{DifferentialOperators}).
Given a deformation $w$, the membrane equation in its matrix form can for example be solved by matrix inversion, yielding the Legendre coefficients of the stress function:
\begin{equation}
{\mathbf {\bar F}} = \left(  {\mathbf M}^{(n)} \right)^{-1} {\mathbf \Delta' \mathbf  {\bar w}} \, ,
\label{matrixinversion}
\end{equation}
An example of the membrane matrix and of its solution is given in Appendix \ref{example} for an expansion of $\bar\alpha$ limited to degree two and a truncation degree equal to~6.
If the thickness variation is symmetric about the equatorial plane, only Legendre polynomials of even degree will contribute to the solution because the deformation is also symmetric (degree zero or two).

Since $\bar\alpha=1+\bar\alpha_2 P_2$ is finite even if $h_E$ or $h_P$ vanishes (though not both), the nondimensional membrane equation can in principle be solved with a vanishing thickness at the equator or at the pole.
Problems of divergence however appear in these extreme cases for the nondimensional stress function and the stress resultants, though the stresses and the strains are well behaved.
The membrane matrix ${\mathbf M}^{(n)}$ is invertible in the physical range of $\nu\in[0,1/2]$ and for $\bar\alpha_2\in[-1,2]$, as can be seen from the examination of its eigenvalues.

If $\bar\alpha$ is limited to degree two, the approximate membrane equation (\ref{flexBmat}) can also be solved by expanding ${\mathbf {\bar F}}$ in $\bar\alpha_2$ and solving order by order:
\begin{equation}
{\mathbf {\bar F}} = \sum_{p=0}^\infty (\bar\alpha_2)^p \, {\mathbf {\bar F}}^{(p)} \, .
\label{alphaseries}
\end{equation}
It is convenient to split the membrane matrix so that the dependence in $\bar\alpha_2$ becomes explicit:
\begin{equation}
{\mathbf M}^{(n)} = {\mathbf M}^{(n)}_0 + \bar\alpha_2 \, {\mathbf M}^{(n)}_2 \, .
\label{membranematrixalpha2}
\end{equation}
The matrix ${\mathbf M}^{(n)}_0$ is diagonal so that its inversion is straightforward.
The solution is then generated order by order in $\bar\alpha_2$ by a recurrence relation:
\begin{equation}
{\mathbf {\bar F}}^{(p+1)} = - \bar\alpha_2 \, \left(  {\mathbf M}^{(n)}_0 \right)^{-1}  {\mathbf M}^{(n)}_2 \, {\mathbf {\bar F}}^{(p)} \, ,
\label{recurrence}
\end{equation}
which is initiated with ${\mathbf {\bar F}}^{(0)}$, the solution of ${\mathbf M}^{(n)}_0 {\mathbf {\bar F}}= {\mathbf \Delta' \mathbf {\bar w}}$.
This method can be generalized to an $\bar\alpha$ of degree higher than two.
Since the largest eigenvalues of the iteration matrix appearing in Eq.~(\ref{recurrence}) tend to~1 (from below) as the truncation degree increases,
the series (\ref{alphaseries}) may diverge for near extremal values of $\bar\alpha_2$.
If $|\bar\alpha_2|<1$, the series converges but the convergence can be very slow if $\bar\alpha_2$ is close to -1.
Even if it is not the best numerical method, the $\bar\alpha_i$-expansion provides an explicit solution which is helpful to understand properties such as the rule governing the decrease of the Legendre coefficients of the solution (see Appendix \ref{example}) or the pseudo-nodes in the stress and strain components (see Appendices \ref{ContractionFirstOrder} and \ref{DespinningFirstOrder}).

\section{Tectonic patterns}
\label{section5}

\subsection{Stress and faulting}

Tectonic patterns can be predicted from the analysis of stresses.
According to Anderson's theory of faulting, the faulting style depends on how the vertical (or radial) stress compares with the horizontal (or tangential) stresses  \citep{melosh1977,jaeger,schultz2009}.
Thrust faults, strike-slip faults and normal faults occur if the radial stress is respectively least compressive, intermediate or most compressive among the principal stresses.
Thrust faults and normal faults strike in the direction of the intermediate principal stress, whereas strike-slip faults strike in a direction at about $30^\circ$ from the direction of the most compressive stress.
The radial stress vanishes since faults occur at the surface (this assumption is criticized by \citet{golombek1985}).
Anderson's theory presupposes that all principal stresses are compressive, at least when including the lithostatic pressure that must be added to the stresses computed from thin shell theory.
I will briefly mention below the possible occurrence of near-surface tensile failure when one (or more) principal stress is extensional.
This point is discussed in more depth by \citet{melosh1977} and \citet{schultz1994}.

\begin{figure}
   \centering
   \includegraphics[width=6.9cm]{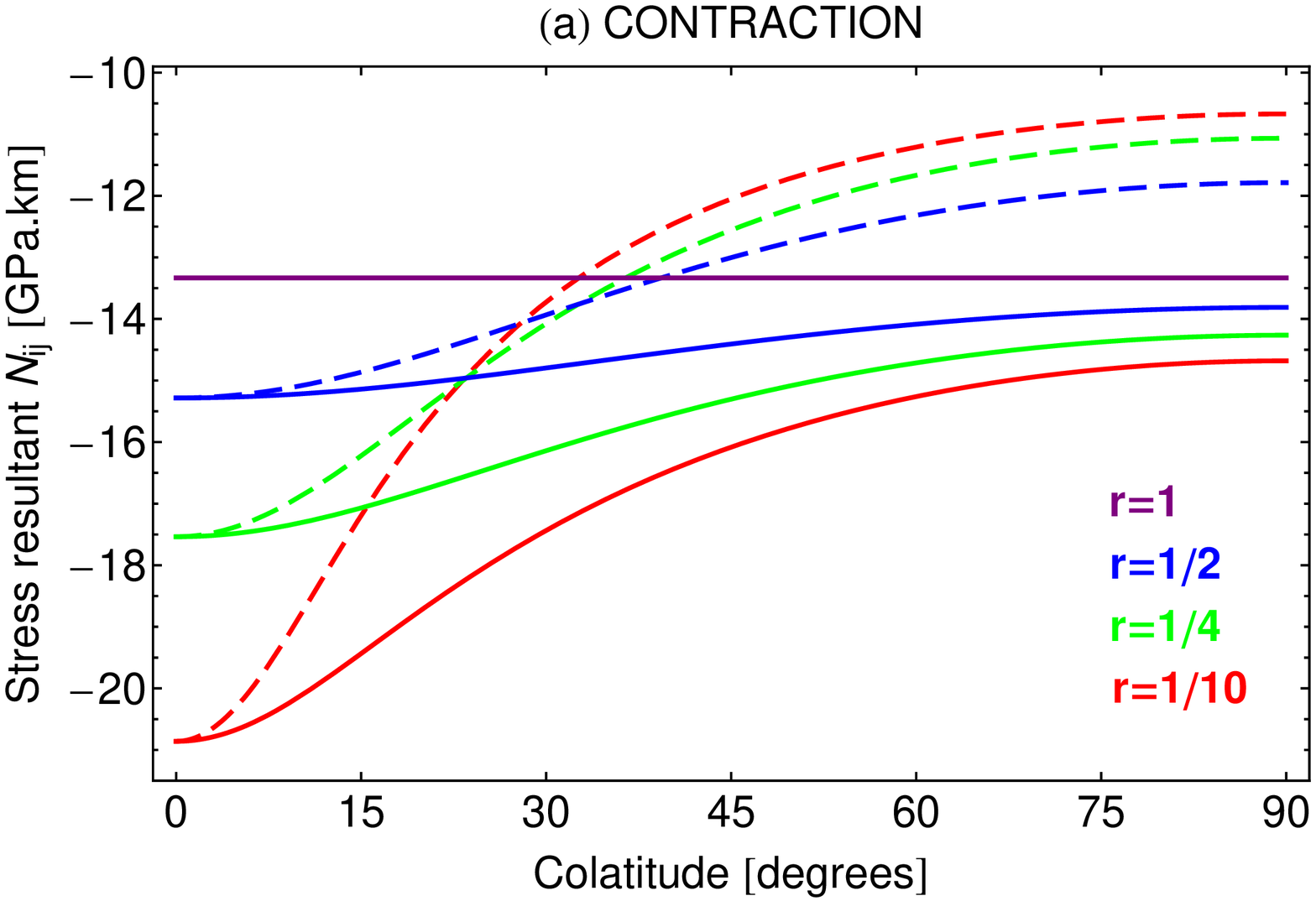}
   \hspace{0.5cm}
    \includegraphics[width=6.9cm]{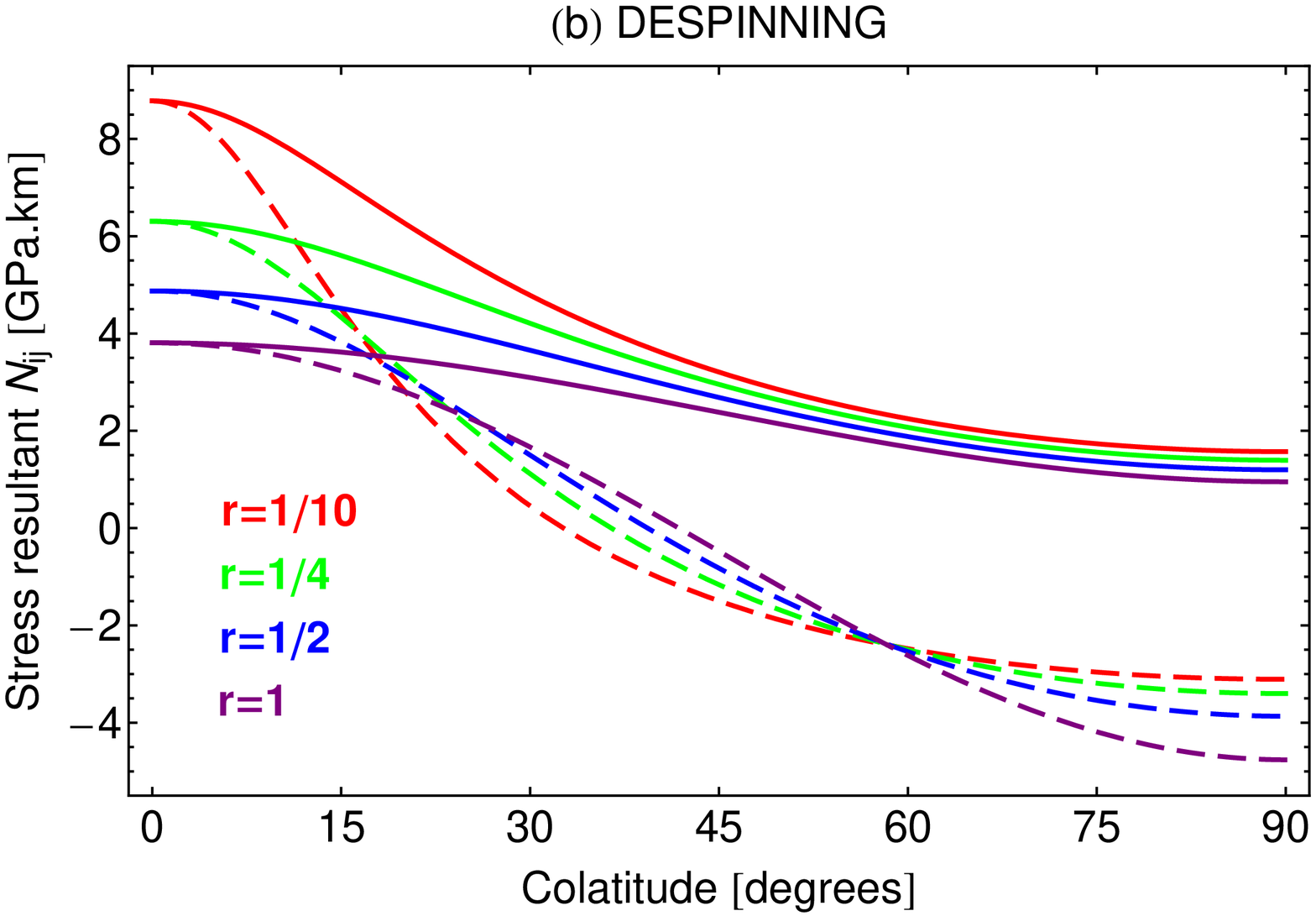} 
   \caption{\footnotesize
   Stress resultant as a function of colatitude when the lithosphere is thinner at the equator:
   (a) contraction of $0.1\,\%$ ($\bar w_0\!=\!-0.001$), (b) despinning with flattening change $\delta f=-0.0015$ ($\bar w_2\!=\!0.001$).
   Continuous curves refer to $N_{\theta\theta}$ while dashed curves refer to $N_{\varphi\varphi}$.
   Elastic parameters are $E=100\rm\,GPa$ and $\nu=0.25$.
   The thickness variation is parameterized by the equator-to-pole thickness ratio $r$ as in Fig.~\ref{AlphaCurves};
   the average inverse thickness is equal to $1/(100\rm\,km)$.
   Truncation degree $n$ is equal to 20. Tensile stress is positive.
   The duality (\ref{relationCD2}) is apparent when comparing contraction and despinning curves.}
   \label{StressIntContraction}
\end{figure}

\begin{figure}
   \centering
   \includegraphics[width=6.9cm]{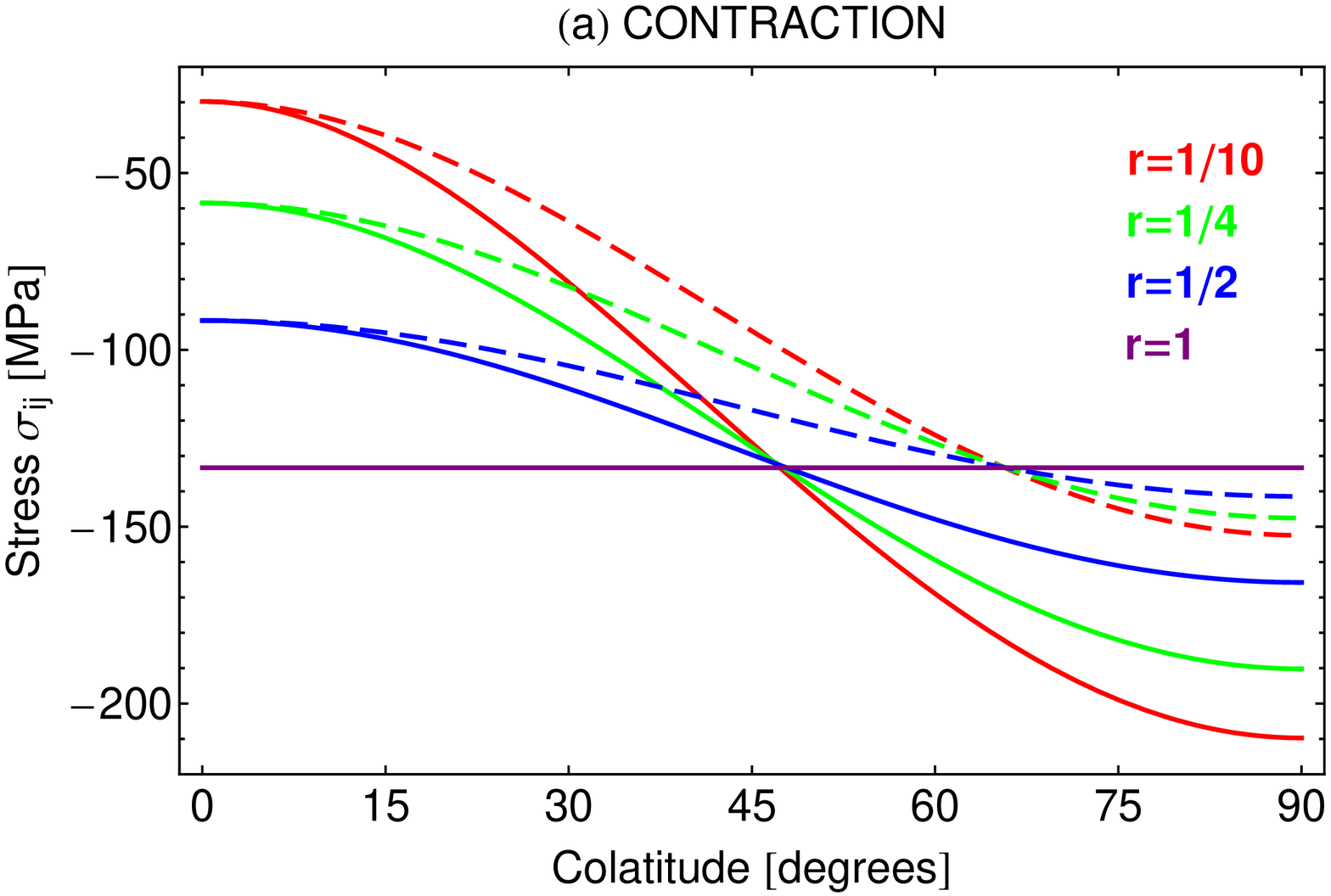} 
   \hspace{0.5cm}
    \includegraphics[width=6.9cm]{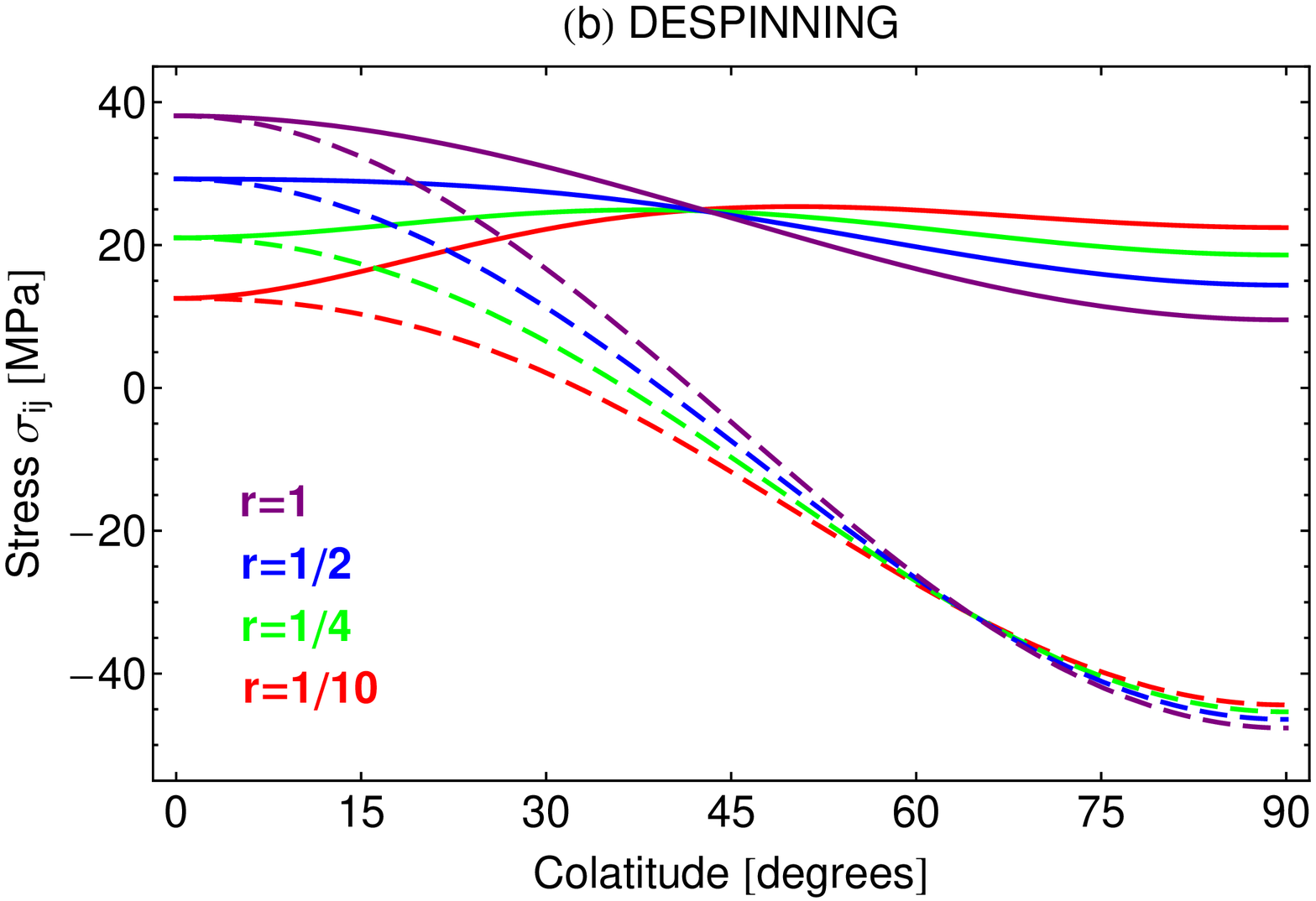}
   \caption{\footnotesize
   Stress as a function of colatitude when the lithosphere is thinner at the equator: (a) contraction, (b) despinning.
    Continuous curves refer to  $\sigma_{\theta\theta}$ while dashed curves refer to $\sigma_{\varphi\varphi}$.
   Other details as in Fig.~\ref{StressIntContraction}.}
   \label{StressSurfContraction}
\end{figure}

\begin{figure}
   \centering
   \includegraphics[width=6.9cm]{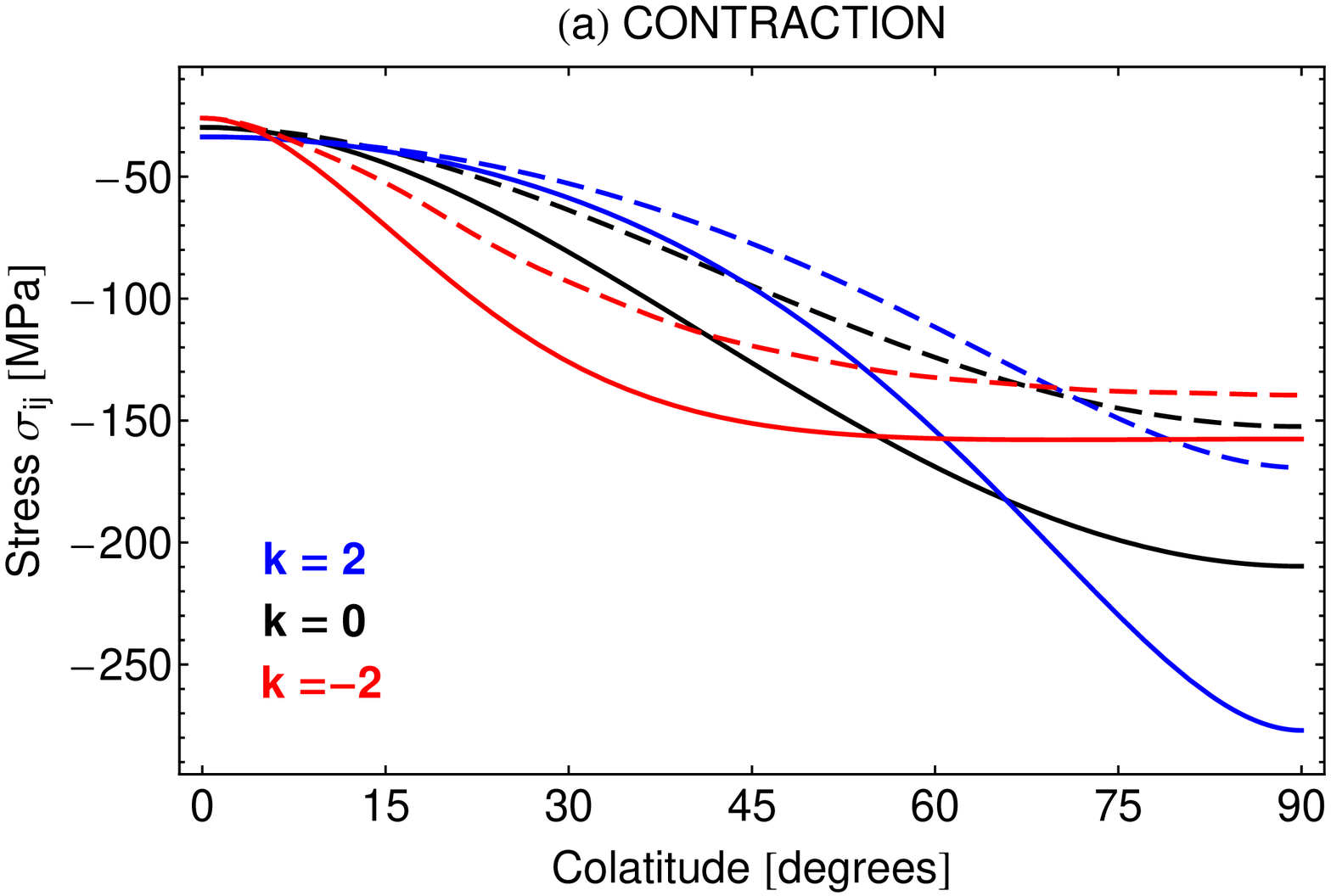} 
   \hspace{0.5cm}
    \includegraphics[width=6.9cm]{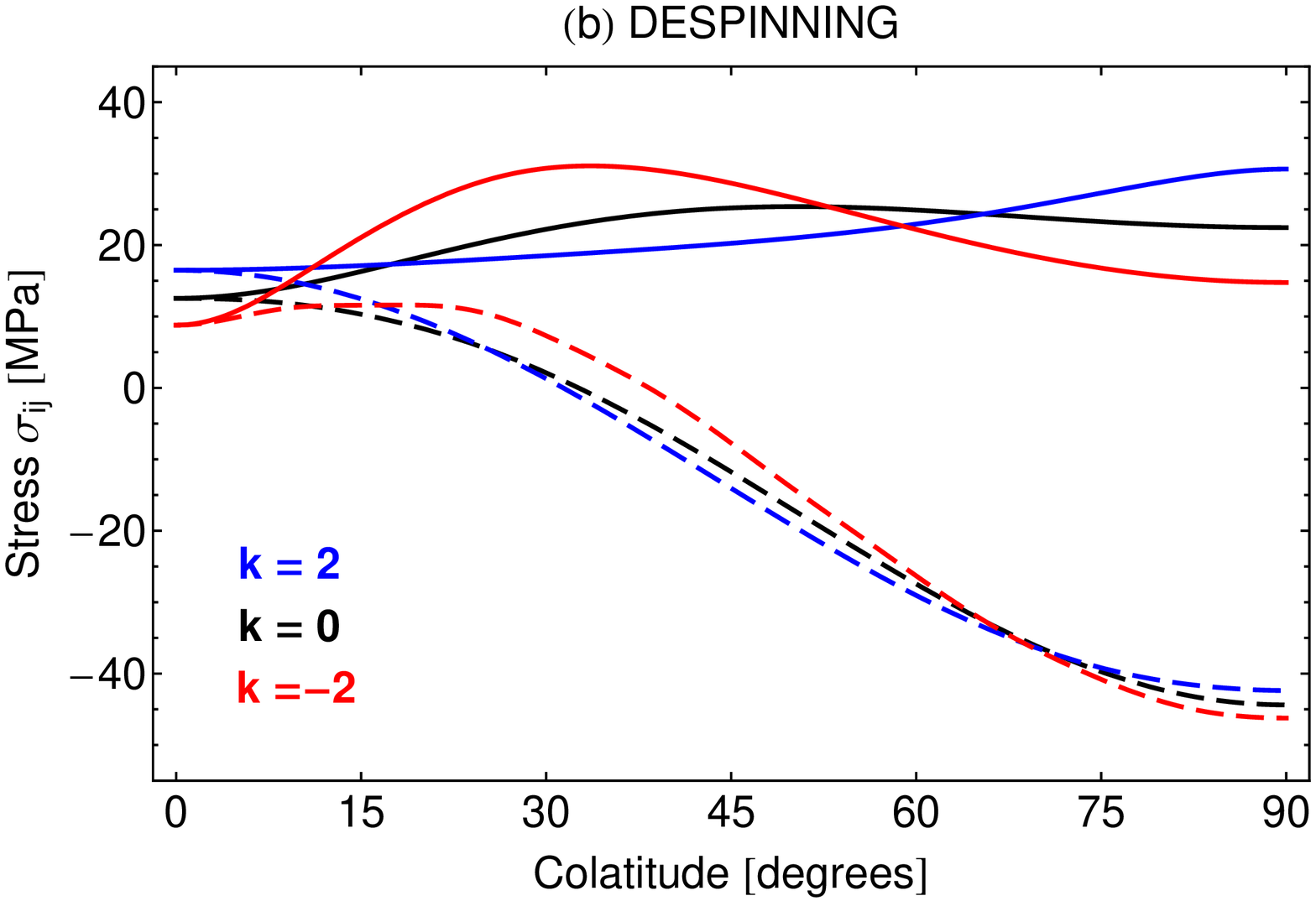}
   \caption{\footnotesize
   Influence of thin zone size on stress: (a) contraction, (b) despinning.
   Continuous curves refer to  $\sigma_{\theta\theta}$ while dashed curves refer to $\sigma_{\varphi\varphi}$.
   The thin zone size varies as in Eq.~(\ref{expansionalphaTer}) (see also Fig.~\ref{DeformedCurves}, but $r$ is different).
   The equator-to-pole thickness ratio $r$ is equal to $1/10$ ($\tilde\alpha_2=-6/7$).
   The curve $k=2$ (resp. $k=-2$) corresponds to a thin zone reduction (resp. enlargement). 
   The curve $k=0$ is the same as the curve $r=1/10$ in Fig.~\ref{StressSurfContraction}.
   Other details as in Fig.~\ref{StressIntContraction}.}
   \label{StressDeformed}
\end{figure}

Once the contraction solution of the membrane equation (\ref{flexB}) has been found with the methods of Section \ref{section4}, it is possible to compute stress resultants, stresses and strains with the formulas of Section \ref{section2}.
The nondimensional stress is sufficient for the determination of tectonic patterns since the physical magnitude of the stress is only useful when comparing the predicted stresses with a failure criterion.
Nonetheless figures represent dimensional quantities for values of parameters typical of terrestrial planets: $E=100\rm\,GPa$, $\nu=0.25$, average inverse thickness equal to $1/(100\rm\,km)$, contraction of $0.1\%$ and flattening reduction of $0.15\%$.
I initially give examples with $\bar\alpha$ limited to harmonic degree two, i.e. $\bar\alpha$ is parameterized by Eq.~(\ref{expansionalphaBis}).
Most basic characteristics of the tectonic patterns are indeed independent of the precise latitudinal variation of the elastic thickness as long as it smoothly varies from the pole to the equator.
Figs.~\ref{StressIntContraction} and \ref{StressSurfContraction} show the stress resultants and the stresses for the contraction and despinning of a lithosphere thinner at the equator.
The duality between contraction and despinning solutions appears clearly on Fig.~\ref{StressIntContraction}.
The stress curves for varying $\bar\alpha_2$ seem to cross at a fixed point on Fig.~\ref{StressSurfContraction} but this is only true at first order in $\bar\alpha_2$ (see Appendices \ref{ContractionFirstOrder} and \ref{DespinningFirstOrder}).
The case of polar thinning is illustrated by Fig.~\ref{StrainSym}, which shows the strains for expansion and despinning (stress curves are very similar).
Besides I will resort to the parameterization (\ref{expansionalphaTer}) in order to demonstrate two effects due to the variation in size of the thin zone: it modifies the position of the extremum of the meridional stress and it affects the orientation of thrust faults when contraction is combined with despinning.
Fig.~\ref{StressDeformed} shows the stresses for contraction and despinning when the thin zone size is extended or reduced.

\subsection{Contraction or expansion}

For the contraction of a shell thinner at the equator, the stresses have the following characteristics:
\begin{itemize}
\item
Stresses are everywhere compressive.
\item
The meridional stress is more compressive than the azimuthal stress (Fig.~\ref{StressSurfContraction}a);
This situation favors the development of thrust faults striking east-west.
\item
The stress is most compressive at the equator (Fig.~\ref{StressSurfContraction}a).
This situation favors the development of thrust faults at the equator.
Note that the stress resultant is most compressive at the poles (Fig.~\ref{StressIntContraction}a).
\end{itemize}
The contraction fault pattern predicted with Anderson's theory is thus significantly modified by a variable elastic thickness: instead of thrust faults with random strikes, the predicted pattern consists of thrust faults striking east-west and preferably formed at the equator.
The third characteristic is not strictly true for all thickness profiles.
The curve $k=-2$ in Fig.~\ref{StressDeformed}a shows that two things happen if the thin zone has a large extension.
First stress curves become nearly flat in that zone, where the stresses approach the value predicted for a lithosphere of constant thickness.
Second the latitude of the most compressive meridional stress moves away from the equator toward the latitude where the thick zone begins.
This peak in compressive stress is however not well-pronounced and for all practical purposes the most compressive stress can be said to occur in the whole thin zone.
If the thin zone is very localized, stress curves become flat in the thick zone and approach the value predicted for a lithosphere of constant thickness (this is not yet apparent on Fig.~\ref{StressDeformed}a because $k$ should be much larger than~2).

If the shell is expanding (see Fig.~\ref{StrainSym}a), stresses change sign, so that Anderson's theory predicts normal faults striking east-west and preferably formed at the equator.
Near the surface tensile failure is also possible because lithostatic pressure might be too small to render all stresses compressive.
It would then lead to the appearance of joints striking east-west, because rocks are much weaker in extension than in compression.

If the contracting shell is thinner at the poles, the azimuthal stress is more compressive than the meridional stress and they are both most compressive at the poles, so that the tectonic pattern consists of thrust faults striking north-south and preferably forming near the poles.
If an expanding shell is thinner at the poles (see Fig.~\ref{StrainSym}a), the azimuthal stress is more extensional than the meridional stress and they are both most extensional at the poles, so that the tectonic pattern consists of normal faults striking north-south and preferably forming near the poles.

\subsection{Despinning}
\label{section53}

For the despinning of a shell thinner at the equator, the stresses have the following characteristics:
\begin{itemize}
\item
The meridional stress is everywhere extensional, whereas the azimuthal stress is extensional near the poles and compressive near the equator.
This situation favors the development of normal faults near the poles and strike-slip faults near the equator.
The limit between the faulting provinces is given by the latitudes for which the azimuthal stress vanishes.
These boundaries are at about $\pm48^\circ$ latitudes (independent of $\nu$) when the elastic thickness is constant, and move toward the poles if the elastic thickness decreases at the equator.
\item
The meridional stress is more extensional - or less compressive - than the azimuthal stress.
This situation either favors normal faults striking east-west, or strike-slip faults striking at about $60^\circ$ from the north (depending on whether the tangential stresses have the same sign or not).
\item
The meridional stress is most extensional at the poles if the elastic thickness is constant.
The maximum moves toward the equator if the elastic thickness variation is large enough (if $\bar\alpha$ is limited to degree two, the maximum moves away from the pole if $\bar\alpha_2<-0.44$).
If the equator-to-pole thickness ratio is small (threshold $\approx1/4$) and the thin zone is not too large (threshold $k\approx1$ if $r=1/10$), the most extensional meridional stress can be at the equator but it is a rather extreme situation (see Fig.~\ref{StressDeformed}b).
\end{itemize}
The despinning fault pattern predicted with Anderson's theory is thus not substantially altered by the variation of the elastic thickness.
There is an equatorial province of strike-slip faults striking at about $60^\circ$ from the north, plus two polar provinces of normal faults striking east-west.
The boundaries between the faulting provinces move toward the poles as the elastic thickness becomes thinner at the equator.
As in the case of expansion, tensile failure may occur near the surface, leading to the production of east-west joints.
In the rather extreme case where the meridional stress is most extensional at the equator, these joints would first form at the equator.

If the shell is spinning up, stresses are the same except for a sign change, so that the tectonic pattern consists of an equatorial province of strike-slip faults striking at about $30^\circ$ from the north and polar provinces of thrust faults striking east-west.

If the shell is thinner at the poles (see Fig.~\ref{StrainSym}b), the tectonic pattern is similar to the one for a constant elastic thickness, except that the boundaries between the tectonic provinces shift toward the equator as the shell becomes thinner at the poles.

\subsection{Contraction plus despinning}

\begin{figure}
   \centering
   \includegraphics[width=6.9cm]{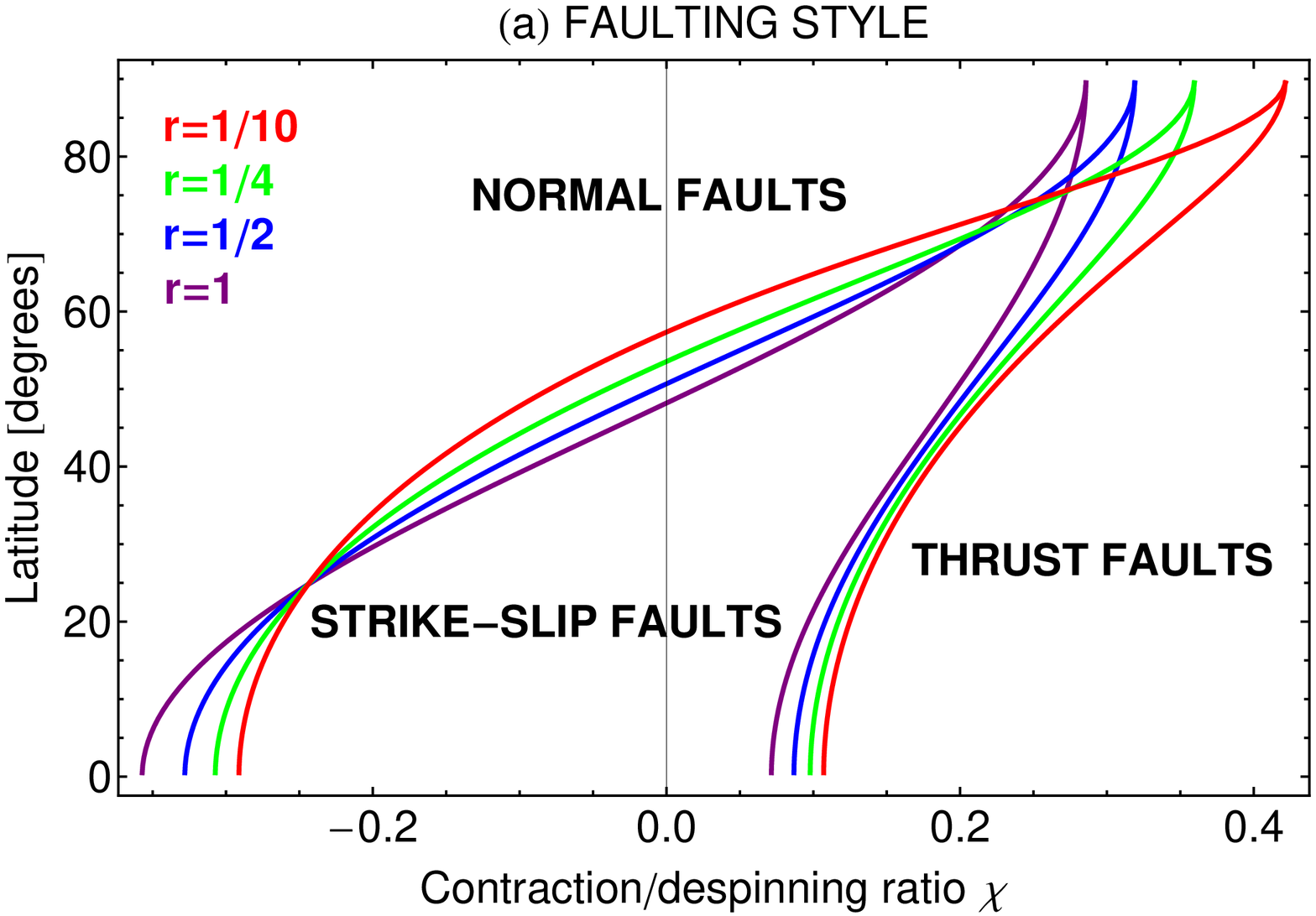}
   \hspace{0.5cm}
    \includegraphics[width=6.9cm]{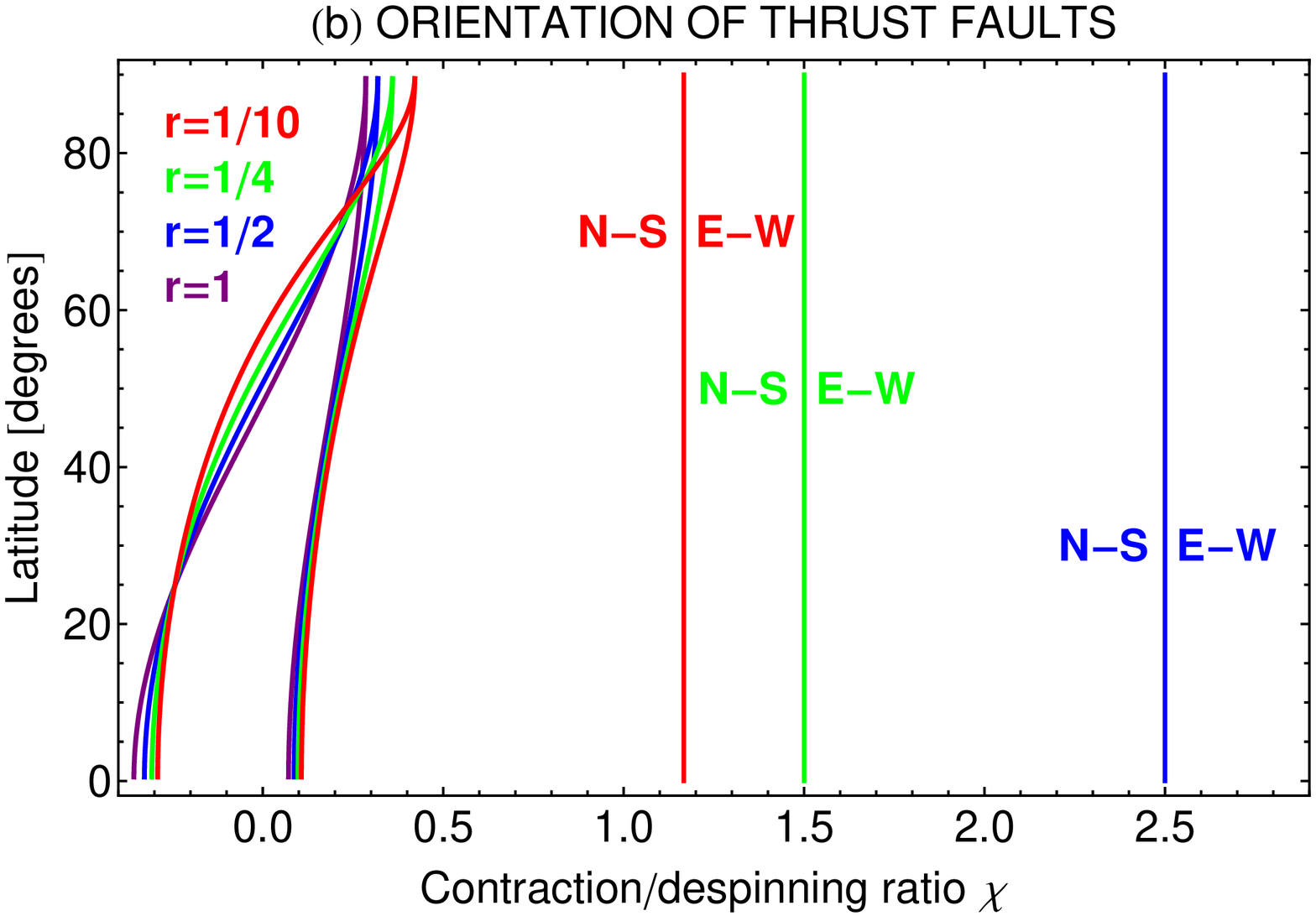}
   \caption{\footnotesize Faulting style and orientation for simultaneous despinning and contraction (or expansion) when the lithosphere is thinner at the equator.
   The thickness variation is parameterized by the equator-to-pole thickness ratio $r$ as on Fig.~\ref{AlphaCurves}.
   The x-axis is the contraction/despinning ratio $\chi$ defined by Eq.~(\ref{defchi}).
   Poisson's ratio is equal to $0.25$.
   (a) Boundaries of the areas with a given fault style.
   For given $\bar\alpha_2$ and $\chi$, the intersections of the curves with a vertical line of abscissa $\chi$ give the latitudes of the boundaries between tectonic provinces.
   (b) Contraction threshold beyond which the orientation of thrust faults changes from north-south to east-west.
   If $\bar\alpha$ is limited to degree two as is the case here, the threshold is independent of latitude and given by $\chi=-1/\bar\alpha_2$.
   The threshold for $r=1$ ($\bar\alpha_2=0$) is at infinity because the orientation of thrust faults never switches from north-south to east-west when the elastic thickness is constant.
   The boundaries of the areas with a given fault style are shown as in (a).}
   \label{Faultstyle}
\end{figure}

\begin{figure}
   \centering
   \includegraphics[width=14cm]{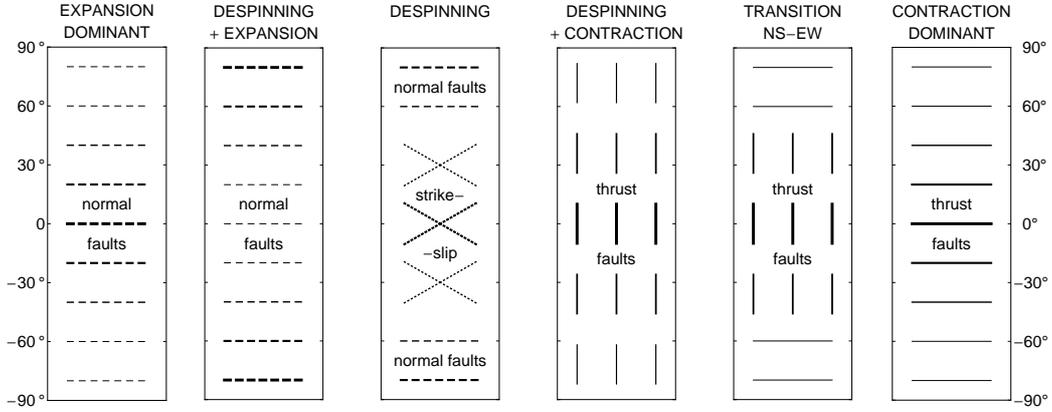}
   \caption{\footnotesize
   Tectonic patterns predicted by Anderson's theory for simultaneous despinning and contraction/expansion when the lithosphere is thinner at the equator.
    This is the pictorial interpretation of Fig.~\ref{Faultstyle}.
    The y-axis is the latitude.
    The third pattern from left is for despinning only; expansion and contraction are respectively added to the left and to the right of it.
    Normal faults, strike-slip faults and thrust faults are respectively represented with dashed, dotted and continuous lines.
    Faulting preferably occurs where the lines are the thickest.
    The figure does not show all possible transition states.
     The second pattern from the left is only realized if the equator-to-pole thickness ratio is close to one.
    The transition pattern EW-NS is shown for the case of an enlarged thin zone ($k<0$).}
   \label{TectonicPattern}
\end{figure}

Let us now consider tectonics due to simultaneous contraction and despinning.
I initially assume that the lithosphere is thinner at the equator and I use the parameterization (\ref{expansionalphaBis}) to start with.
Results for the case of polar thinning are given at the end.
Fig.~\ref{Faultstyle} (inspired by Fig.~5 of \citet{melosh1977}) represents the latitudes of the boundaries between tectonic provinces when a varying amount of contraction (or expansion) is added to despinning (Poisson's ratio is chosen to be $0.25$ but only weakly affects the position of the boundaries).
The proportion between contraction and despinning is parameterized by the contraction/despinning ratio $\chi$:
\begin{equation}
\chi= - \frac{\bar w_0}{\bar w_2} \, ,
\label{defchi}
\end{equation}
which is zero if there is only despinning, positive if there is additional contraction, negative if there is additional expansion.
Fig.~\ref{TectonicPattern} illustrates some possible tectonic patterns.
The predicted tectonic pattern has the following features:
\begin{itemize}
\item
If there is no contraction or expansion, strike-slip faults are predicted near the equator and east-west normal faults near the poles.
The boundaries of tectonic provinces move by a few degrees with respect to the case of constant elastic thickness.
\item
If the despinning planet also contracts, the strike-slip fault province extends toward the poles.
Beyond a first threshold of contraction, an area of north-south thrust faults appears near the equator; this area gets larger if contraction increases, whereas the strike-slip fault province is split in two smaller parts which are displaced toward the poles.
Beyond a second threshold of contraction, the provinces of normal and strike-slip faults vanish.
Thrust faults are then predicted over the whole planet.
The two thresholds have a moderate dependence on the equator-to-pole thickness ratio and a weak dependence on Poisson's ratio.
\item
When contraction exceeds the first threshold, thrust faults striking north-south start to form.
Since thrust faults for pure contraction strike east-west, the orientation of the faults must change for a large enough contraction.
If $\bar\alpha$ is limited to degree two, Eqs.~(\ref{threshold})-(\ref{isotropicstress}) show that the orientation of thrust faults switches everywhere from north-south to east-west if $\bar w_2=\bar\alpha_2\bar w_0$, that is if $\chi=-1/\bar\alpha_2$.
\item  If the despinning planet also expands, the strike-slip faulting province becomes smaller and vanishes for a large enough expansion.
East-west normal faults are then predicted all over the planet.
If the thickness variation is small, normal faults preferably form in polar areas.
Beyond some expansion threshold, normal faults preferably form in the equatorial region, except if the thickness is constant.
\end{itemize}

\begin{figure}
   \centering
   \includegraphics[width=6.9cm]{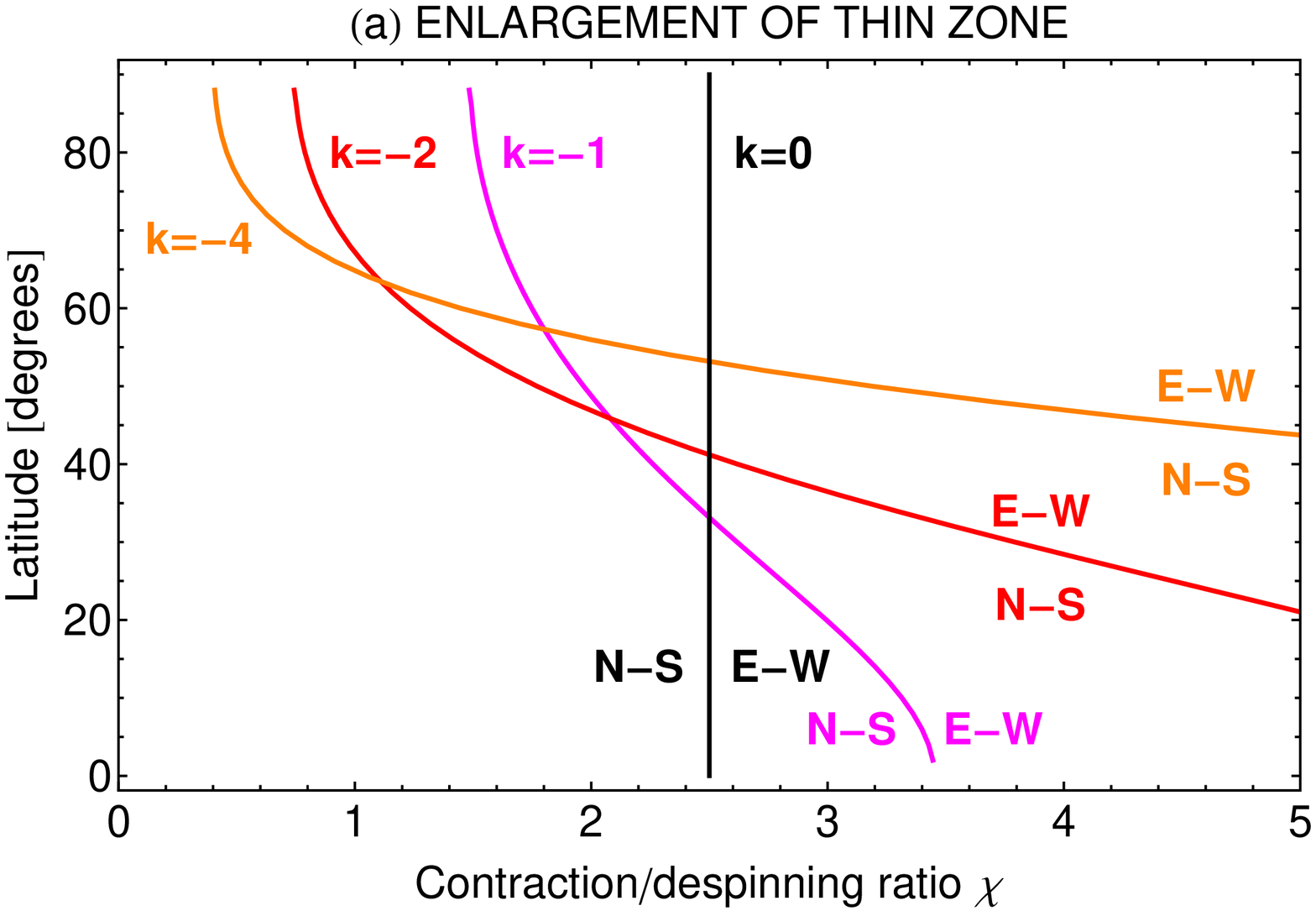}
   \hspace{0.5cm}
    \includegraphics[width=6.9cm]{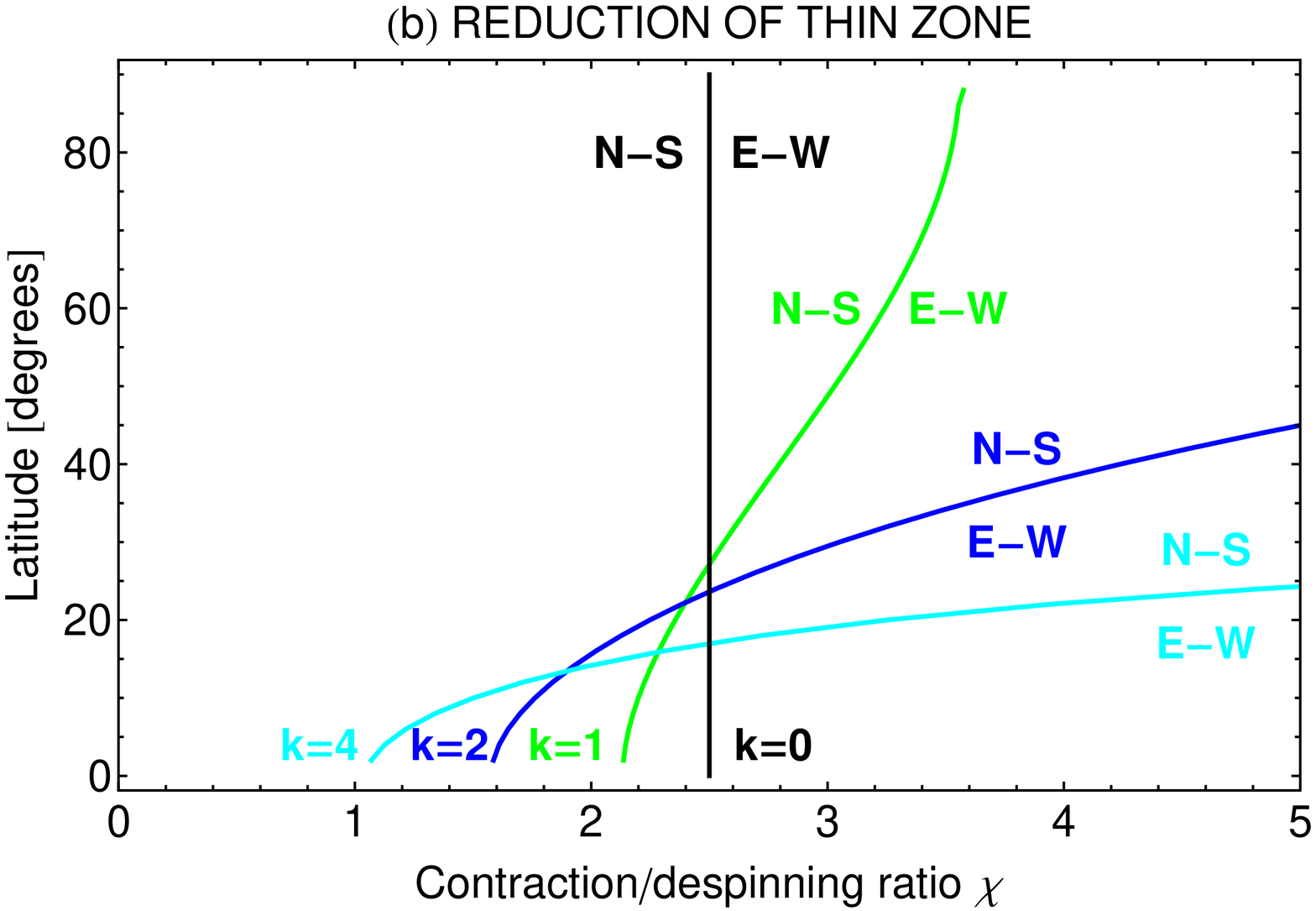}
   \caption{\footnotesize Prediction of thrust fault orientation for simultaneous despinning and contraction when the extension of the thin zone varies as on Fig.~\ref{DeformedCurves} ($r=1/2$; supplementary curves are drawn for $k=\pm1$).
   These figures generalize Fig.~\ref{Faultstyle}b: the vertical line in both graphs corresponds to the vertical line $r=1/2$ in Fig.~\ref{Faultstyle}b.
    Poisson's ratio is equal to $0.25$.
   (a) Thin zone enlargement: latitude of the transition between the polar province of east-west faults and the equatorial province of north-south faults.
   (b) Thin zone reduction: latitude of the transition between the polar province of north-south faults and the equatorial province of east-west faults.
   }
   \label{ThrustFaultOrientation}
\end{figure}

If $\bar\alpha$ is not limited to degree two, the orientation of thrust faults may depend on the latitude.
With the assumptions $\bar\alpha_2,\bar\alpha_4\ll1$ and other $\bar\alpha_i=0$ ($i>4$), Eq.~(\ref{stressdiff}) gives the sign of the stress difference at the threshold $\bar w_2=\bar\alpha_2\bar w_0$.
If $\bar\alpha_4<0$, the change from north-south to east-west thrust faults first occurs in the polar region.
As contraction increases, the frontier between provinces of thrust faults with different orientations shifts from the pole to the equator.
Conversely, the change from north-south to east-west thrust faults first occurs in the equatorial region if $\bar\alpha_4>0$ and the frontier moves from the equator to the pole as contraction increases.
These two cases are illustrated by Figs.~\ref{ThrustFaultOrientation}a,b where the parameterization (\ref{expansionalphaTer}) is used:  negative values of $\bar\alpha_4$ correspond to negative values of $k$ (enlargement of thin zone), whereas positive values of $\bar\alpha_4$ correspond to positive values of $k$ (reduction of thin zone).
For a given $k$, one can select a preferred latitude for the boundary between these provinces and then read on the x-axis the required contraction/despinning ratio $\chi$.
Using the same symbols  as in Eq.~(\ref{physicalthreshold}), I can relate the contraction to the rotation variation:
\begin{equation}
\delta R = - \chi \,h_2^T \, \frac{(\Omega_f^2-\Omega_i^2) R^2}{3g} \, .
\label{thresholdk}
\end{equation}
The above equation results from the substitution of Eqs.~(\ref{w2flat}) and (\ref{flat}) into Eq.~(\ref{defchi}).
The important conclusion is that a particular combination of contraction and despinning leads to tectonic provinces of thrust faults differing in orientation.

Finally I consider simultaneous contraction and despinning on a lithosphere thinner at the poles.
In contrast with the case of lithospheric thinning at the equator, the tectonic pattern when contraction (resp. expansion) is dominant consists of north-south thrust faults (resp. normal faults) preferably formed near the poles.
Hence thrust faults do not change in orientation as the amount of contraction increases but the area where they preferably form is displaced from the equator to the poles.
The change in orientation occurs in the expansion regime: normal faults switch from striking east-west to north-south when $\bar w_2=\bar\alpha_2\bar w_0$ (see Eqs.~(\ref{threshold})-(\ref{isotropicstress}) with $\bar\alpha_2>0$).
Near the threshold, provinces of normal faults with different orientations (north-south or east-west) can coexist.

\section{Applications}
\label{section6}

\subsection{Iapetus' ridge}

Cassini images revealed in December 2004 an extraordinary feature on Saturn's satellite Iapetus, a high ridge running along the equator spanning more than half of its circumference \citep{porco2005}.
High-resolution imaging is not available on the whole surface, but it seems that impact basins are present where the ridge has not been detected \citep{denk2008}.
It is thus reasonable to assume that the ridge originally spanned the whole circumference.
In well-imaged areas, it has an average width of $60-70\rm\,km$ and heights up to $18\rm\,km$ \citep{giese2008,denk2008}.

Several theories have been proposed to explain the origin of the ridge, including deposition of ring remnants, convection, despinning and compaction.
The morphologic analysis of the ridge by \citet{giese2008} excluded an exogenous origin due to the accretion of a now-vanished ring \citep{ip2006}.
Convection \citep{czechowski2008} cannot create a sufficiently narrow topographic rise, even if a width of $100-200\rm\,km$ is generously attributed to the ridge \citep{roberts2009}.
Despinning \citep{porco2005,melosh2009} cannot account for equatorial faults striking east-west, even if the lithospheric thickness is variable (see Section \ref{section53}; dikes are discussed below).
\citet{castillo2007} suggested that the ridge is due to surface reduction caused either by despinning or by compaction, without explaining how this surface reduction could be localized along the equator.
 
The tectonic patterns studied in this paper provide new possibilities.
In particular, the contraction of a lithosphere thinner at the equator leads to a tectonic pattern of compressional faults striking east-west and preferably formed at the equator.
The two assumptions underlying this mechanism are not arbitrary.
First, the lithosphere must be thinner at the equator because of the latitudinal variation in solar insolation, as shown in Figs.~13 and 14 of \citet{howett2010}.
Second, a collapse of porosity during the first $15\rm\,Myr$ of Iapetus occurs in the interior model of \citet{castillo2007}, leading to a global contraction of about $40\rm\,km$ during which the average lithospheric thickness is less than $20\rm\,km$.
Though it is not known whether faulting occurred at the ridge, other tectonic processes such as buckling or folding will result in the same orientation and location since the meridional stress is more compressive than the azimuthal stress and the maximum compression is at the equator.

How does despinning affect the contraction hypothesis?
Iapetus is at present locked in a 1:1 resonance with Saturn: its rotation period and orbital period are both equal to 79~days.
This synchronicity is generic of all large satellites which initially rotated much faster.
Moreover Iapetus has the peculiarity of being very flattened (equatorial and polar radii differ by $35\rm\,km$), though it should be nearly spherical if it were hydrostatic.
Its present shape is well fitted by a homogeneous hydrostatic ellipsoid with a rotation period of about 16 hours \citep{castillo2007}.
The most obvious explanation is that Iapetus had an initial rotation period of 16 hours or less and that its shape froze while despinning.
The lithosphere must have been several hundred kilometer thick when the rotation period reached 16 hours, otherwise the body would not have retained its flattened shape.
In the evolution model of \citet{castillo2007}, despinning is a rapid process occurring $800\rm\,Myr$ after formation, at which time the lithosphere has a thickness of about $240\rm\,km$.
In this scenario, contraction happens long before despinning, when the lithosphere is thin enough to be easily deformed.
In such a case the east-west faulting pattern due to contraction is not directly affected by despinning.

If contraction occurred before despinning, a question to be addressed is the absence of a superimposed tectonic pattern due to despinning.
This problem is not unique to Iapetus.
Though all large satellites underwent despinning, none exhibits unambiguous evidence of it in its global tectonic grid.
As for Iapetus, one way out is to postulate an initial period not much shorter than 16 hours.
If the lithosphere is thick enough, the flattening change is small and the associated stresses not large enough to cause faulting.

\begin{figure}
   \centering
   \includegraphics[width=6.9cm]{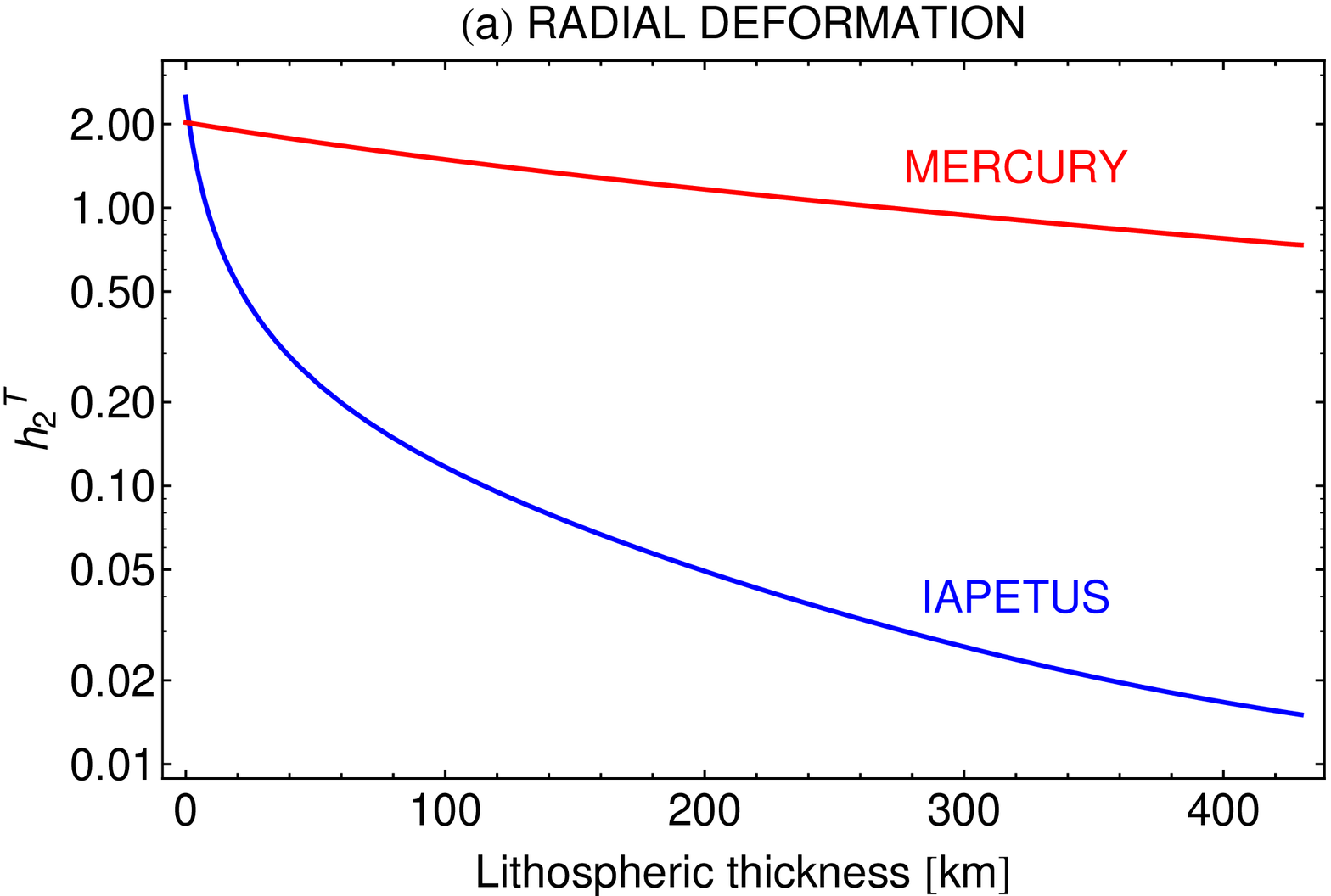}
   \hspace{0.5cm}
    \includegraphics[width=6.9cm]{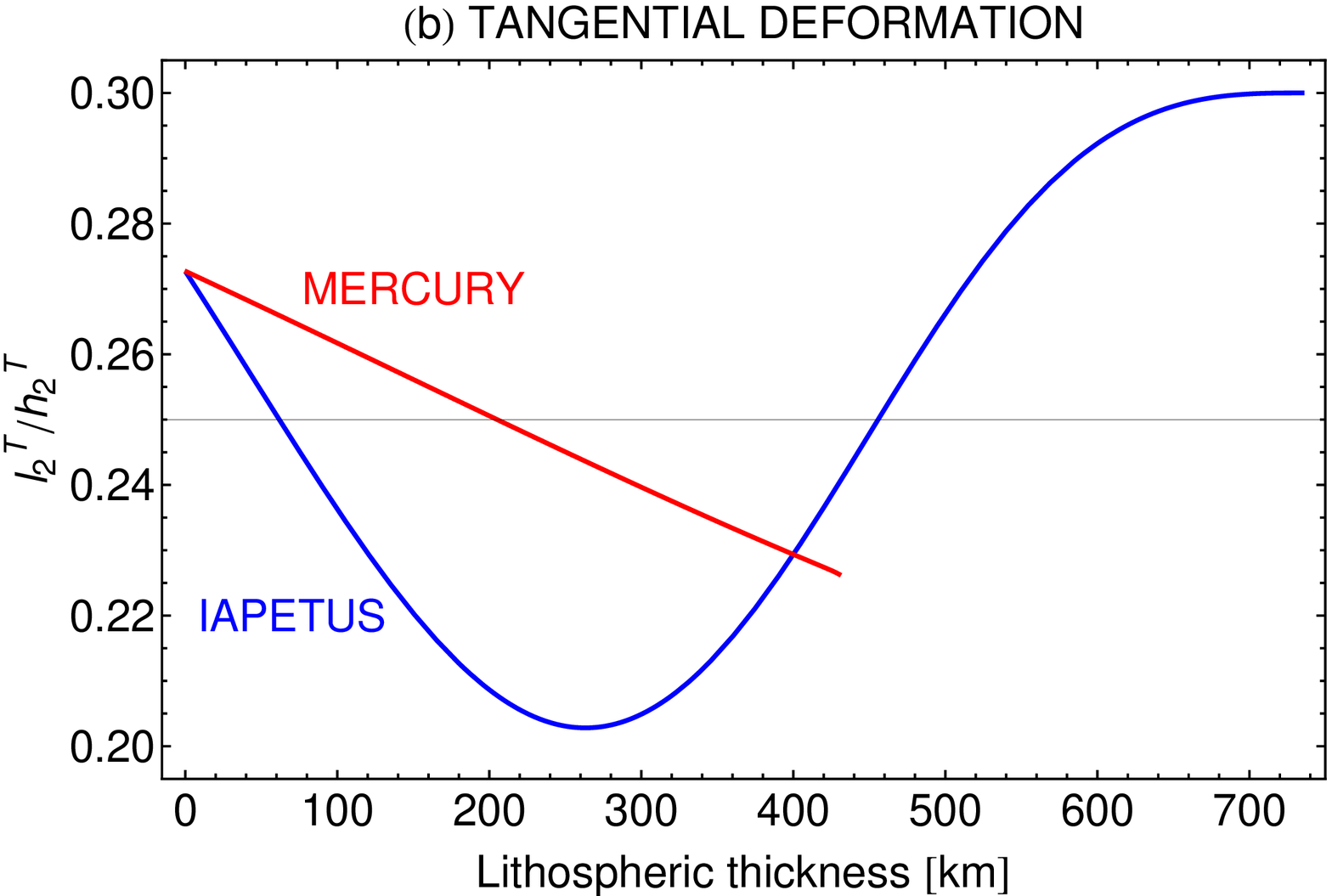}
   \caption{\footnotesize
   Secular degree-two tidal Love numbers for Mercury and Iapetus as a function of the lithospheric thickness:
   (a) $h_2^T$, (b) $l_2^T/h_2^T$.
   The model is described in Appendix~\ref{Lovenumbers}.
   For Mercury, the lithospheric thickness is limited by the size of the core.
   On an incompressible planet, despinning leads to north-south thrust faults in the equatorial region when $l_2^T/h_2^T<1/4$ (see Eq.~(\ref{thresholdthrust}) with $\nu=0.5$); this threshold is indicated by the horizontal line in graph (b).}
   \label{LoveFig}
\end{figure}

How thick should the lithosphere be in order to resist despinning stresses?
The simplest failure criterion is that of Coulomb \citep{jaeger}, which consists of a linear relation between the extreme principal stresses.
For thrust and normal faults, this criterion can be rewritten as a constraint between the maximum differential stress $\sigma_f=|\sigma_1-\sigma_3|$ ($\sigma_1$ and $\sigma_3$ are the most and least compressive stresses, respectively) and the vertical stress $\sigma_z=\rho{g}z$:
\begin{equation}
\sigma_f = a + b \, \sigma_z \, ,
\end{equation}
where $a$ is the yield strength of ice at zero pressure and $\sigma_z$ is positive.
If $\sigma_z\leq12\rm\,MPa$,  the constants $(a,b)$ for ice at a temperature of about $100\rm\,K$ are equal to $(3.4\rm\,MPa,1.86)$ and $(1.2\rm\,MPa,0.65)$ for compression and extension, respectively \citep{beeman1988}.
Failure starts either at the poles or at the equator since the differential stress is maximum at these locations.
At the poles, normal faults start to form when the meridional stress exceeds the yield strength of ice for extension.
At the equator, thrust faults form instead of strike-slip faults when the shell is thick (see Appendix \ref{Lovenumbers}): faulting begins when the azimuthal stress exceeds (in absolute value) the yield strength of ice for compression.
As discussed in Section \ref{section2}, a thick lithosphere can usually be approximated by a shell of constant thickness so that Eqs.~(\ref{despinningConst1Ter})-(\ref{despinningConst2Ter}) become valid.
Iapetus' mean radius is $735.6\rm\,km$, its mean density is $1083\rm\,kg/m^3$ and thus its surface gravity is $0.22\rm\,m/s^2$ \citep{thomas2007}.
Young's modulus is $11.7\rm\,GPa$ and Poisson's ratio is $0.325$ (see Appendix \ref{Lovenumbers}), but I compute Love numbers with $\nu=0.5$ (see Fig.~\ref{LoveFig}).
If the initial period is 16 hours, normal faults form at the poles if the thickness is smaller than $400\rm\,km$ whereas thrust faults form at the equator if the thickness is smaller than $220\rm\,km$.
These thresholds are respectively reduced to $325\rm\,km$ and $150\rm\,km$ if faulting is initiated at $5\rm\,km$ depth.
There is also a lot of uncertainty in the experimental value of the yield strength of ice at zero pressure: \citet{beeman1988} consider alternative fits leading to yield strength ranges of $[0,2.8]\rm\,MPa$ for extension and $[0,6.8]\rm\,MPa$ for compression.
The thickness thresholds associated with the upper values of these ranges are $270\rm\,km$ for normal faults and $125\rm\,km$ for thrust faults.
Combining the effect of fault initiation at depth with high yield strength values, one can obtain a thickness threshold for normal faults lower than $240\rm\,km$, which is the lithospheric thickness at the time of despinning in the model of \citet{castillo2007} (note that the initial rotation period is 7 hours in that model).
Therefore, it is conceivable that Iapetus' lithosphere resisted failure during despinning if the initial period was not much shorter than 16 hours.

Let us now examine what happens if contraction and despinning are simultaneous.
As shown in Figs.~\ref{Faultstyle} and \ref{TectonicPattern}, the combination of contraction and despinning leads to three possible faulting patterns at the equator, given in order of increasing contraction: strike-slip faults, north-south thrust faults, or east-west thrust faults.
The amount of contraction required to yield the third outcome is given by Eq.~(\ref{physicalthreshold}), at least if the lithospheric thickness varies as in Eq.~(\ref{expansionalphaBis}).
As a starting point, I set the initial rotation period to 16 hours, the equator-to-pole thickness ratio to $1/2$ (i.e. $\bar\alpha_2=-2/5$) and the Love number $h_2^T$ to $5/2$ (limit of vanishing lithospheric thickness).
The contraction threshold between north-south and east-west thrust faults is then $\delta R\approx60\rm\,km$ which is large, though of the same order of magnitude as the one due to porosity collapse in the model of \citet{castillo2007}.
Of course, the amount of contraction due to porosity depends on the initial porosity profile which is unknown.
Other values of the parameters strongly affect the result.
On the one hand, the threshold increases with the square of the initial rotation rate.
On the other hand the threshold decreases as the lithospheric thickness increases, because the flattening variation is then smaller.
The two-layer incompressible model discussed in Appendix \ref{Lovenumbers} yields $\delta R\approx13\rm\,km$ and $\delta R\approx0.9\rm\,km$ if the lithospheric thickness is equal to $20\rm\,km$ and $240\rm\,km$, respectively.
Furthermore the threshold not only decreases with the equator-to-pole thickness ratio (see Fig.~\ref{Faultstyle}b), but is also affected by the precise form of the thickness variation: Fig.~\ref{ThrustFaultOrientation}b shows that a reduction of the thin zone can decrease the value of the threshold at the equator by a factor of two.
Therefore, simultaneous contraction and despinning may lead, or not, to east-west thrust faults.
For a given amount of contraction, the outcome sensitively depends on the initial rotation rate and on the lithospheric thickness (average thickness, equator-to-pole thickness ratio, thin zone size).

Besides the contraction hypothesis, an episode of expansion in Iapetus' evolution due to differentiation \citep{squyres1986} would lead to equatorial normal faults striking east-west and forming first at the equator.
Iapetus' ridge does not look like normal faulting, but the same extensional stress field is compatible with the intrusive dike origin for the ridge proposed by \citet{melosh2009}.
In that scenario, the simultaneous occurrence of despinning is generally not a problem because it is only required that the meridional stress be most extensional at the equator.
This is true for moderate amounts of expansion, the value of the threshold depending on the thickness profile.
The expansion threshold is for example $7.5$ times smaller in magnitude than the contraction threshold discussed above, assuming an equator-to-pole thickness ratio of $1/2$ (it coincides in this particular case with the expansion threshold for normal faults at the equator).
Note that despinning can also be favorable to the formation of an east-west dike at the equator, but the conditions are rather stringent: the equator-to-pole thickness ratio must be smaller than $1/4$ and the thin zone not too large (see Fig.~\ref{StressDeformed}b).

\subsection{Mercury's scarps}

Lobate scarps on Mercury are linear or arcuate escarpments varying in length from about $20\rm\,km$ to $600\rm\,km$ and in height from a few hundred meters to about $1.5\rm\,km$ \citep{strom1975,watters2009}.
They are the most common tectonic features on Mercury and are thought to be the result of thrust faulting.
The predominant north-south orientation of lobate scarps was first attributed to illumination bias but \citet{watters2004} showed that this preferred orientation is real.
The combination of planetary contraction and despinning is a possible explanation for this pattern \citep{melosh1978,pechmann1979,dombardhauck2008}.
Another interesting observation is that lobate scarps in the southern polar region are as often east-west than north-south (see Fig.~2.23 in \citet{wattersnimmo2009}).
East-west lobate scarps are much rarer in the northern polar region, but the total length of scarps of any orientation is also much smaller than in the south.
Let us suppose that lobate scarps predominantly strike north-south from $50^\circ\rm\,S$ to $50^\circ\rm\,N$ latitude and east-west in the polar regions, as done by \citet{king2008}.
Fig.~\ref{ThrustFaultOrientation}a shows that simultaneous contraction and despinning may lead to this pattern if the thin zone is rather extended.
As mentioned in the introduction, the lithosphere could be thinner by a factor of two at the equator in comparison with the poles \citep{mckinnon1981,melosh1988}.
In that case, the required contraction/despinning ratio should be close to two, assuming that the shape parameter $k$ is between $-1$ and $-2$.
Substituting $\chi=2$ in Eq.~(\ref{thresholdk}), I obtain $\delta R\approx20\rm\,km$ if $h_2^T=5/2$ and the initial rotation period is 20 hours ($g=3.7\rm\,m/s^2$, $R=2440\rm\,km$).
The contraction threshold is somewhat reduced by more realistic values of $h_2^T$.
The thickness of the lithosphere at the time of the formation of lobate scarps is difficult to pinpoint: arguments have been made for values of $25\!-\!30\rm\,km$ and $100\!-\!200\rm\,km$ (see a summary in \citet{wattersnimmo2009}).
In any case, Fig.~\ref{LoveFig}a shows that the secular Love number $h_2^T$ is not very sensitive to the lithospheric thickness.
The contraction threshold is only reduced to $15\rm\,km$ and $12\rm\,km$ if the lithospheric thickness is equal to $30\rm\,km$ and $100\rm\,km$, respectively (see Fig.~\ref{LoveFig} and Appendix \ref{Lovenumbers} for the value of $h_2^T$).
The radius contraction at the origin of lobate scarps likely being in the range $1\!-\!2\rm\,km$ \citep{strom1975,watters2009}, it is far too small to make possible the existence of east-west thrust faults.

One kilometer of contraction, combined with despinning, is also too small to generate north-south thrust faults all over the planet, though two kilometers are enough if the yield strength of rock is neglected (use Eq.~(\ref{thresholdk}) with $\chi$ determined from Fig.~\ref{Faultstyle}a).
Yet another problem is that lobate scarps are more abundant in the south than near the equator, contrary to the prediction of contraction plus despinning.
\citet{dombardhauck2008} thus suggested that a larger amount of contraction occurred during despinning in the early history of Mercury
(their contraction threshold of $3\!-\!5.5\rm\,km$ is however significantly reduced if the factor $h_2^T$ is taken into account).
The Late Heavy Bombardment then erased the faults and new scarps were produced along the old fault lines by a later global contraction event of smaller magnitude, when the planet had despun.
Such a scenario widens the choice of the relative weights of despinning and contraction.
On the one hand despinning needs not be finished when the early tectonic grid forms.
On the other hand early contraction can be rather large in thermal evolution models: $2.5\!-\!7.5\rm\,km$ in \citet{dombardhauck2008}, with an upper bound of $17\rm\,km$ if the core completely solidifies \citep{solomon1976} (measurements of Mercury's librations however indicate that the core is at least partially molten, see \citet{margot2007}).
In conclusion, it is possible to imagine a sequence of events in which various events of contraction and despinning lead to thrust faults that strike north-south near the equator and east-west near the poles.
However the observation that the total scarp length increases from north to south remains to be explained.
East-west thrust faults could instead be the result of the reactivation of early normal faults due to despinning \citep{wattersnimmo2009}.
The various orientations of lobate scarps could also be due to a combination of despinning, contraction and true polar wander \citep{matsuyama2009}.

\section{Conclusions}

The main result of the paper can be very simply stated: the contraction or expansion of a planet with lithospheric thinning at the equator results in tectonic features striking east-west, preferably formed in the equatorial region.
If the lithosphere is thinner at the poles, contraction or expansion generates tectonic features striking north-south.
In general, despinning alone cannot produce east-west features near the equator.
One exception could be the appearance of east-west joints or dikes at the equator if the thickness variation is large and strongly peaked at the equator.

A spectacular and rather unique illustration of an east-west structure is the equatorial ridge on Iapetus, which can be explained by the theory developed in this paper.
The mountainous structure of the ridge is suggestive of a compressional event but an extensional process is not excluded.
Since despinning inhibits the formation of east-west compressional tectonics, the ridge probably formed long before Iapetus had despun, but this is not required if the contraction is large enough.

Contraction and despinning stresses can be computed either numerically or with a semi-analytical method, which yields a perturbation expansion in the parameters describing the thickness variation.
First order formulas for stresses are sufficient if the thickness varies by a factor of two.
In the simplest case (inverse thickness of degree two), contraction stresses are well approximated by the elementary formulas (\ref{stressC1dim})-(\ref{stressC2dim}),
while Eqs.~(\ref{stressC1bis})-(\ref{stressC2bis}) are valid at first order for an arbitrary thickness profile.

The combination of radius change and despinning generates new tectonic patterns when the lithospheric thickness is variable.
On the one hand, contraction plus despinning makes possible the coexistence of tectonic provinces of thrust faults differing in orientation if the lithosphere is thinner at the equator.
Lobate scarps on Mercury roughly follow such a pattern but the small amount of contraction associated with the scarps makes it necessary to resort to rather complicated models involving fault reactivation.
On the other hand, expansion plus despinning may lead to the coexistence of tectonic provinces of normal faults differing in orientation if the lithosphere is thinner at the poles.
In both cases the magnitude of the contraction/despinning ratio must be large otherwise the predicted patterns are similar to those valid for a lithosphere of constant thickness.

Global contraction or expansion occur in many models of interior evolution and are often assumed to be the cause of planetary tectonics.
Since lithospheric thinning due to the latitudinal variation in solar insolation must be a generic phenomenon, it is surprising that an east-west orientation of tectonic features at the planetary scale is so rare in the solar system.
One reason could be that the thinning is too small to have a significant effect.
Also, tidal heating might counteract the effect of the variation in solar insolation.
Another explanation is the simultaneous occurrence of despinning.
Finally, it is possible that global contraction or expansion events generally happen very early in the history of the planet so that either faults do not appear because the lithosphere is not yet formed, or faults do form but the associated tectonic pattern is subsequently erased.
Further inferences about the origin of tectonic features depend on more complete tectonic mapping, which is underway for Mercury with Messenger data and for Saturn's icy satellites with Cassini data.

\subsection*{Acknowledgments}
This work was financially supported by the Belgian PRODEX program managed by the European Space Agency in collaboration with the Belgian Federal Science Policy Office.
I thank Isamu Matsuyama for suggesting the inclusion of Love numbers and Jay Melosh for discussions on the numerical solution.
I benefited from exchanges with Surdas Mohit, Francis Nimmo and James Roberts.
I also thank Attilio Rivoldini and Antony Trinh for discussions on Love numbers.

\section{Appendices}

\subsection{Differential operators on the sphere}
\label{DifferentialOperators}

The operators ${\cal O}_i$ are linear differential operators of the second degree on the sphere:
\begin{eqnarray}
{\cal O}_1 &=& \frac{\partial^2}{\partial \theta^2} + 1 \, ,
\label{opO1} \\
{\cal O}_2 &=&  \csc^2 \theta \, \frac{\partial^2}{\partial \varphi^2} + \cot \theta \, \frac{\partial}{\partial \theta} + 1 \, ,
\label{opO2} \\
{\cal O}_3 &=& \csc \theta  \left( \frac{\partial^2}{\partial \theta \partial \varphi} - \cot \theta \, \frac{\partial}{\partial \varphi} \right) \, ,
\label{opO3}
\end{eqnarray}
where $\theta$ is the colatitude and $\varphi$ is the longitude.
They give zero when applied on spherical surface harmonics of degree one.

The scalar differential operator $\Delta'$ is defined by \citep{beuthe2008}:
\begin{eqnarray}
\Delta' &=& {\cal O}_1 + {\cal O}_2
\label{defD1} \\
&=& \Delta + 2 \, ,
\label{defD3}
\end{eqnarray}
where
$\Delta$ is the spherical Laplacian (called surface Laplacian in \citet{beuthe2008}).
Spherical surface harmonics of degree $\ell$ are eigenfunctions of $\Delta'$ with eigenvalues \citep[][pp.~121-122]{blakely}
\begin{equation}
\delta'_\ell = - \ell(\ell+1) + 2 \, .
\label{eigendelta}
\end{equation}

The scalar differential operators ${\cal C}$ and ${\cal A}$ appearing in the membrane equation (\ref{flexB}) are defined by
\begin{eqnarray}
{\cal C}(a\,;b) &=& \Delta' \left( a \, \Delta' \,  b \right) \, ,
\label{defC}
\\
{\cal A}(a\,;b) &=& ({\cal O}_1 \, a)({\cal O}_2 \, b) +  ({\cal O}_2 \, a)({\cal O}_1 \, b) - 2 \, ({\cal O}_3 \, a)({\cal O}_3 \, b) \, .
\label{defA1}
\end{eqnarray}
Since ${\cal C}(a\,;b)$, ${\cal A}(a\,;b)$ and $\Delta'b$ give zero if $b$ is a spherical surface harmonic of degree one, the membrane equation (\ref{flexB}) does not constrain the degree-one components of the stress function $F$ and of the transverse displacement $w$ (translation invariance).

${\cal C}$ and ${\cal A}$ are linear in $a$ and in $b$.
If $a$ is constant ${\cal A}(a\,;b)=a\Delta'b$ whereas ${\cal A}(a\,;b)=b\Delta'a$ if $b$ is constant, so that the following identities hold:
\begin{eqnarray}
{\cal C} \left( 1 \,; b \right) - (1+\nu) \, {\cal A} \left( 1 \,; b \right) &=&  \left( \Delta' - 1 - \nu \right) \Delta' b \, .
 \label{flexureidentity1}
 \\
{\cal C} \left( a \,; 1 \right) - (1+\nu) \, {\cal A} \left( a \,; 1 \right) &=&  \left(1-\nu\right) \Delta' a \, .
 \label{flexureidentity2}
\end{eqnarray}

\subsection{Operators ${\cal O}_i$ on Legendre polynomials}
\label{OiOnPoly}

Legendre polynomials $P_\ell(x)$ are defined in many books \citep{whittaker,blakely}.
I only give those that appear in the text, in the form $P_\ell(\cos\theta)$:
\begin{eqnarray}
P_0 &=& 1 \, ,
\label{defP0} \\
P_2 &=& \frac{1}{4} \left( 3 \cos 2 \theta + 1 \right)  \, ,
\label{defP2} \\
P_4 &=& \frac{1}{64} \left( 35 \cos 4 \theta + 20 \cos 2\theta + 9 \right) \, ,
\label{defP4} \\
P_6 &=& \frac{1}{512} \left( 231 \cos 6 \theta + 126 \cos 4 \theta + 105 \cos 2 \theta + 50 \right) \, .
\label{defP6}
\end{eqnarray}

The action of the operators ${\cal O}_i$ defined by Eqs.~(\ref{opO1})-(\ref{opO3}) on Legendre polynomials $P_\ell(\cos\theta)$ is
\begin{eqnarray}
{\cal O}_1 \, P_\ell &=& \left( 1-\ell^2 \right) P_\ell + \Sigma_\ell \, ,
\label{Opoly1} \\
{\cal O}_2 \, P_\ell &=& \left( 1-\ell \right) P_\ell - \Sigma_\ell \, ,
\label{Opoly2} \\
{\cal O}_3 \, P_\ell &=& 0 \, ,
\label{Opoly3} 
\end{eqnarray}
where
\begin{eqnarray}
\Sigma_\ell &=& \sum_{r=1}^p \left( 2\, \ell - 4 \, r + 1 \right) P_{\ell-2r} \, ,
\nonumber \\
&=& (2\ell-3) \, P_{\ell-2} + (2\ell-7) P_{\ell-4} + ...
\label{Sigmal}
\end{eqnarray}
with $p=\ell/2$ for even $\ell$ and $p=(\ell-1)/2$ for odd $\ell$.
The action of ${\cal O}_2$ is obtained from an expression for the first derivative of a Legendre polynomial found in \citet[][p.~330]{whittaker}.
The action of ${\cal O}_1$ is then obtained from Eqs.~(\ref{defD1}) and (\ref{eigendelta}).

The values of ${\cal O}_1P_\ell$ and ${\cal O}_2P_\ell$ are equal at the poles:
\begin{eqnarray}
{\cal O}_i P_\ell |_{\theta=0} &=& \frac{1}{2} \, \delta'_\ell \hspace{5mm} (i=1,2) \, ,
\label{OiN} \\
{\cal O}_i P_\ell |_{\theta=\pi} &=&  \frac{(-1)^\ell}{2} \, \delta'_\ell \hspace{5mm} (i=1,2) \, ,
\label{OiS}
\end{eqnarray}
where $\delta'_\ell$ is defined by Eq.~(\ref{eigendelta}).

Since the coefficients multiplying the Legendre polynomials in Eqs.~(\ref{Opoly1})-(\ref{Opoly2}) are independent of $\theta$, the operators ${\cal O}_i$ can be represented as matrices acting on vectors of Legendre coefficients.
The action of ${\cal O}_1$ on a vector of Legendre coefficients of degrees 0, 2, 4 and 6 is represented by the matrix
\begin{equation}
{\mathbf O}_1^{(6)} =
\left(
\begin{array}{rrrr}
1 & 1 & 1 & 1 \\
0 & -3 & 5 & 5 \\
0 & 0 & -15 & 9 \\
0 & 0 & 0 & -35
\end{array}
\right)
\, .
\label{O1mat}
\end{equation}
The action of ${\cal O}_2$ on a vector of Legendre coefficients of degrees 0, 2, 4 and 6 is represented by the matrix
\begin{equation}
{\mathbf O}_2^{(6)} =
\left(
\begin{array}{rrrr}
1 & -1 & -1 & -1 \\
0 & -1 & -5 & -5 \\
0 & 0 & -3 & -9 \\
0 & 0 & 0 & -5
\end{array}
\right)
\, .
\label{O2mat}
\end{equation}

\subsection{Constant thickness: thin shell}
\label{ConstantElasticThickness}

When the shell thickness $h$ is constant, the membrane equation (\ref{flexB}) becomes
\begin{equation}
\left( \Delta' - 1 - \nu \right) \Delta' F = \frac{Eh}{R} \, \Delta' w \, ,
\label{flexBconst}
\end{equation}
where I used the identity (\ref{flexureidentity1}).
Eq.~(\ref{flexBconst}) is diagonalized by expanding $F$ and $w$ in spherical surface harmonics whose coefficients ($\ell\neq1$) satisfy (see Eq.~(\ref{eigendelta}))
\begin{equation}
F_{\ell m} = \frac{Eh}{R} \frac{1}{-\ell(\ell+1) +1-\nu} \, w_{\ell m} \, ,
\label{Flm}
\end{equation}
The index $m$ refers to the order which is equal to zero under the assumption of axial symmetry.
The component of degree one remains indeterminate.
The stress function generates the stresses according to Eqs.~(\ref{stressInt})-(\ref{stressSurf}).
The action of ${\cal O}_i$ on Legendre polynomials of degree 0, 2, 4, 6 is given by Eqs.~(\ref{O1mat})-(\ref{O2mat}).

For a contracting shell, $w=w_0 P_0$ and the stress function is constant ($P_0=1$):
\begin{equation}
F^C = \frac{Eh}{1-\nu} \, \frac{w_0}{R} \, P_0 \, .
\label{FCconst}
\end{equation}
The non-zero stresses are
\begin{equation}
\sigma^C_{\theta\theta} = \sigma^C_{\varphi\varphi} = \frac{E}{1-\nu} \, \frac{w_0}{R} \, .
\end{equation}

For a despinning shell,  $w=w_2 P_2$ and the stress function reads
\begin{equation}
F^D = - \frac{Eh}{5+\nu} \, \frac{w_2}{R} \, P_2 \, .
\label{FDconst}
\end{equation}
The non-zero stresses are
\begin{eqnarray}
\sigma^D_{\theta\theta} &=&  - \frac{E}{5+\nu} \, \frac{w_2}{R} \left( - P_2 - 1  \right) \, ,
\label{despinningConst1} \\
\sigma^D_{\varphi\varphi} &=&  - \frac{E}{5+\nu} \, \frac{w_2}{R} \left( - 3 P_2+ 1 \right) \, .
\label{despinningConst2}
\end{eqnarray}
The coefficient $w_2$ can be related to the flattening variation $\delta\!f$ ($\delta\!f\!<\!0$ for despinning) by
\begin{equation}
w_2 = - \frac{2}{3} \, R \, \delta\!f \, .
\label{w2flat}
\end{equation}
After substitution of Eqs.~(\ref{defP2}) and (\ref{w2flat}), despinning stresses \citep{vening1947,melosh1977} read
\begin{eqnarray}
\sigma^D_{\theta\theta} &=& - \frac{E}{5+\nu} \, \frac{\delta\!f }{6} \left( 3 \cos 2 \theta + 5 \right)  \, ,
\label{despinningConst1Bis} \\
\sigma^D_{\varphi\varphi} &=& - \frac{E}{5+\nu} \, \frac{\delta\!f }{6} \left( 9 \cos 2 \theta -1 \right)  \, .
\label{despinningConst2Bis}
\end{eqnarray}
The flattening is related to the angular rate $\Omega$ by
\begin{equation}
f = h_2^T \, \frac{\Omega^2R}{2g} \, ,
\label{flat}
\end{equation}
where $g$ is the surface gravity and $h_2^T$ is the degree-two tidal Love number for radial displacement \citep{matsuyama2008}.

For a deformation of degree four, $w=w_4P_4$ and the stress function reads
\begin{equation}
F^E = - \frac{Eh}{19+\nu} \, \frac{w_4}{R} \, P_4 \, .
\label{FEconst}
\end{equation}
The non-zero stresses are
\begin{eqnarray}
\sigma^E_{\theta\theta} &=& - \frac{E}{19+\nu} \, \frac{w_4}{R} \left( - 3 P_4 - 5 P_2 -1 \right) \, ,
\label{degree4Const1} \\
 \sigma^E_{\varphi\varphi} &=& - \frac{E}{19+\nu} \, \frac{w_4}{R} \left( -15 P_4 + 5 P_2 + 1 \right) \, .
 \label{degree4Const2}
\end{eqnarray}

\subsection{Constant thickness:  thick shell}
\label{Lovenumbers}

In a tidal or centrifugal potential, the radial deformation of the surface of a spherically stratified planet is parameterized by the Love number $h_2^T$, as in Eq.~(\ref{flat}).
Other Love numbers noted $l_2^T$ and $k_2^T$ parameterize the tangential deformation of the surface and the gravity perturbation at the surface induced by a tidal or centrifugal potential \citep{lambeck}.
Love numbers are complicated functions of the internal planetary structure.
A simple formula exists for an incompressible body of uniform density and shear modulus (see below), but this model is not appropriate for long-term processes such as despinning.
Stresses indeed relax by viscous flow or creep everywhere within the planet except in the lithosphere, which is by definition rigid on geologic time scales.
Internal layers in which stresses have relaxed are modeled as fluid, inertial forces are neglected (static limit) and the resulting Love numbers are called secular \citep{matsuyama2008}.
The simplest interior model including a lithosphere is a two-layer spherically stratified planet made of an incompressible fluid core surrounded by an incompressible elastic lithosphere \citep{melosh1977}.
A more realistic density profile is obtained by dividing the planet into several homogeneous layers, each of constant thickness.
Tidal Love numbers for such models are readily computed with the propagator matrix technique \citep{sabadini}.
The assumption of incompressibility is not realistic but satisfactory for our purpose, since the main uncertainty regarding secular Love numbers lies in the unknown thickness of the lithosphere.

In the limit of vanishing core radius, the two-layer model gives the degree-two tidal Love numbers for a homogeneous incompressible body \citep[pp.~257-259]{love}:
\begin{eqnarray}
h_2^T &=& \frac{5}{2} \left(1+\frac{19}{2}\frac{\mu}{\rho g R} \right)^{-1} \, ,
\label{lovenumbers11} \\
l_2^T &=& \frac{3}{10} \, h_2^T \, ,
\label{lovenumbers12} \\
k_2^T &=& \frac{3}{5} \, h_2^T \, .
\label{lovenumbers13}
\end{eqnarray}
$R$ is the radius, $\rho$ is the density, $g$ is the surface gravity and $\mu$ is the shear modulus.
Note that \citet{munkmacdonald} give an incorrect formula for $l_2^T$ while \citet{lambeck} gets it right.
Love numbers for a hydrostatic homogeneous planet are obtained by setting $\mu=0$: $h_2^T=5/2$ and $k_2^T=3/2$.
The Love number $l_2^T$ does not exist for a hydrostatic body since displacements in a fluid are indeterminate in the static limit, apart from the radial displacement at the surface.

In the limit of vanishing elastic thickness, the two-layer incompressible model gives the following degree-two tidal Love numbers:
\begin{eqnarray}
h_2^T &=& \frac{5}{2} \, ,
\label{lovenumbers21} \\
l_2^T &=& \frac{3}{11} \, h_2^T \, ,
\label{lovenumbers22} \\
k_2^T &=& \frac{3}{5} \, h_2^T \, .
\label{lovenumbers23}
\end{eqnarray}
As expected, $h_2^T$ and $k_2^T$ coincide with the values for a homogeneous hydrostatic body.
More generally, a lithosphere of vanishing thickness neither affects the shape of the body nor its gravitational field, whatever the internal density distribution.
However $l_2^T$ differs from the value for a homogeneous incompressible body since the presence of a membrane modifies tangential displacements at the surface.

I now show that the value of $l_2^T$ given by Eq.~(\ref{lovenumbers22}) agrees with the membrane limit of thin shell theory.
Toroidal displacement is absent when the shell thickness is constant (axial symmetry is actually sufficient), assuming that there is no toroidal external potential.
The tangential displacement ${\bf v}$ can then be expressed as the surface gradient of a potential $S$ \citep{beuthe2008}:
\begin{equation}
{\bf v} = {\bf \bar \nabla} \, S \, .
\end{equation}
The displacement potential $S$ is related to the stress function by Eq.~(71) of \citet{beuthe2008}, which becomes in the membrane limit
\begin{equation}
\Delta \, S = R \alpha \, (1-\nu) \Delta' F - 2 w \, .
\end{equation}
The substitution in this equation of the stress function (\ref{Flm}) yields the degree-$\ell$ component of $S$ in terms of $w_\ell$:
\begin{equation}
S_\ell = \frac{1+\nu}{\ell(\ell+1)-1+\nu} \, w_\ell \, .
\end{equation}
If $V_2=-\Omega^2R^2/3$ is the degree-two component of the external potential at the surface,  $S_2$ and $w_2$ are related to degree-two tidal Love numbers by 
\begin{eqnarray}
w_2 &=& \frac{h_2^T}{g} V_2 \, ,
\label{w2Love} \\
S_2 &=& \frac{l_2^T}{g} V_2 \, .
\label{S2Love}
\end{eqnarray}
In the membrane limit, thin shell theory thus predicts that
\begin{equation}
l_2^T = \frac{1+\nu}{5+\nu} \, h_2^T \, .
\label{l2h2}
\end{equation}
This relation holds for any density distribution within the planet, as long as it is spherically symmetric.
If the lithosphere is incompressible ($\nu=1/2$), the ratio $l_2^T/h_2^T$ is equal to $3/11$, that is Eq.~(\ref{lovenumbers22}).
If the lithosphere is made of ice ($\nu\approx1/3$), the same ratio is equal to $1/4$ (this result has been noted by \citet{wahr2009} regarding tidal stresses).

The insertion of Eq.~(\ref{w2flat}) and Eqs.~(\ref{w2Love})-(\ref{S2Love}) into stress-strain relations yields the following expressions for the tangential stresses at the surface of a spherically stratified planet:
\begin{eqnarray}
\sigma^D_{\theta\theta} &=&  - \frac{E}{1-\nu^2} \, \frac{2}{3} \, \delta \! f \left( (1+\nu) \, P_2 + \frac{l_2^T}{h_2^T} \left( 1-\nu-2(2+\nu)P_2 \right)  \right) \, ,
\label{despinningConst1Ter} \\
\sigma^D_{\varphi\varphi} &=&  - \frac{E}{1-\nu^2} \, \frac{2}{3} \, \delta \! f \left( (1+\nu) \, P_2 + \frac{l_2^T}{h_2^T} \left( -1+\nu-2(1+2\nu)P_2 \right)  \right) \, ,
\label{despinningConst2Ter}
\end{eqnarray}
where I used the fact that the $rr$ component of the stress vanishes at the surface.
In the membrane limit of thin shell theory, Love numbers satisfy Eq.~(\ref{l2h2}) so that Eqs.~(\ref{despinningConst1Ter})-(\ref{despinningConst2Ter}) become equivalent to the stresses (\ref{despinningConst1Bis})-(\ref{despinningConst2Bis}).
Moreover the substitution of Eq.~(\ref{lovenumbers12}) into Eqs.~(\ref{despinningConst1Ter})-(\ref{despinningConst2Ter}) yields the same stresses as those found by \citet{melosh1977} for a homogeneous incompressible body (see his Eq.~(20)).

Once Love numbers are known for some interior model, it becomes possible to compute surface stresses and thus predict despinning tectonic patterns using Anderson's theory of faulting (see Section \ref{section5}).
This is a convenient way to compute thick shell effects on faulting style as was first done by \citet{melosh1977}, since several codes for the computation of Love numbers are now available.
The boundaries of tectonic provinces depend only on two numbers: the ratio $l_2^T/h_2^T$ and Poisson's ratio $\nu$.
For example, despinning induces an equatorial zone of north-south thrust faults if the meridional stress is compressional at the equator, that is if
\begin{equation}
\frac{l_2^T}{h_2^T} < \frac{1+\nu}{6} \, .
\label{thresholdthrust}
\end{equation}
This inequality is satisfied neither by thin shell theory (see Eq.~(\ref{l2h2})) nor by the homogeneous incompressible model (see Eq.~(\ref{lovenumbers12})), but is verified for the incompressible models shown on Fig.~\ref{LoveFig}b if the shell thickness is between $8.4\%$ and $62\%$ of the radius.
Though $h_2^T$ and $l_2^T$ are very sensitive to the internal structure (see Fig.~\ref{LoveFig}a), their ratio varies much less.
That is why the despinning tectonic pattern is not much affected by the shell thickness.

For Iapetus, I adopt an incompressible two-layer model with a radius of $735.6\rm\,km$ and a uniform density of $1083\rm\,kg/m^3$ in both layers \citep{thomas2007}.
The shear modulus $\mu$ of ice at $257\rm\,K$ is $3.52\rm\,GPa$ so that its value at $80\rm\,K$ (approximately the mean temperature of Iapetus' surface) is about $4.4\rm\,GPa$ \citep{gammon1983}.
Poisson's ratio $\nu$ for ice is $0.325$ (but $\nu=0.5$ in the model) and Young's modulus $E=2(1+\nu)\mu$ is $11.7\rm\,GPa$.
The model yields $h_2^T=0.53$ and $h_2^T=0.038$ when the lithospheric thickness is equal to $20\rm\,km$ and $240\rm\,km$, respectively (for other values, see Fig.~\ref{LoveFig}).

For Mercury, I adopt an incompressible three-layer model with a fluid core, fluid mantle and elastic lithosphere.
Parameters are chosen to be in the middle range of admissible values \citep{rivoldini2009}: core radius of $2000\rm\,km$, density of mantle and lithosphere equal to $3100\rm\,kg/m^3$ (thus core density equal to $7325\rm\,kg/m^3$), seismic wave velocities in the lithosphere given by $v_p=7900\rm\,m/s$ and $v_s=4550\rm\,m/s$.
The corresponding elastic parameters for the lithosphere are:  $\mu=64\rm\,GPa$, $E=160\rm\,GPa$, $\nu=0.25$ (but $\nu=0.5$ in the model).
The model yields $h_2^T=1.83$ and $h_2^T=1.49$ when the lithospheric thickness is equal to $30\rm\,km$ and $100\rm\,km$, respectively (for other values, see Fig.~\ref{LoveFig}).

\subsection{Two properties of the strain}
\label{twoproperties}

The tangential displacement in the azimuthal direction vanishes if there is axial symmetry, i.e. if the elastic thickness and the radial deformation of the shell only depend on the latitude.
In that case, the strain-displacement equations \citep[e.g.][] {beuthe2008} read
\begin{eqnarray}
\epsilon_{\theta\theta} &=& \frac{1}{R} \left( \frac{\partial v_\theta}{\partial\theta} + w \right) \, ,
\\
\epsilon_{\varphi\varphi} &=& \frac{1}{R} \left( \cot\theta \, v_\theta + w \right) \, ,
\\
\epsilon_{\theta\varphi} &=& 0 \, ,
\end{eqnarray}
where $w$ is the radial displacement and $v_\theta$ the tangential displacement in the meridional direction.

By symmetry, $v_\theta$ vanishes at the poles.
The average of $\epsilon_{\theta\theta}$ on $[0,\pi]$ is thus related to the average of $\bar w=w/R$:
\begin{equation}
< \epsilon_{\theta\theta} > \, = \, < \bar w > \, ,
\end{equation}
where
\begin{equation}
< f > = \frac{1}{\pi} \, \int_0^{\pi} \, f \, d\theta \, .
\end{equation}
There is equatorial symmetry if the elastic thickness and the radial deformation are symmetric about the equatorial plane.
In that case, $v_\theta$ vanishes at the equator so that the average can be done on $[0,\pi/2]$.
For contraction plus despinning, $\bar w=\bar w_0+\bar w_2 P_2$, so that
\begin{equation}
< \epsilon_{\theta\theta} > \, =  \bar w_0 + \frac{\bar w_2}{4} \, . 
\hspace{1cm} \mbox{(property 1)}
\label{average}
\end{equation}

Another consequence of axial symmetry is that the azimuthal strain becomes independent of the elastic thickness at the equator:
\begin{equation}
\epsilon_{\varphi\varphi} |_{\theta=\frac{\pi}{2}} = \bar w |_{\theta=\frac{\pi}{2}}
\end{equation}
This property also directly results from the variation in length of the equator, $\delta L=2\pi\bar w|_{\theta=\frac{\pi}{2}}$, which is uniformly distributed along the equator.
For contraction plus despinning,
\begin{equation}
\epsilon_{\varphi\varphi} |_{\theta=\frac{\pi}{2}} = \bar w_0 - \frac{\bar w_2}{2}  \, .
\hspace{1cm} \mbox{(property 2)}
\label{straineq}
\end{equation}

\subsection{Contraction solution at first order}
\label{ContractionFirstOrder}

Given the despinning solution on a shell of constant thickness, the dualities of Section \ref{duality} generate a first approximation of the contraction solution on a shell of variable thickness.
If $\bar\alpha$ depends only on the parameter $\bar\alpha_2$ as in Eq.~(\ref{expansionalpha}), the contraction and despinning solutions can be expanded in $\bar\alpha_2$ about the solution for a shell of constant thickness given in Appendix \ref{ConstantElasticThickness}.
The dualities (\ref{relationCD1})-(\ref{relationCD4}) then allow to compute the contraction solution at order $\bar\alpha_2^{n+1}$ in terms of the despinning solution at order $\bar\alpha_2^{n}$.
The nondimensional despinning stress function at zeroth order is obtained from Eq.~(\ref{FDconst}):
\begin{equation}
\frac{\bar F^D}{ \bar w_2} = - \frac{P_2}{5+\nu} + {\cal O}(\bar\alpha_2) \, .
\end{equation}
Nondimensionalization has been carried out with Eqs.~(\ref{adim1})-(\ref{adim5}).
The duality (\ref{relationCD1}) yields the contraction stress function at first order in $\bar\alpha_2$:
\begin{equation}
\frac{\bar F^C}{\bar w_0} =  \frac{1}{1-\nu} + \bar \alpha_2 \frac{P_2}{5+\nu} + {\cal O}(\bar\alpha_2^2) \, .
\end{equation}

The duality (\ref{relationCD3}) acting on Eqs.~(\ref{despinningConst1})-(\ref{despinningConst2}) yields the contraction stresses at first order in $\bar\alpha_2$ (an alternative is to use Eqs.~(\ref{stressIntAdim})-(\ref{stressSurfAdim}) on $\bar F^C$):
\begin{eqnarray}
\frac{\bar\sigma^C_{\theta\theta}}{\bar w_0} &\cong&  \frac{1}{1-\nu} + \bar\alpha_2 \frac{2(2+\nu) P_2 -1 + \nu}{(1-\nu)(5+\nu)} \, ,
\label{stressC1}
\\
\frac{\bar\sigma^C_{\varphi\varphi}}{\bar w_0} &\cong&  \frac{1}{1-\nu} + \bar\alpha_2 \frac{2(1+2\nu) P_2 +1 - \nu}{(1-\nu)(5+\nu)}  \, .
\label{stressC2}
\end{eqnarray}
Fig.~\ref{StressFirstOrder}a compares Eqs.~(\ref{stressC1})-(\ref{stressC2}) with the stresses computed at high order in $\bar\alpha_2$, that is with a truncation degree $n=20$.
The first order approximation is good for small $\bar\alpha_2$ ($\bar\alpha_2=-2/5$) but is bad for large $\bar\alpha_2$ ($\bar\alpha_2=-6/7$).

The strains at first order in $\bar\alpha_2$ can be computed with Eqs.~(\ref{strainSurfAdim1})-(\ref{strainSurfAdim3}):
\begin{eqnarray}
\frac{\epsilon^C_{\theta\theta}}{\bar w_0} &\cong& 1 + \bar\alpha_2 \, \frac{1+\nu}{5+\nu} \left( 4 P_2 -1 \right)  \, ,
\label{strainC1}
\\
\frac{\epsilon^C_{\varphi\varphi}}{\bar w_0} &\cong&  1 + \bar\alpha_2 \, \frac{1+\nu}{5+\nu} \left( 2 P_2 +1 \right)  \, .
\label{strainC2}
\end{eqnarray}
With the values $P_2|_{\theta=0}=1$, $P_2|_{\theta=\pi/2}=-1/2$ and $<P_2>=1/4$, one can check that the stresses and strains (\ref{stressC1})-(\ref{strainC2}) satisfy the properties enumerated at the beginning of Section~\ref{symmetry}.
Besides the meridional strain as given by Eq.~(\ref{strainC1}) is equal to its average value at $\theta=45^\circ$ (where $P_2=1/4$), so that the meridional strain curves for varying $\bar\alpha_2$ seem to have a unique crossing point (see Fig.~\ref{StrainSym}a).
Since this property is only valid at first order in $\bar\alpha_2$, I call this point a {\it pseudo-node}.
Pseudo-nodes for stresses are obtained by setting to zero the term of order $\bar\alpha_2$ in Eqs.~(\ref{stressC1})-(\ref{stressC2}), yielding respectively $\theta\cong48.2^\circ$ and $\theta\cong65.9^\circ$ ($\nu=0.25$) for the meridional and azimuthal stresses (see Fig.~\ref{StressSurfContraction}a).

The same method works for $\bar\alpha$ not limited to degree two.
The generalization of the duality (\ref{relationCD1}) to an $\bar\alpha$ depending on an arbitrary number of parameters $\bar\alpha_\ell$ reads
\begin{equation}
\frac{\bar F^C}{\bar w_0}  = \frac{1}{1-\nu} - \sum_{L\geq2} \bar\alpha_L \, \frac{\bar F^L}{\bar w_L} \, ,
\label{GeneralizedDuality}
\end{equation}
where $\bar F^L$ denotes the solution of the membrane equation (\ref{flexBadim}) for $\bar w=\bar w_L P_L$.
At first order in $\bar\alpha_\ell$, the contraction stress function can thus be computed from the solutions (\ref{Flm}) on a shell of constant thickness:
\begin{equation}
\frac{\bar F^C}{\bar w_0} \cong  \frac{1}{1-\nu} + \sum_{\ell\geq2} \bar \alpha_\ell \frac{P_\ell}{f_{\ell\nu}}  \, ,
\label{FCfirstorder}
\end{equation}
where
\begin{equation}
f_{\ell\nu}=\ell(\ell+1)-1+\nu \, .
\end{equation}
There is no contribution of $\bar\alpha_1$ to the contraction stress function (\ref{FCfirstorder}), but this is only valid at first order in $\bar\alpha_\ell$ (terms in $\bar\alpha_1^n$ vanish, but mixed terms like $\bar\alpha_1\bar\alpha_2$ do not).
At first order in $\bar\alpha_\ell$, the contraction stresses read
\begin{eqnarray}
\frac{\bar\sigma^C_{\theta\theta}}{\bar w_0} &\cong&  \frac{1}{1-\nu} + \sum_{\ell\geq1} \bar\alpha_\ell \left( \frac{{\cal O}_2 P_\ell}{f_{\ell\nu}} + \frac{P_\ell}{1-\nu} \right) \, ,
\label{stressC1bis}
\\
\frac{\bar\sigma^C_{\varphi\varphi}}{\bar w_0} &\cong&  \frac{1}{1-\nu} + \sum_{\ell\geq1} \bar\alpha_\ell \left( \frac{{\cal O}_1 P_\ell}{f_{\ell\nu}} + \frac{P_\ell}{1-\nu} \right)  \, ,
\label{stressC2bis}
\end{eqnarray}
with the action of ${\cal O}_i$ on $P_\ell$ defined by Eqs.~(\ref{Opoly1})-(\ref{Opoly2}).
The sum now starts at $\ell=1$ because of the multiplication by $\bar\alpha$ when computing the stresses from the stress resultants, but the first term in the parentheses vanishes for $\ell=1$.
At first order in $\bar\alpha_\ell$, the contraction strains read
\begin{eqnarray}
\frac{\epsilon^C_{\theta\theta}}{\bar w_0} &\cong& 1 - (1+\nu) \sum_{\ell\geq1} \frac{\bar\alpha_\ell}{f_{\ell\nu}} \left( {\cal O}_1 -1 \right) P_\ell  \, ,
\label{strainC1bis}
\\
\frac{\epsilon^C_{\varphi\varphi}}{\bar w_0} &\cong&  1 - (1+\nu) \sum_{\ell\geq1} \frac{\bar\alpha_\ell}{f_{\ell\nu}} \left( {\cal O}_2 - 1 \right) P_\ell  \, .
\label{strainC2bis}
\end{eqnarray}
The stresses (\ref{stressC1bis})-(\ref{stressC2bis}) and the strains (\ref{strainC1bis})-(\ref{strainC2bis}) satisfy the properties enumerated at the beginning of Section~\ref{symmetry} because of Eqs.~(\ref{OiN})-(\ref{OiS}) and since Eqs.~(\ref{opO1})-(\ref{opO2}) with axial symmetry imply
\begin{eqnarray}
< \left( {\cal O}_1 - 1 \right) P_\ell >  &=& 0 \, ,
\\
\left( {\cal O}_2 - 1 \right) P_\ell \, |_{\theta=\frac{\pi}{2}}  &=& 0 \, .
\end{eqnarray}

\subsection{Operators ${\cal C}$ and ${\cal A}$ on spherical surface harmonics}
\label{OperatorsOnHarmonics}

If the shell thickness is constant, the membrane equation reduces to Eq.~(\ref{flexBconst}) which depends only on the operator $\Delta'$ and can be solved with spherical surface harmonics as in Appendix \ref{ConstantElasticThickness}.
If the thickness is variable, the membrane equation is not diagonal in the basis of spherical surface harmonics.
This basis remains however useful if the action of the operators ${\cal C}$ and ${\cal A}$ on spherical surface harmonics generates a linear combination of the same functions, so that the operators can be represented as square matrices.
The action of the operators ${\cal O}_i$ - from which ${\cal C}$ and ${\cal A}$ are built - on Legendre polynomials is known (see Eqs.~(\ref{Opoly1})-(\ref{Opoly3})) and could be used to derive a formula for  the action of ${\cal C}$ and ${\cal A}$ on Legendre polynomials. I will however obtain a more compact formula for the action of ${\cal A}$ by rewriting it exclusively in terms of the scalar operator $\Delta'$.
The operator ${\cal A}$ can be written in the equivalent form \citep{beuthe2008}:
\begin{equation}
{\cal A}(a\,;b) = (\Delta \, a)(\Delta \, b) - (\nabla_i \nabla_j \, a)(\nabla^i \nabla^j \, b) + (\Delta \, a) \, b + a \, (\Delta \, b) + 2 \, a \, b \, ,
\label{defA2}
\end{equation}
where $\nabla_i$ is the covariant derivative on the sphere and summation on repeated indices is implicit (indices are raised with the inverse metric $g^{ij}=diag(1,\sin^{-2}\theta)$).
Repeated covariant differentiation yields the following identities:
\begin{eqnarray}
\Delta(ab) &=& (\Delta \, a) \, b + 2 \, (\nabla_i \, a)(\nabla^i \, b) + a \, (\Delta \, b)
\label{id1}  \\
\Delta \Delta (ab) &=& ( \Delta \Delta \, a) \, b + 2 \, ( \Delta \, a)(\Delta \, b) + a \, (\Delta \Delta \, b)  + \, 4 \, (\nabla_i \nabla_j \, a)(\nabla^i \nabla^j b) \nonumber \\
			&  &  + \, 4 \, (\nabla_i  \Delta \, a)(\nabla^i b)  + 4 \, (\nabla_i \, a)(\nabla^i \Delta  b)  + 4 \, (\nabla_i a)(\nabla^i b)  \, .  
\label{id4} 
\end{eqnarray}
In the derivation of the last identity, I used the commutation relation for the spherical Laplacian and the covariant derivative on the surface of the sphere,
\begin{equation}
\Delta \nabla_i \, a - \nabla_i \Delta \, a = \nabla_i \, a \, ,
\end{equation}
which can be derived from the general commutation relation of covariant derivatives of a vector \citep[see Eq.~(G1) in][]{beuthe2008}.
Other useful identities are obtained by substituting $a\rightarrow\Delta\,a$ or $b\rightarrow\Delta\,b$ in Eq.~(\ref{id1}).
With the above identities, the operator ${\cal A}$ can be rewritten as an expression where the only intervening operator is $\Delta'$: 
\begin{eqnarray}
{\cal A}(a\,;b) &=& \frac{1}{4}  \left[
\, - \Delta' \Delta' (ab) - (\Delta' \Delta' \, a) \, b - a \, (\Delta' \Delta' \, b)
\right. \nonumber \\
&& \hspace{5mm} + \, 2 \, (\Delta' \, a)(\Delta' \, b) + 2 \, \Delta' ( (\Delta' \, a) \, b  +  a \, (\Delta' \, b) )
\nonumber \\
&&\hspace{5mm}   \left. - \, 2 \, \Delta' (ab) - 2 \, (\Delta' \, a) \, b - 2 \, a \, ( \Delta' \, b) + 8 \, ab \, \right] \, .
\end{eqnarray}
The action of the operators ${\cal C}$ and ${\cal A}$ can be partially evaluated if $a$ and $b$ are spherical surface harmonics of degree $m$ and $n$ (noted $a_m$ and $b_n$):
\begin{eqnarray}
{\cal C}(a_m\,;b_n) &=&
\delta'_n \, \Delta' \, (a_m b_n)  \, ,
\label{Charmonics}\\
{\cal A}(a_m\,;b_n) &=& 
- \frac{1}{4} \left( \, \Delta' \Delta' + \kappa_{mn} \, \Delta'  + \lambda_{mn} \right) (a_m b_n) \, .
\label{Aharmonics}
\end{eqnarray}
where $\delta'_n$ is given by Eq.~(\ref{eigendelta}) and
\begin{eqnarray}
\kappa_{mn} &=& 2 \left(1 - \delta'_m - \delta'_n \right) \, ,
\label{kappa} \\
\lambda_{mn} &=& \left( \delta'_m - \delta'_n \right)^2 + 2 \left( \delta'_m + \delta'_n \right) - 8  \, .
\label{lambda}
\end{eqnarray}
The action of the operators ${\cal C}$ and ${\cal A}$ can only be fully evaluated if the product $a_m b_n$ is expanded in spherical surface harmonics.

\subsection{Operators ${\cal C}$ and ${\cal A}$ on Legendre polynomials}
\label{OperatorsOnPoly}

The action of the operators ${\cal C}$ and ${\cal A}$ on spherical surface harmonics $a_m$ and $b_n$ is given by Eqs.~(\ref{Charmonics})-(\ref{Aharmonics}) in which products $a_m\,b_n$ appear.
I am now going to evaluate these products with the assumption that the spherical surface harmonics are zonal or, equivalently, Legendre polynomials.

The product of two Legendre polynomials can be expanded into Legendre polynomials with the Adams-Neumann formula \citep[][p.~331]{whittaker}:
\begin{equation}
P_m(x) \, P_n(x) = \sum_{j=0}^{min(m,n)} k_{mnj} \, P_{m+n-2j}(x) \, ,
\label{adamsneumannformula}
\end{equation}
with the coefficients of the expansion being given by
\begin{equation}
k_{mnj} = \frac{m+n+\frac{1}{2}-2j}{m+n+\frac{1}{2}-j} \, \frac{K_j K_{m-j} K_{n-j}}{K_{m+n-j}} \, ,
\label{adamsneumanncoeff}
\end{equation}
where
\begin{equation}
K_m = \frac{(2m-1)!!}{m!} = \frac{1.3...(2m-1)}{m!} \, ,
\end{equation}
with $K_0=1$.
These coefficients $k_{mnj}$ are symmetric in $(m,n)$ and $k_{m00}=k_{0m0}=1$.

The action of the operators ${\cal C}$ and ${\cal A}$ on Legendre polynomials of degree $m$ and $n$ is thus (the dependence on $x$ is omitted):
\begin{eqnarray}
{\cal C}(P_m\,;P_n) &=& \sum_{j=0}^{min(m,n)} k_{mnj} \, \chi_{mnj} \, P_{m+n-2j} \, ,
\label{Cpoly} \\
{\cal A}(P_m\,;P_n) &=& \sum_{j=0}^{min(m,n)} k_{mnj} \, \psi_{mnj} \, P_{m+n-2j} \, ,
\label{Apoly}
\end{eqnarray}
where
\begin{eqnarray}
\chi_{mnj} &=& \delta'_n \, \delta'_{m+n-2j} \, ,
\label{chimnj}
\\
\psi_{mnj} &=& - \frac{1}{4} \left(  \left(\delta'_{m+n-2j}\right)^2 +  \kappa_{mn} \, \delta'_{m+n-2j} + \lambda_{mn} \right) \, ,
\label{psimnj}
\end{eqnarray}
with $\delta'_n$ given by Eq.~(\ref{eigendelta}) and coefficients ($\kappa_{mn}$, $\lambda_{mn}$) given by Eqs.~(\ref{kappa})-(\ref{lambda}).
If $m=0$ (constant elastic thickness), the only relevant coefficients are
\begin{eqnarray}
\chi_{0n0} &=& \left( \delta'_n \right)^2 \, ,
\label{chimnj0}
\\
\psi_{0n0} &=&  \delta'_n \, .
\label{psimnj0}
\end{eqnarray}

The Legendre polynomial of degree one does not belong to the images of the operators ${\cal C}$ and ${\cal A}$: if $m+n-2j=1$ (meaning that $|m-n|=1$), then $\delta'_{m+n-2j}=0$ and $\lambda_{mn}=0$, so that $\chi_{mnj}=\psi_{mnj}=0$. The more general proof that spherical surface harmonics of degree one are not included in the images of ${\cal C}$ and ${\cal A}$ is given in \citet{beuthe2008}.
Since the coefficients multiplying the Legendre polynomials in Eqs.~(\ref{Cpoly})-(\ref{Apoly}) are independent of $\theta$, the operators ${\cal C}$ and ${\cal A}$ can be represented as matrices acting on vectors of Legendre coefficients.

\subsection{Membrane matrix for $n=6$}
\label{example}

The membrane matrix formalism described in Section \ref{section4} is illustrated here with a simple example:
$\bar\alpha$ is limited to degree two and the truncation degree $n$ is equal to~6.
The action of the operator ${\cal C}$ on the vector ${\mathbf {\bar F}}=\left( \bar F_0,\bar F_2,\bar F_4,\bar F_6\right)^T$ can be computed with Eq.~(\ref{Cpoly}).
The matrix ${\mathbf C}^{(6)}$ appearing in Eq.~(\ref{membranematrix}) is given by the sum 
${\mathbf C}^{(6)} = {\mathbf C}^{(6)}_0 + \bar\alpha_2 \, {\mathbf C}^{(6)}_2$, where
\begin{equation}
{\mathbf C}_0^{(6)} =
\left(
\begin{array}{cccc}
 4\,k_{000} &0 & 0 & 0 \\
0 & 16\,k_{020} & 0 & 0 \\
 0 & 0 & 324\,k_{040} & 0 \\
 0 & 0 & 0 & 1600\,k_{060}
\end{array}
\right)
\, ,
\end{equation}
\begin{equation}
{\mathbf C}_2^{(6)} =
\left(
\begin{array}{cccc}
0 & -8\,k_{222} & 0 & 0 \\
 -8\,k_{200} & 16\,k_{221} & 72\,k_{242} & 0 \\
 0 & 72\,k_{220} & 324\,k_{241} & 720\,k_{262} \\
 0 & 0 & 720\,k_{240} & 1600\,k_{261}
\end{array}
\right)
\, .
\end{equation}
The action of the operator ${\cal A}$ on the vector ${\mathbf {\bar F}}$ is computed with Eq.~(\ref{Apoly}).
The matrix ${\mathbf A}^{(6)}$ appearing in Eq.~(\ref{membranematrix}) is given by the sum 
${\mathbf A}^{(6)} = {\mathbf A}^{(6)}_0 + \bar\alpha_2 \, {\mathbf A}^{(6)}_2$, where
\begin{equation}
{\mathbf A}_0^{(6)}= 
\left(
\begin{array}{cccc}
 2\,k_{000} & 0 & 0 & 0 \\
 0 & -4\,k_{020} & 0 & 0 \\
 0 & 0 & -18\,k_{040} & 0 \\
 0 & 0 & 0 & -40\,k_{060}
\end{array}
\right)
\, ,
\end{equation}
\begin{equation}
{\mathbf A}_2^{(6)}= 
\left(
\begin{array}{cccc}
0 & -4\,k_{222} & 0 & 0 \\
 -4\,k_{200} & 20\,k_{221} & 6\,k_{242} & 0 \\
 0 & 6\,k_{220} & 90\,k_{241} & 24\,k_{262} \\
 0 & 0 & 24\,k_{240} & 200\,k_{261}
\end{array}
\right)
\, .
\end{equation}
The four rows are associated with the projection of the membrane equation on harmonic degrees 0, 2, 4, 6.
The upper index on the matrices denotes that the expansion is limited to degree~6.
The integer coefficients are generated by the functions $\chi_{mnj}$ and $\psi_{mnj}$ defined by Eqs.~(\ref{chimnj})-(\ref{psimnj}).
In particular, the integer coefficients in the matrices ${\mathbf C}_0^{(6)}$ and ${\mathbf A}_0^{(6)}$ are given by Eqs.~(\ref{chimnj0})-(\ref{psimnj0}).
The Adams-Neumann coefficients $k_{mnj}$, given by Eq.~(\ref{adamsneumanncoeff}), are in symbolic form so that their origin is apparent:
the $mn$ indices indicate that the term comes from the coefficients of $\bar\alpha$ and $\bar F$ of degree $m$ and $n$, respectively, whereas $m\!+\!n\!-\!2j$ is equal to the harmonic degree associated with the row of the matrix.
Apart from the trivial values $k_{m00}=k_{0m0}=1$, the values relevant to the example above are associated with the products $P_2P_2$, $P_2P_4$ and $P_2P_6$:
\begin{eqnarray}
\left( k_{220} , k_{221} , k_{222} \right) &=& \left( \frac{18}{35} , \frac{2}{7} , \frac{1}{5} \right) \, ,
\\
\left( k_{240} , k_{241} , k_{242} \right) &=& \left( \frac{5}{11} , \frac{20}{77} , \frac{2}{7} \right) \, ,
\\
\left( k_{260} , k_{261}, k_{262} \right) &=& \left( \frac{28}{65} , \frac{14}{55} , \frac{45}{143} \right) \, .
\end{eqnarray}
The coefficient $k_{260}$ does not appear in ${\mathbf C}_2^{(6)}$ and ${\mathbf A}_2^{(6)}$ because it is associated with a polynomial of degree~8.

The approximate membrane equation (\ref{flexBmat}) now takes the following form:
\begin{equation}
{\mathbf M}^{(6)} \left( \bar F_0,\bar F_2,\bar F_4,\bar F_6\right)^T = \left( 2 \bar w_0 , -4 \bar w_2 , 0 , 0 \right)^T \, .
\end{equation}

The coefficients $\bar F_\ell^C$ of the contraction solution up to $\bar\alpha_2^3$ are generated with the recurrence relation (\ref{recurrence}), starting with
${\mathbf {\bar F}}^{(0)}=\left(\bar w_0/(1-\nu),0,0,0\right)$:
\begin{eqnarray}
\frac{\bar F_0^C}{\bar w_0} &\cong&
   \frac{1}{1-\nu}
+ \frac{2}{5} \, \frac{1}{5+\nu} \, \bar\alpha_2^2
+ \frac{4}{35} \, \frac{1+5\nu}{(5+\nu)^2} \, \bar\alpha_2^3 \, ,
\label{F0C} \\
\frac{\bar F_2^C}{\bar w_0} &\cong&
   \frac{1}{5+\nu} \, \bar\alpha_2
+ \frac{2}{7} \, \frac{1+5\nu}{(5+\nu)^2} \, \bar\alpha_2^2
+ \frac{12}{7} \, \frac{49-7\nu+17\nu^2+\nu^3}{(5+\nu)^3(19+\nu)}  \, \bar\alpha_2^3 \, ,
\label{F2C} \\
\frac{\bar F_4^C}{\bar w_0} &\cong&
   \frac{6}{35} \, \frac{\nu-11}{(5+\nu)(19+\nu)} \, \bar\alpha_2^2
+ \frac{36}{385} \, \frac{(\nu-11)(-21+56\nu+5\nu^2)}{(5+\nu)^2(19+\nu)^2}  \, \bar\alpha_2^3 \, ,
\label{F4C} \\
\frac{\bar F_6^C}{\bar w_0} &\cong&
   \frac{18}{385} \, \frac{(\nu-11)(\nu-29)}{(5+\nu)(19+\nu)(41+\nu)}  \, \bar\alpha_2^3 \, .
\label{F6C}
\end{eqnarray}
Coefficients $\bar F_\ell^C$ have a leading term of order $\bar\alpha_2^{\ell/2}$.
Thus the truncation at harmonic degree $n$ can be understood as an expansion in $\bar\alpha_2$: harmonic coefficients that are neglected are at most of order $\bar\alpha_2^{1+n/2}$: if $n=6$, $\bar F_\ell^C$ with $\ell\geq 8$ are at most of order ${\cal O}(\bar\alpha_2^4)$ and are ignored.
However the truncation at degree $n$ is not equivalent to an expansion at order $\bar\alpha_2^{n/2}$ since the lower the degree of the harmonic coefficient, the better it is known.
More precisely, the band-diagonal structure of the membrane matrix has the effect that, for a given truncation degree $n$, the expansion of a term $\bar F_\ell^C$ can be done up to order $\bar\alpha_2^{n-\ell/2+1}$ included: if $n=6$, $(\bar F_0^C,\bar F_2^C,\bar F_4^C,\bar F_6^C)$ can be expanded up to order $(\bar\alpha_2^7,\bar\alpha_2^6,\bar\alpha_2^5,\bar\alpha_2^4)$ respectively.

The coefficients $\bar F_\ell^D$ of the despinning solution are related to the coefficients of the contraction solution by Eq.~(\ref{relationCD1}):
\begin{eqnarray}
\frac{\bar F_0^D}{\bar w_2} &=& - \frac{1}{\bar\alpha_2}  \left( \frac{\bar F_0^C}{\bar w_0} - \frac{1}{1-\nu} \right) \, ,
\label{relationCD5a} \\
\frac{\bar F_i^D}{\bar w_2} &=& - \frac{1}{\bar\alpha_2} \, \frac{\bar F_i^C}{\bar w_0}
\hspace{1cm} (i=2,4,6) \, .
\label{relationCD5d}
\end{eqnarray}
Coefficients $\bar F_\ell^D$ thus have a leading term of order $\bar\alpha_2^{\ell/2-1}$, except $\bar F_0^D$ which begins at order $\bar\alpha_2$.
Furthermore, for a given truncation degree $n$,  coefficients $\bar F_\ell^D$ can be expanded up to order $\bar\alpha_2^{n-\ell/2}$.

\begin{figure}
   \centering
   \includegraphics[width=6.9cm]{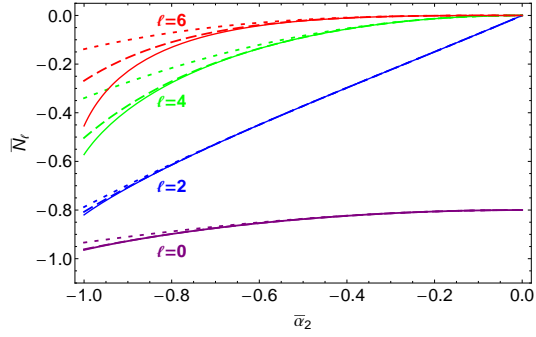}
   \caption{\footnotesize
    Influence of the truncation degree $n$ on the Legendre coefficients of the contraction solution.
    The coefficients $\bar N_\ell$ (see Eq.~(\ref{deftraceN})) are plotted as a function of $\bar\alpha_2$ for $\ell=0,2,4,6$.
    The contraction is $\bar w_0\!=\!-1$ and Poisson's ratio is equal to $0.25$. 
    The curve for $\ell=0$ has been shifted, its real y-intercept being $2/(\nu-1)=-8/3$ instead of $-0.8$.
    Three different approximations to the solution are plotted: n=6 with expansion limited to $\bar\alpha_2^3$ (dotted curve), n=6 with no expansion in $\bar\alpha_2$ (dashed curve), n=20 with no expansion in $\bar\alpha_2$ (continuous curve).}
   \label{TruncationEffect}
\end{figure}

\begin{figure}
   \centering
   \includegraphics[width=6.9cm]{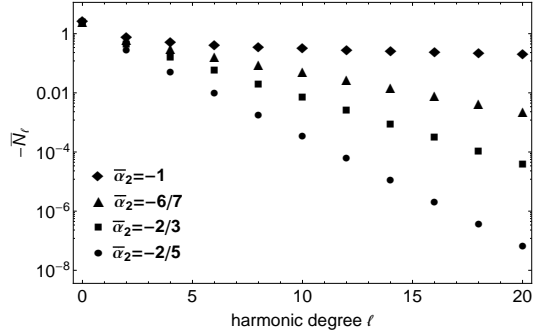} 
   \caption{\footnotesize
   Decrease with harmonic degree of the Legendre coefficients of the contraction solution ($\bar w_0\!=\!-1$).
   The coefficients $\bar N_\ell$ (see Eq.~(\ref{deftraceN})) are plotted as a function of the harmonic degree (up to $\ell=20$) for different values of $\bar\alpha_2$.
  The symbols overlap when $\ell=0$. 
  Truncation degree $n$ is equal to 50.}
   \label{StressTraceLog}
\end{figure}

Figs.~\ref{TruncationEffect} and \ref{StressTraceLog} show the effect of the truncation degree $n$ on the coefficients of the contraction solution.
The effects on the despinning solution are similar since they are related by Eqs.~(\ref{relationCD5a})-(\ref{relationCD5d}).
Since the stresses are the quantities of physical interest, the figures do not show the Legendre coefficients of the stress function $\bar F$ but rather those of the sum of the nondimensional stress resultants:
\begin{equation}
\bar N_\ell \equiv \left( \bar N_{\theta\theta} + \bar N_{\varphi\varphi} \right)_\ell = \delta'_\ell \, \bar F_\ell \, .
\label{deftraceN}
\end{equation}
Truncation effects can be large close to $\bar\alpha_2=2$, but I only consider negative values of $\bar\alpha_2$ relevant to equatorial thinning.
Fig.~\ref{TruncationEffect} shows the effect of the truncation degree and of the expansion in $\bar\alpha_2$ on the first four Legendre coefficients of the solution.
The comparison of truncation degrees $n=6$ and $n=20$ indicates that that the truncation degree $n=6$ already yields a precise solution except when $\bar\alpha_2$ is close to~-1.
The solution obtained by direct matrix inversion is preferable to the solution obtained by expansion in $\bar\alpha_2$.
Fig.~\ref{StressTraceLog} shows the decrease in the values of the Legendre coefficients of the contraction solution up to $\ell=20$ for various values of $\bar\alpha_2$.
The decrease is approximately exponential for $|\bar\alpha_2|\ll1$, but very slow for $\bar\alpha_2=-1$.
The convergence of the harmonic expansion of the stress resultants is thus not guaranteed for extremal values of $\bar\alpha_2$.
Since the stresses and the strains include a supplementary factor $\bar\alpha$ vanishing at the poles when $\bar\alpha_2=-1$, they are not affected by the convergence problem.

Poisson's ratio has generally a weak effect on the solutions for contraction and despinning, except for the coefficients $\ell\!=\!0$ for contraction and to a lesser extent $\ell\!=\!2$ for despinning.
The other coefficients are mainly affected by $\nu$ at near extremal values of $\bar\alpha_2$.

The computation of the stresses completes the example.
The nondimensional stress resultants are computed with Eq.~(\ref{stressIntAdim}) and Eqs.~(\ref{Opoly1})-(\ref{Opoly3}):
\begin{eqnarray}
\bar N_{\theta\theta} &=& \bar F_0 + \bar F_2 \, {\cal O}_2 P_2 + \bar F_4 \, {\cal O}_2 P_4  + \bar F_6 \, {\cal O}_2 P_6 \, ,
\\
\bar N_{\varphi\varphi} &=& \bar F_0 + \bar F_2 \, {\cal O}_1 P_2 + \bar F_4 \, {\cal O}_1 P_4 + \bar F_6 \, {\cal O}_1 P_6 \, .
\end{eqnarray}
The Legendre coefficients of ${\cal O}_iP_j$ are given by the columns of the matrices (\ref{O1mat})-(\ref{O2mat}).
$\bar N_{\theta\varphi}$ is zero.
The nondimensional stress is related to $\bar N_{ij}$ by Eq.~(\ref{stressSurfAdim}):
\begin{equation}
\bar\sigma_{ij} = \left( 1 + \bar\alpha_2 \, P_2 \right) \, \bar N_{ij} \, .
\end{equation}
The product of Legendre polynomials can be computed with the Adams-Neumann formula (\ref{adamsneumannformula}) but I will not give the explicit formulas.
The stresses and the strains for the contraction solution at first order in $\bar\alpha_2$ have already been given in Appendix \ref{ContractionFirstOrder} whereas those for the despinning solution are computed in Appendix \ref{DespinningFirstOrder}.

\subsection{Despinning solution at first order}
\label{DespinningFirstOrder}

If $\bar\alpha$ depends only on the parameter $\bar\alpha_2$, the despinning stress function at first order in $\bar\alpha_2$ can be obtained from the contraction solution (\ref{F0C})-(\ref{F4C}) and the dualities (\ref{relationCD5a})-(\ref{relationCD5d}):
\begin{equation}
\frac{\bar F^D}{\bar w_2} \cong - \frac{1}{5+\nu} \left( P_2 + \bar \alpha_2 \left( \frac{2}{5} + \frac{2}{7} \frac{1+5\nu}{5+\nu} \, P_2 + \frac{6}{35} \, \frac{\nu-11}{19+\nu} \, P_4 \right) \right)  \, .
\end{equation}
The despinning stresses are computed at first order in $\bar\alpha_2$ from $\bar F^D$ with Eqs.~(\ref{stressIntAdim})-(\ref{stressSurfAdim}):
\begin{eqnarray}
\frac{\bar\sigma^D_{\theta\theta}}{\bar w_2} &\cong& \frac{P_2 +1}{5+\nu} + \bar\alpha_2 \, \frac{s_{\theta\theta}(\nu,P_2)}{(5+\nu)^2 (19+\nu)} \, .
\label{stressD1}
\\
\frac{\bar\sigma^D_{\varphi\varphi}}{\bar w_2} &\cong& \frac{3 P_2 -1}{5+\nu} + \bar\alpha_2 \, \frac{s_{\varphi\varphi}(\nu,P_2)}{(5+\nu)^2 (19+\nu)} \, .
\label{stressD2}
\end{eqnarray}
where the auxiliary functions defined by
\begin{eqnarray}
s_{\theta\theta}(\nu,x) &=& - \, 31+18\nu+\nu^2 + \left( 69+48\nu+3\nu^2\right) x + \left( 40+18\nu+2\nu^2\right) x^2 \, ,
\\
s_{\varphi\varphi}(\nu,x) &=& 21-30\nu-3\nu^2 + \left( 47+72\nu+\nu^2\right) x + \left( 10+42\nu+8\nu^2\right) x^2 \, ,
\end{eqnarray}
have the property
\begin{equation}
s_{\theta\theta}(\nu,1)=s_{\varphi\varphi}(\nu,1) \, ,
\end{equation}
which ensures the equality of the stresses at the poles.
In Eqs.~(\ref{stressD1})-(\ref{stressD2}), $P_4$ has been expressed in terms of $P_2$:
\begin{equation}
P_4 = \frac{35}{18} \left( (P_2)^2-\frac{2}{7}P_2-\frac{1}{5} \right) \, .
\end{equation}

The strains can be computed at first order in $\bar\alpha_2$ with Eqs.~(\ref{strainSurfAdim1})-(\ref{strainSurfAdim3}):
\begin{eqnarray}
\frac{\epsilon^D_{\theta\theta}}{\bar w_2} &\cong& \frac{(1-3\nu) P_2 +1+\nu}{5+\nu} + \bar\alpha_2 \, \frac{(1-\nu^2) \, e_{\theta\theta}(\nu,P_2)}{(5+\nu)^2 (19+\nu)} \, .
\label{strainD1}
\\
\frac{\epsilon^D_{\varphi\varphi}}{\bar w_2} &\cong& \frac{(3-\nu) P_2 -1-\nu}{5+\nu} + \bar\alpha_2 \, \frac{(1-\nu^2) \, e_{\varphi\varphi}(\nu,P_2)}{(5+\nu)^2 (19+\nu)} \, .
\label{strainD2}
\end{eqnarray}
where the auxiliary functions defined by
\begin{eqnarray}
e_{\theta\theta}(\nu,x) &=& - \, 31- 3\nu + \left( 69+\nu \right) x + \left( 40+8\nu \right) x^2 \, ,
\\
e_{\varphi\varphi}(\nu,x) &=& \left( 2 x + 1 \right) \left( 21 + \nu + \left( 5+\nu \right) x \right) \, ,
\end{eqnarray}
have the properties
\begin{eqnarray}
e_{\theta\theta}(\nu,1) &=& e_{\varphi\varphi}(\nu,1) \, ,
\\
<e_{\theta\theta}(\nu,P_2)> &=& 0 \, ,
\\
e_{\varphi\varphi}(\nu,-1/2) &=& 0 \, ,
\end{eqnarray}
which ensure that the properties enumerated at the beginning of Section~\ref{symmetry} are satisfied (note that $<P_2>=1/4$ and $<(P_2)^2>=11/32$).

At zeroth order in $\bar\alpha_2$, the meridional strain is equal to its average value at $\theta=45^\circ$ (where $P_2=1/4$), but this is not true at first order in $\bar\alpha_2$ since $e_{\theta\theta}(\nu,1/4)\neq0$.
If $\nu=0.25$, the pseudo-nodes defined in Section \ref{ContractionFirstOrder} occur at $\theta\cong40.3^\circ$ for the meridional strain (see Fig.~\ref{StrainSym}b), $\theta=43.8^\circ$ for the meridional stress and $\theta=64.4^\circ$ for the azimuthal stress (see Fig.~\ref{StressSurfContraction}b).

\begin{figure}
   \centering
   \includegraphics[width=6.9cm]{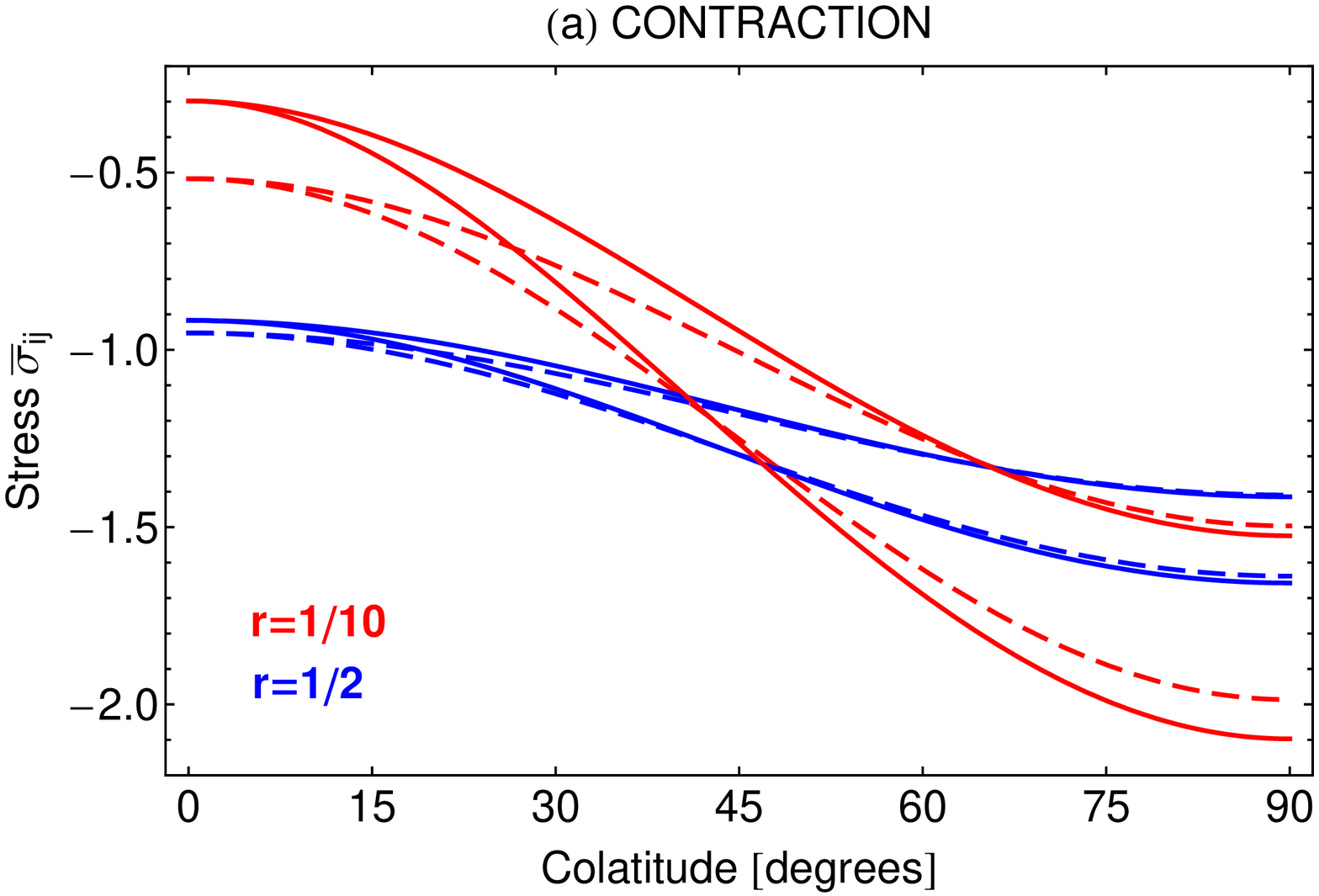}
    \hspace{0.5cm}
     \includegraphics[width=6.9cm]{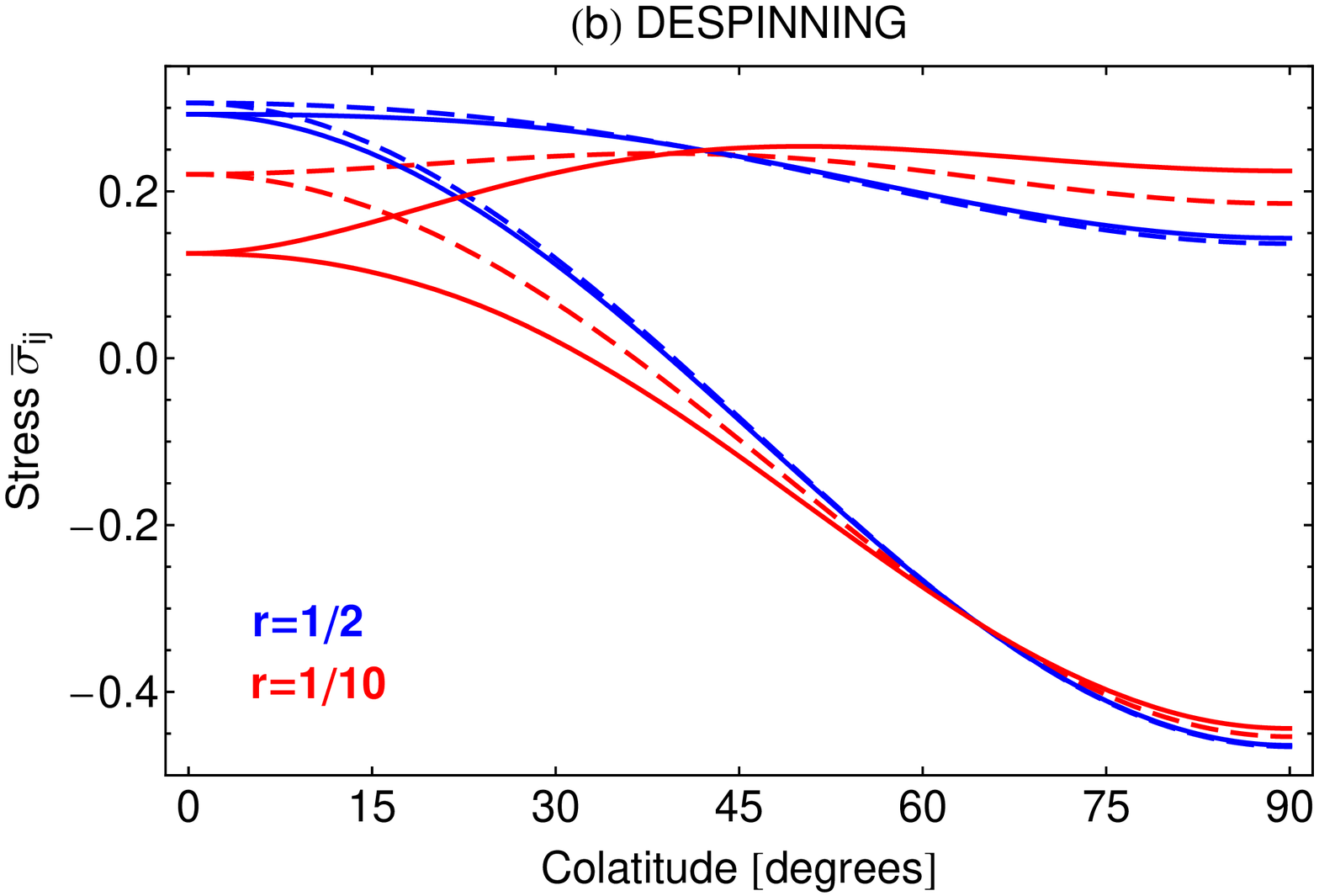} 
   \caption{\footnotesize
   Comparison of the stresses at first order in $\bar\alpha_2$ (dashed curves) and at truncation degree $n=20$ (continuous curves) : (a) contraction case ($\bar w_0\!=\!-1$), (b) despinning case ($\bar w_2\!=\!1$).
   The first order approximation is good if the equator-to-pole thickness ratio $r=1/2$ ($\bar\alpha_2=-2/5$) and rather bad if $r=1/10$ ($\bar\alpha_2=-6/7$).
   Poisson's ratio is equal to $0.25$.
   The meridional and azimuthal stresses are not distinguished but can be identified by looking at Fig.~\ref{StressSurfContraction}.}
   \label{StressFirstOrder}
\end{figure}

Fig.~\ref{StressFirstOrder}b compares Eqs.~(\ref{stressD1})-(\ref{stressD2}) with the stresses computed at high order in $\bar\alpha_2$, that is with truncation degree $n=20$.
The first order approximation is good for small $\bar\alpha_2$ ($\bar\alpha_2=-2/5$) but is rather bad for large $\bar\alpha_2$ ($\bar\alpha_2=-6/7$).

\footnotesize
\begin{thebibliography}{80}
\providecommand{\natexlab}[1]{#1}
\expandafter\ifx\csname urlstyle\endcsname\relax
  \providecommand{\doi}[1]{doi:\discretionary{}{}{}#1}\else
  \providecommand{\doi}{doi:\discretionary{}{}{}\begingroup
  \urlstyle{rm}\Url}\fi

\bibitem[{\textit{{Anderson} et~al.}(2001)\textit{{Anderson}, {Dohm},
  {Golombek}, {Haldemann}, {Franklin}, {Tanaka}, {Lias}, and
  {Peer}}}]{anderson2001}
{Anderson}, R.~C., J.~M. {Dohm}, M.~P. {Golombek}, A.~F.~C. {Haldemann}, B.~J.
  {Franklin}, K.~L. {Tanaka}, J.~{Lias}, and B.~{Peer} (2001), {Primary centers
  and secondary concentrations of tectonic activity through time in the western
  hemisphere of Mars}, \textit{J.~Geophys.~Res.}, \textit{106}, 20,563--20,586,
  \doi{10.1029/2000JE001278}.

\bibitem[{\textit{{Anderson} et~al.}(2008)\textit{{Anderson}, {Dohm},
  {Haldemann}, {Pounders}, {Golombek}, and {Castano}}}]{anderson2008}
{Anderson}, R.~C., J.~M. {Dohm}, A.~F.~C. {Haldemann}, E.~{Pounders},
  M.~{Golombek}, and A.~{Castano} (2008), {Centers of tectonic activity in the
  eastern hemisphere of Mars}, \textit{Icarus}, \textit{195}, 537--546,
  \doi{10.1016/j.icarus.2007.12.027}.

\bibitem[{\textit{{Andrews-Hanna} et~al.}(2008)\textit{{Andrews-Hanna},
  {Zuber}, and {Hauck}}}]{andrewshanna2008}
{Andrews-Hanna}, J.~C., M.~T. {Zuber}, and S.~A. {Hauck} (2008), {Strike-slip
  faults on Mars: Observations and implications for global tectonics and
  geodynamics}, \textit{J.~Geophys.~Res.}, \textit{113}, E08002,
  \doi{10.1029/2007JE002980}.

\bibitem[{\textit{{Beeman} et~al.}(1988)\textit{{Beeman}, {Durham}, and
  {Kirby}}}]{beeman1988}
{Beeman}, M., W.~B. {Durham}, and S.~H. {Kirby} (1988), {Friction of ice},
  \textit{J.~Geophys.~Res.}, \textit{93}, 7625--7633,
  \doi{10.1029/JB093iB07p07625}.

\bibitem[{\textit{{Beuthe}}(2008)}]{beuthe2008}
{Beuthe}, M. (2008), {Thin elastic shells with variable thickness for
  lithospheric flexure of one-plate planets}, \textit{Geophys.~J.~Int.},
  \textit{172}, 817--841, \doi{10.1111/j.1365-246X.2007.03671.x}.

\bibitem[{\textit{{Blakely}}(1995)}]{blakely}
{Blakely}, R.~J. (1995), \textit{{Potential Theory in Gravity and Magnetic
  Applications}}, Cambridge University Press, Cambridge.

\bibitem[{\textit{{Burns}}(1976)}]{burns1976}
{Burns}, J.~A. (1976), {Consequences of the tidal slowing of Mercury},
  \textit{Icarus}, \textit{28}, 453--458, \doi{10.1016/0019-1035(76)90118-4}.

\bibitem[{\textit{{Castillo-Rogez} et~al.}(2007)\textit{{Castillo-Rogez},
  {Matson}, {Sotin}, {Johnson}, {Lunine}, and {Thomas}}}]{castillo2007}
{Castillo-Rogez}, J.~C., D.~L. {Matson}, C.~{Sotin}, T.~V. {Johnson}, J.~I.
  {Lunine}, and P.~C. {Thomas} (2007), {Iapetus' geophysics: Rotation rate,
  shape, and equatorial ridge}, \textit{Icarus}, \textit{190}, 179--202,
  \doi{10.1016/j.icarus.2007.02.018}.

\bibitem[{\textit{{Collins} et~al.}(2009)\textit{{Collins}, {McKinnon},
  {Moore}, {Nimmo}, {Pappalardo}, {Prockter}, and {Schenk}}}]{collins2009}
{Collins}, G.~C., W.~B. {McKinnon}, J.~M. {Moore}, F.~{Nimmo}, R.~T.
  {Pappalardo}, L.~M. {Prockter}, and P.~M. {Schenk} (2009), {Tectonics of the
  Outer Planet Satellites}, in \textit{Planetary Tectonics}, edited by
  {{Watters}, T.~R. and {Schultz}, R.~A.}, pp. 264--350, Cambridge University
  Press, Cambridge.

\bibitem[{\textit{{Consolmagno}}(1985)}]{consolmagno1985}
{Consolmagno}, G.~J. (1985), {Resurfacing Saturn's satellites - Models of
  partial differentiation and expansion}, \textit{Icarus}, \textit{64},
  401--413, \doi{10.1016/0019-1035(85)90064-8}.

\bibitem[{\textit{{Czechowski} and
  {Leliwa-Kopysty{\'n}ski}}(2008)}]{czechowski2008}
{Czechowski}, L., and J.~{Leliwa-Kopysty{\'n}ski} (2008), {The Iapetus's ridge:
  Possible explanations of its origin}, \textit{Adv. Space Res.}, \textit{42},
  61--69, \doi{10.1016/j.asr.2007.08.008}.

\bibitem[{\textit{{Denk} et~al.}(2008)\textit{{Denk}, {Neukum}, {Schmedemann},
  {Roatsch}, {Wagner}, {Giese}, {Perry}, {Helfenstein}, {Turtle}, and
  {Porco}}}]{denk2008}
{Denk}, T., G.~{Neukum}, N.~{Schmedemann}, T.~{Roatsch}, R.~J. {Wagner},
  B.~{Giese}, J.~E. {Perry}, P.~{Helfenstein}, E.~P. {Turtle}, and C.~C.
  {Porco} (2008), {Iapetus Imaging During the Targeted Flyby of the Cassini
  Spacecraft}, in \textit{Lunar Planet. Sci. 39}, abstract 2533.

\bibitem[{\textit{{Dombard} and {Hauck}}(2008)}]{dombardhauck2008}
{Dombard}, A.~J., and S.~A. {Hauck} (2008), {Despinning plus global contraction
  and the orientation of lobate scarps on Mercury: Predictions for MESSENGER},
  \textit{Icarus}, \textit{198}, 274--276, \doi{10.1016/j.icarus.2008.06.008}.

\bibitem[{\textit{{Gammon} et~al.}(1983)\textit{{Gammon}, {Kiefte}, {Clouter},
  and {Denner}}}]{gammon1983}
{Gammon}, P.~H., H.~{Kiefte}, M.~J. {Clouter}, and W.~W. {Denner} (1983),
  {Elastic constants of artificial and natural ice samples by Brillouin
  spectroscopy}, \textit{J. Glaciol.}, \textit{29}, 433--460.

\bibitem[{\textit{{Giese} et~al.}(2008)\textit{{Giese}, {Denk}, {Neukum},
  {Roatsch}, {Helfenstein}, {Thomas}, {Turtle}, {McEwen}, and
  {Porco}}}]{giese2008}
{Giese}, B., T.~{Denk}, G.~{Neukum}, T.~{Roatsch}, P.~{Helfenstein}, P.~C.
  {Thomas}, E.~P. {Turtle}, A.~{McEwen}, and C.~C. {Porco} (2008), {The
  topography of Iapetus' leading side}, \textit{Icarus}, \textit{193},
  359--371, \doi{10.1016/j.icarus.2007.06.005}.

\bibitem[{\textit{{Goff-Pochat} and {Collins}}(2009)}]{goffpochat2009}
{Goff-Pochat}, N., and G.~C. {Collins} (2009), {Strain Measurement Across Fault
  Scarps on Dione}, in \textit{Lunar Planet. Sci. 40}, abstract 2111.

\bibitem[{\textit{{Goldreich} and {Peale}}(1968)}]{goldreich1968}
{Goldreich}, P., and S.~J. {Peale} (1968), {The Dynamics of Planetary
  Rotations}, \textit{Annu.~Rev.~Astron.~Astrophys.}, \textit{6}, 287--320,
  \doi{10.1146/annurev.aa.06.090168.001443}.

\bibitem[{\textit{{Golombek}}(1985)}]{golombek1985}
{Golombek}, M.~P. (1985), {Fault type predictions from stress distributions on
  planetary surfaces - Importance of fault initiation depth},
  \textit{J.~Geophys.~Res.}, \textit{90}, 3065--3074,
  \doi{10.1029/JB090iB04p03065}.

\bibitem[{\textit{{Greenberg} et~al.}(1998)\textit{{Greenberg}, {Geissler},
  {Hoppa}, {Tufts}, {Durda}, {Pappalardo}, {Head}, {Greeley}, {Sullivan}, and
  {Carr}}}]{greenberg1998}
{Greenberg}, R., P.~{Geissler}, G.~{Hoppa}, B.~R. {Tufts}, D.~D. {Durda},
  R.~{Pappalardo}, J.~W. {Head}, R.~{Greeley}, R.~{Sullivan}, and M.~H. {Carr}
  (1998), {Tectonic Processes on Europa: Tidal Stresses, Mechanical Response,
  and Visible Features}, \textit{Icarus}, \textit{135}, 64--78,
  \doi{10.1006/icar.1998.5986}.

\bibitem[{\textit{{Helfenstein} and {Parmentier}}(1983)}]{helfenstein1983}
{Helfenstein}, P., and E.~M. {Parmentier} (1983), {Patterns of fracture and
  tidal stresses on Europa}, \textit{Icarus}, \textit{53}, 415--430,
  \doi{10.1016/0019-1035(83)90206-3}.

\bibitem[{\textit{{Helfenstein} and {Parmentier}}(1985)}]{helfenstein1985}
{Helfenstein}, P., and E.~M. {Parmentier} (1985), {Patterns of fracture and
  tidal stresses due to nonsynchronous rotation - Implications for fracturing
  on Europa}, \textit{Icarus}, \textit{61}, 175--184,
  \doi{10.1016/0019-1035(85)90099-5}.

\bibitem[{\textit{{Howett} et~al.}(2010)\textit{{Howett}, {Spencer}, {Pearl},
  and {Segura}}}]{howett2010}
{Howett}, C.~J.~A., J.~R. {Spencer}, J.~{Pearl}, and M.~{Segura} (2010),
  {Thermal inertia and bolometric Bond albedo values for Mimas, Enceladus,
  Tethys, Dione, Rhea and Iapetus as derived from Cassini/CIRS measurements},
  \textit{Icarus}, \textit{206}, 573--593, \doi{10.1016/j.icarus.2009.07.016}.

\bibitem[{\textit{{Ip}}(2006)}]{ip2006}
{Ip}, W.-H. (2006), {On a ring origin of the equatorial ridge of Iapetus},
  \textit{Geophys.~Res.~Lett.}, \textit{33}, L16203,
  \doi{10.1029/2005GL025386}.

\bibitem[{\textit{{Jaeger} et~al.}(2007)\textit{{Jaeger}, {Cook}, and
  {Zimmerman}}}]{jaeger}
{Jaeger}, J.~C., N.~G.~W. {Cook}, and R.~W. {Zimmerman} (2007),
  \textit{{Fundamentals of Rock Mechanics}}, 4th edition, Blackwell.

\bibitem[{\textit{{King}}(2008)}]{king2008}
{King}, S.~D. (2008), {Pattern of lobate scarps on Mercury's surface reproduced
  by a model of mantle convection}, \textit{Nat. Geosci.}, \textit{1},
  229--232, \doi{10.1038/ngeo152}.

\bibitem[{\textit{{Knapmeyer} et~al.}(2006)\textit{{Knapmeyer}, {Oberst},
  {Hauber}, {W{\"a}hlisch}, {Deuchler}, and {Wagner}}}]{knapmeyer2006}
{Knapmeyer}, M., J.~{Oberst}, E.~{Hauber}, M.~{W{\"a}hlisch}, C.~{Deuchler},
  and R.~{Wagner} (2006), {Working models for spatial distribution and level of
  Mars' seismicity}, \textit{J.~Geophys.~Res.}, \textit{111}, E11006,
  \doi{10.1029/2006JE002708}.

\bibitem[{\textit{{Kraus}}(1967)}]{kraus}
{Kraus}, H. (1967), \textit{{Thin Elastic Shells}}, John Wiley, New York.

\bibitem[{\textit{{Lambeck}}(1980)}]{lambeck}
{Lambeck}, K. (1980), \textit{{The earth's variable rotation: Geophysical
  causes and consequences}}, Cambridge University Press, Cambridge.

\bibitem[{\textit{{Leith} and {McKinnon}}(1996)}]{leith1996}
{Leith}, A.~C., and W.~B. {McKinnon} (1996), {Is There Evidence for Polar
  Wander on Europa?}, \textit{Icarus}, \textit{120}, 387--398,
  \doi{10.1006/icar.1996.0058}.

\bibitem[{\textit{{Love}}(1944)}]{love}
{Love}, A.~E.~H. (1944), \textit{{A Treatise on the Mathematical Theory of
  Elasticity}}, 4th edition, Dover, New York.

\bibitem[{\textit{{Mangold} et~al.}(2000)\textit{{Mangold}, {Allemand},
  {Thomas}, and {Vidal}}}]{mangold2000}
{Mangold}, N., P.~{Allemand}, P.~G. {Thomas}, and G.~{Vidal} (2000),
  {Chronology of compressional deformation on Mars: evidence for a single and
  global origin}, \textit{Planet. Space Sci.}, \textit{48}, 1201--1211.

\bibitem[{\textit{{Margot} et~al.}(2007)\textit{{Margot}, {Peale}, {Jurgens},
  {Slade}, and {Holin}}}]{margot2007}
{Margot}, J.~L., S.~J. {Peale}, R.~F. {Jurgens}, M.~A. {Slade}, and I.~V.
  {Holin} (2007), {Large Longitude Libration of Mercury Reveals a Molten Core},
  \textit{Science}, \textit{316}, 710--714, \doi{10.1126/science.1140514}.

\bibitem[{\textit{{Matsuyama} and {Nimmo}}(2008)}]{matsuyama2008}
{Matsuyama}, I., and F.~{Nimmo} (2008), {Tectonic patterns on reoriented and
  despun planetary bodies}, \textit{Icarus}, \textit{195}, 459--473,
  \doi{10.1016/j.icarus.2007.12.003}.

\bibitem[{\textit{{Matsuyama} and {Nimmo}}(2009)}]{matsuyama2009}
{Matsuyama}, I., and F.~{Nimmo} (2009), {Gravity and tectonic patterns of
  Mercury: Effect of tidal deformation, spin-orbit resonance, nonzero
  eccentricity, despinning, and reorientation}, \textit{J.~Geophys.~Res.},
  \textit{114}, E01010, \doi{10.1029/2008JE003252}.

\bibitem[{\textit{{McKinnon}}(1981)}]{mckinnon1981}
{McKinnon}, W.~B. (1981), {Application of ring tectonic theory to Mercury and
  other solar system bodies}, in \textit{Multi-ring basins: Formation and
  Evolution}, edited by R.~B. {Merill} and P.~H. {Schultz}, pp. 259--273.

\bibitem[{\textit{{Melosh}}(1977)}]{melosh1977}
{Melosh}, H.~J. (1977), {Global tectonics of a despun planet}, \textit{Icarus},
  \textit{31}, 221--243, \doi{10.1016/0019-1035(77)90035-5}.

\bibitem[{\textit{{Melosh}}(1980{\natexlab{a}})}]{melosh1980a}
{Melosh}, H.~J. (1980{\natexlab{a}}), {Tectonic patterns on a tidally distorted
  planet}, \textit{Icarus}, \textit{43}, 334--337,
  \doi{10.1016/0019-1035(80)90178-5}.

\bibitem[{\textit{{Melosh}}(1980{\natexlab{b}})}]{melosh1980b}
{Melosh}, H.~J. (1980{\natexlab{b}}), {Tectonic patterns on a reoriented planet
  - Mars}, \textit{Icarus}, \textit{44}, 745--751,
  \doi{10.1016/0019-1035(80)90141-4}.

\bibitem[{\textit{{Melosh} and {Dzurisin}}(1978)}]{melosh1978}
{Melosh}, H.~J., and D.~{Dzurisin} (1978), {Mercurian global tectonics - A
  consequence of tidal despinning}, \textit{Icarus}, \textit{35}, 227--236,
  \doi{10.1016/0019-1035(78)90007-6}.

\bibitem[{\textit{{Melosh} and {McKinnon}}(1988)}]{melosh1988}
{Melosh}, H.~J., and W.~B. {McKinnon} (1988), {The tectonics of Mercury}, in
  \textit{Mercury}, edited by {{Vilas}, F. and {Chapman}, C.~R. and {Matthews},
  M.~S.}, pp. 374--400, University of Arizona Press, Tucson.

\bibitem[{\textit{{Melosh} and {Nimmo}}(2009)}]{melosh2009}
{Melosh}, H.~J., and F.~{Nimmo} (2009), {An Intrusive Dike Origin for Iapetus'
  Enigmatic Ridge?}, in \textit{Lunar Planet. Sci. 40}, abstract 2478.

\bibitem[{\textit{{Moore}}(1984)}]{moore1984}
{Moore}, J.~M. (1984), {The tectonic and volcanic history of Dione},
  \textit{Icarus}, \textit{59}, 205--220, \doi{10.1016/0019-1035(84)90024-1}.

\bibitem[{\textit{{Moore} and {Schenk}}(2007)}]{moore2007}
{Moore}, J.~M., and P.~M. {Schenk} (2007), {Topography of Endogenic Features on
  Saturnian Mid-Sized Satellites}, in \textit{Lunar Planet. Sci. 38}, abstract
  2136.

\bibitem[{\textit{{Moore} et~al.}(1985)\textit{{Moore}, {Horner}, and
  {Greeley}}}]{moore1985}
{Moore}, J.~M., V.~M. {Horner}, and R.~{Greeley} (1985), {The geomorphology of
  Rhea: implications for geologic history and surface processes.},
  \textit{J.~Geophys.~Res.}, \textit{90}, C785--C795.

\bibitem[{\textit{{Munk} and {MacDonald}}(1960)}]{munkmacdonald}
{Munk}, W.~H., and G.~J.~F. {MacDonald} (1960), \textit{{The rotation of the
  earth; a geophysical discussion}}, Cambridge University Press, Cambridge.

\bibitem[{\textit{{Murray} and {Dermott}}(1999)}]{murraydermott}
{Murray}, C.~D., and S.~F. {Dermott} (1999), \textit{{Solar System Dynamics}},
  Cambridge University Press, Cambridge.

\bibitem[{\textit{{Nahm} and {Schultz}}(2010)}]{nahm2010}
{Nahm}, A.~L., and R.~A. {Schultz} (2010), {A test of global contraction models
  for Mars using observations}, in \textit{Lunar Planet. Sci. 41}, abstract
  2086.

\bibitem[{\textit{{Nimmo} et~al.}(2007)\textit{{Nimmo}, {Thomas}, {Pappalardo},
  and {Moore}}}]{nimmo2007}
{Nimmo}, F., P.~C. {Thomas}, R.~T. {Pappalardo}, and W.~B. {Moore} (2007), {The
  global shape of Europa: Constraints on lateral shell thickness variations},
  \textit{Icarus}, \textit{191}, 183--192, \doi{10.1016/j.icarus.2007.04.021}.

\bibitem[{\textit{{Nyffenegger} et~al.}(1997)\textit{{Nyffenegger}, {Davis},
  and {Consolmagno}}}]{nyffenegger1997}
{Nyffenegger}, P., D.~M. {Davis}, and G.~J. {Consolmagno} (1997), {Tectonic
  lineations and frictional faulting on a relatively simple body (Ariel)},
  \textit{Planet. Space Sci.}, \textit{45}, 1069--1080.

\bibitem[{\textit{{Ojakangas} and {Stevenson}}(1989)}]{ojakangas1989}
{Ojakangas}, G.~W., and D.~J. {Stevenson} (1989), {Thermal state of an ice
  shell on Europa}, \textit{Icarus}, \textit{81}, 220--241,
  \doi{10.1016/0019-1035(89)90052-3}.

\bibitem[{\textit{{Pappalardo} et~al.}(2004)\textit{{Pappalardo}, {Collins},
  {Head}, {Helfenstein}, {McCord}, {Moore}, {Prockter}, {Schenk}, and
  {Spencer}}}]{pappalardo2004}
{Pappalardo}, R.~T., G.~C. {Collins}, J.~W. {Head}, III, P.~{Helfenstein},
  T.~B. {McCord}, J.~M. {Moore}, L.~M. {Prockter}, P.~M. {Schenk}, and J.~R.
  {Spencer} (2004), {Geology of Ganymede}, in \textit{Jupiter.~The Planet,
  Satellites and Magnetosphere}, edited by F.~{Bagenal}, T.~E. {Dowling}, and
  W.~B. {McKinnon}, pp. 363--396, Cambridge University Press, Cambridge.

\bibitem[{\textit{{Peale}}(1999)}]{peale1999}
{Peale}, S.~J. (1999), {Origin and Evolution of the Natural Satellites},
  \textit{Annu.~Rev.~Astron.~Astrophys.}, \textit{37}, 533--602,
  \doi{10.1146/annurev.astro.37.1.533}.

\bibitem[{\textit{{Pechmann} and {Melosh}}(1979)}]{pechmann1979}
{Pechmann}, J.~B., and H.~J. {Melosh} (1979), {Global fracture patterns of a
  despun planet - Application to Mercury}, \textit{Icarus}, \textit{38},
  243--250, \doi{10.1016/0019-1035(79)90181-7}.

\bibitem[{\textit{{Plescia}}(1987)}]{plescia1987}
{Plescia}, J.~B. (1987), {Geological terrains and crater frequencies on Ariel},
  \textit{Nature}, \textit{327}, 201--204, \doi{10.1038/327201a0}.

\bibitem[{\textit{{Porco} et~al.}(2005)\textit{{Porco}, {Baker}, {Barbara},
  {Beurle}, {Brahic}, {Burns}, {Charnoz}, {Cooper}, {Dawson}, {Del Genio},
  {Denk}, {Dones}, {Dyudina}, {Evans}, {Giese}, {Grazier}, {Helfenstein},
  {Ingersoll}, {Jacobson}, {Johnson}, {McEwen}, {Murray}, {Neukum}, {Owen},
  {Perry}, {Roatsch}, {Spitale}, {Squyres}, {Thomas}, {Tiscareno}, {Turtle},
  {Vasavada}, {Veverka}, {Wagner}, and {West}}}]{porco2005}
{Porco}, C.~C., E.~{Baker}, J.~{Barbara}, K.~{Beurle}, A.~{Brahic}, J.~A.
  {Burns}, S.~{Charnoz}, N.~{Cooper}, D.~D. {Dawson}, A.~D. {Del Genio},
  T.~{Denk}, L.~{Dones}, U.~{Dyudina}, M.~W. {Evans}, B.~{Giese}, K.~{Grazier},
  P.~{Helfenstein}, A.~P. {Ingersoll}, R.~A. {Jacobson}, T.~V. {Johnson},
  A.~{McEwen}, C.~D. {Murray}, G.~{Neukum}, W.~M. {Owen}, J.~{Perry},
  T.~{Roatsch}, J.~{Spitale}, S.~{Squyres}, P.~C. {Thomas}, M.~{Tiscareno},
  E.~{Turtle}, A.~R. {Vasavada}, J.~{Veverka}, R.~{Wagner}, and R.~{West}
  (2005), {Cassini Imaging Science: Initial Results on Phoebe and Iapetus},
  \textit{Science}, \textit{307}, 1237--1242, \doi{10.1126/science.1107981}.

\bibitem[{\textit{{Rivoldini} et~al.}(2009)\textit{{Rivoldini}, {van Hoolst},
  and {Verhoeven}}}]{rivoldini2009}
{Rivoldini}, A., T.~{van Hoolst}, and O.~{Verhoeven} (2009), {The interior
  structure of Mercury and its core sulfur content}, \textit{Icarus},
  \textit{201}, 12--30, \doi{10.1016/j.icarus.2008.12.020}.

\bibitem[{\textit{{Roberts} and {Nimmo}}(2009)}]{roberts2009}
{Roberts}, J.~H., and F.~{Nimmo} (2009), {Tidal Dissipation Due to Despinning
  and the Equatorial Ridge on Iapetus}, in \textit{Lunar Planet. Sci. 40},
  abstract 1927.

\bibitem[{\textit{{Sabadini} and {Vermeersen}}(2004)}]{sabadini}
{Sabadini}, R., and B.~{Vermeersen} (2004), \textit{{Global Dynamics of the
  Earth}}, Kluwer Academic Publishers, Dordrecht.

\bibitem[{\textit{{Schultz} and {Zuber}}(1994)}]{schultz1994}
{Schultz}, R.~A., and M.~T. {Zuber} (1994), {Observations, models, and
  mechanisms of failure of surface rocks surrounding planetary surface loads},
  \textit{J.~Geophys.~Res.}, \textit{99}, 14,691--14,702,
  \doi{10.1029/94JE01140}.

\bibitem[{\textit{{Schultz} et~al.}(2009)\textit{{Schultz}, {Soliva}, {Okubo},
  and {M\`ege}}}]{schultz2009}
{Schultz}, R.~A., R.~{Soliva}, C.~H. {Okubo}, and D.~{M\`ege} (2009), {Fault
  populations}, in \textit{Planetary Tectonics}, edited by {{Watters}, T.~R.
  and {Schultz}, R.~A.}, pp. 457--510, Cambridge University Press, Cambridge.

\bibitem[{\textit{{Singer} and {McKinnon}}(2008)}]{singer2008}
{Singer}, K.~N., and W.~B. {McKinnon} (2008), {A Search for Despinning
  Fractures on Iapetus}, in \textit{Lunar Planet. Sci. 39}, abstract 2415.

\bibitem[{\textit{{Solomon}}(1976)}]{solomon1976}
{Solomon}, S.~C. (1976), {Some aspects of core formation in Mercury},
  \textit{Icarus}, \textit{28}, 509--521, \doi{10.1016/0019-1035(76)90124-X}.

\bibitem[{\textit{{Solomon} et~al.}(2008)\textit{{Solomon}, {McNutt},
  {Watters}, {Lawrence}, {Feldman}, {Head}, {Krimigis}, {Murchie}, {Phillips},
  {Slavin}, and {Zuber}}}]{solomon2008}
{Solomon}, S.~C., R.~L. {McNutt}, T.~R. {Watters}, D.~J. {Lawrence}, W.~C.
  {Feldman}, J.~W. {Head}, S.~M. {Krimigis}, S.~L. {Murchie}, R.~J. {Phillips},
  J.~A. {Slavin}, and M.~T. {Zuber} (2008), {Return to Mercury: A Global
  Perspective on MESSENGER's First Mercury Flyby}, \textit{Science},
  \textit{321}, 59--62, \doi{10.1126/science.1159706}.

\bibitem[{\textit{{Squyres} and {Croft}}(1986)}]{squyres1986}
{Squyres}, S.~W., and S.~K. {Croft} (1986), {The tectonics of icy satellites},
  in \textit{Satellites}, edited by J.~A. {Burns} and M.~S. {Matthews}, pp.
  293--341, University of Arizona Press, Tucson.

\bibitem[{\textit{{Stephan} et~al.}(2010)\textit{{Stephan}, {Jaumann},
  {Wagner}, {Clark}, {Cruikshank}, {Hibbitts}, {Roatsch}, {Hoffmann}, {Brown},
  {Filacchione}, {Buratti}, {Hansen}, {McCord}, {Nicholson}, and
  {Baines}}}]{stephan2010}
{Stephan}, K., R.~{Jaumann}, R.~{Wagner}, R.~N. {Clark}, D.~P. {Cruikshank},
  C.~A. {Hibbitts}, T.~{Roatsch}, H.~{Hoffmann}, R.~H. {Brown},
  G.~{Filacchione}, B.~J. {Buratti}, G.~B. {Hansen}, T.~B. {McCord}, P.~D.
  {Nicholson}, and K.~H. {Baines} (2010), {Dione's spectral and geological
  properties}, \textit{Icarus}, \textit{206}, 631--652,
  \doi{10.1016/j.icarus.2009.07.036}.

\bibitem[{\textit{{Strom} et~al.}(1975)\textit{{Strom}, {Trask}, and
  {Guest}}}]{strom1975}
{Strom}, R.~G., N.~J. {Trask}, and J.~E. {Guest} (1975), {Tectonism and
  volcanism on Mercury}, \textit{J.~Geophys.~Res.}, \textit{80}, 2478--2507,
  \doi{10.1029/JB080i017p02478}.

\bibitem[{\textit{{Thomas} et~al.}(2007)\textit{{Thomas}, {Burns},
  {Helfenstein}, {Squyres}, {Veverka}, {Porco}, {Turtle}, {McEwen}, {Denk},
  {Giese}, {Roatsch}, {Johnson}, and {Jacobson}}}]{thomas2007}
{Thomas}, P.~C., J.~A. {Burns}, P.~{Helfenstein}, S.~{Squyres}, J.~{Veverka},
  C.~{Porco}, E.~P. {Turtle}, A.~{McEwen}, T.~{Denk}, B.~{Giese}, T.~{Roatsch},
  T.~V. {Johnson}, and R.~A. {Jacobson} (2007), {Shapes of the saturnian icy
  satellites and their significance}, \textit{Icarus}, \textit{190}, 573--584,
  \doi{10.1016/j.icarus.2007.03.012}.

\bibitem[{\textit{{Thomas}}(1988)}]{thomas1988a}
{Thomas}, P.~G. (1988), {The tectonic and volcanic history of Rhea as inferred
  from studies of scarps, ridges, troughs, and other lineaments},
  \textit{Icarus}, \textit{74}, 554--567, \doi{10.1016/0019-1035(88)90121-2}.

\bibitem[{\textit{{Thomas} et~al.}(1988)\textit{{Thomas}, {Masson}, and
  {Fleitout}}}]{thomas1988b}
{Thomas}, P.~G., P.~{Masson}, and L.~{Fleitout} (1988), {Tectonic history of
  Mercury}, in \textit{Mercury}, edited by {{Vilas}, F. and {Chapman}, C.~R.
  and {Matthews}, M.~S.}, pp. 401--428, University of Arizona Press, Tucson.

\bibitem[{\textit{{Tobie} et~al.}(2003)\textit{{Tobie}, {Choblet}, and
  {Sotin}}}]{tobie2003}
{Tobie}, G., G.~{Choblet}, and C.~{Sotin} (2003), {Tidally heated convection:
  Constraints on Europa's ice shell thickness}, \textit{J.~Geophys.~Res.},
  \textit{108}, 5124, \doi{10.1029/2003JE002099}.

\bibitem[{\textit{{Turcotte} et~al.}(1981)\textit{{Turcotte}, {Willemann},
  {Haxby}, and {Norberry}}}]{turcotte1981}
{Turcotte}, D.~L., R.~J. {Willemann}, W.~F. {Haxby}, and J.~{Norberry} (1981),
  {Role of membrane stresses in the support of planetary topography},
  \textit{J.~Geophys.~Res.}, \textit{86}, 3951--3959.

\bibitem[{\textit{{Vening-Meinesz}}(1947)}]{vening1947}
{Vening-Meinesz}, F.~A. (1947), {Shear patterns of the Earth's crust},
  \textit{Trans. Amer. Geophys. Union}, \textit{28}, 1--61.

\bibitem[{\textit{{Wagner} et~al.}(2007)\textit{{Wagner}, {Neukum}, {Giese},
  {Roatsch}, and {Wolf}}}]{wagner2007}
{Wagner}, R.~J., G.~{Neukum}, B.~{Giese}, T.~{Roatsch}, and U.~{Wolf} (2007),
  {The Global Geology of Rhea: Preliminary Implications from the Cassini ISS
  Data}, in \textit{Lunar Planet. Sci. 38}, abstract 1958.

\bibitem[{\textit{{Wagner} et~al.}(2009)\textit{{Wagner}, {Neukum}, {Stephan},
  {Roatsch}, {Wolf}, and {Porco}}}]{wagner2009}
{Wagner}, R.~J., G.~{Neukum}, K.~{Stephan}, T.~{Roatsch}, U.~{Wolf}, and C.~C.
  {Porco} (2009), {Stratigraphy of Tectonic Features on Saturn's Satellite
  Dione Derived from Cassini ISS Camera Data}, in \textit{Lunar Planet. Sci.
  40}, abstract 2142.

\bibitem[{\textit{{Wahr} et~al.}(2009)\textit{{Wahr}, {Selvans}, {Mullen},
  {Barr}, {Collins}, {Selvans}, and {Pappalardo}}}]{wahr2009}
{Wahr}, J., Z.~A. {Selvans}, M.~E. {Mullen}, A.~C. {Barr}, G.~C. {Collins},
  M.~M. {Selvans}, and R.~T. {Pappalardo} (2009), {Modeling stresses on
  satellites due to nonsynchronous rotation and orbital eccentricity using
  gravitational potential theory}, \textit{Icarus}, \textit{200}, 188--206,
  \doi{10.1016/j.icarus.2008.11.002}.

\bibitem[{\textit{{Watters} and {Nimmo}}(2009)}]{wattersnimmo2009}
{Watters}, T.~R., and F.~{Nimmo} (2009), {The Tectonics of Mercury}, in
  \textit{Planetary Tectonics}, edited by {{Watters}, T.~R. and {Schultz},
  R.~A.}, pp. 15--80, Cambridge University Press, Cambridge.

\bibitem[{\textit{{Watters} et~al.}(2004)\textit{{Watters}, {Robinson}, {Bina},
  and {Spudis}}}]{watters2004}
{Watters}, T.~R., M.~S. {Robinson}, C.~R. {Bina}, and P.~D. {Spudis} (2004),
  {Thrust faults and the global contraction of Mercury},
  \textit{Geophys.~Res.~Lett.}, \textit{31}, L04701,
  \doi{10.1029/2003GL019171}.

\bibitem[{\textit{{Watters} et~al.}(2009)\textit{{Watters}, {Solomon},
  {Robinson}, {Head}, {Andr\'e}, {Hauck~II}, and {Murchie}}}]{watters2009}
{Watters}, T.~R., S.~C. {Solomon}, M.~S. {Robinson}, J.~W. {Head}, S.~L.
  {Andr\'e}, S.~A. {Hauck~II}, and S.~L. {Murchie} (2009), {The tectonics of
  Mercury: The view after MESSENGER's first flyby},
  \textit{Earth~Planet.~Sci.~Lett.}, \textit{285}, 283--296,
  \doi{10.1016/j.epsl.2009.01.025}.

\bibitem[{\textit{{Whittaker} and {Watson}}(1935)}]{whittaker}
{Whittaker}, E.~T., and G.~N. {Watson} (1935), \textit{{A Course of Modern
  Analysis}}, 4th edition, Cambridge University Press, Cambridge.

\bibitem[{\textit{{Wolfram Research}}(2008)}]{wolfram}
{Wolfram Research}, I. (2008), \textit{{Mathematica, Version 7.0}}, Champaign,
  Illinois.

\end{thebibliography}
\end{document}